%% file: bayescalib.tex
\documentclass{article}
 
\usepackage[
backend=biber,
style=numeric-comp,
date=year,
sorting=none, 
doi=true,
url=false,
isbn=false,
giveninits=true,
maxbibnames=99,
maxcitenames=2,
]{biblatex}
\renewbibmacro{in:}{\adddot\addspace}

\AtEveryBibitem{%
  \ifentrytype{book}{%
    \clearfield{url}%
    \clearfield{urldate}%
  }{}%
}
\AtEveryBibitem{%
  \ifentrytype{incollection}{%
    \clearfield{url}%
    \clearfield{urldate}%
  }{}%
}
\AtEveryBibitem{\clearfield{pagetotal}}
\AtEveryBibitem{%
  \ifentrytype{misc}{}{%
    \clearfield{note}%
  }%
}

\usepackage{color}
\usepackage{xcolor}
\usepackage[colorlinks=true, 
linkcolor=blue, 
urlcolor=blue,
bookmarks=true,
bookmarksopen=true,
bookmarksopenlevel=1,
bookmarksnumbered=true]{hyperref}
\usepackage{graphicx}
\usepackage[utf8]{inputenc}
\usepackage{amsmath}
\usepackage{amsfonts}
\usepackage[a4paper]{geometry}
\usepackage{caption,subfig}
\usepackage{booktabs}
\usepackage{array}
\usepackage{import}
\usepackage[nameinlink]{cleveref}
\usepackage{placeins}

\usepackage{algorithm}
\usepackage{algpseudocode}
\usepackage{soul}

\usepackage{pgfplots}
\pgfplotsset{compat=1.5}
\usepgfplotslibrary[external]

\usepackage{authblk}
\newtheorem{remark}{Remark}

\graphicspath{{./figures/}}

\hypersetup
{
    pdfauthor={Harald Willmann},
    pdfsubject={BayesianIA},
    pdftitle={Bayesian calibration of coupled computational mechanics models under uncertainty based on interface deformation},
    pdfkeywords={inverse anaylsis, biofilm}
}

\input{symbols.tex}

\DeclareUnicodeCharacter{2212}{-}
\newcommand{\keywords}[1]{\textbf{Keywords: }#1}
\newcommand*{\email}[1]{\normalsize\href{mailto:#1}{#1}\par}

\begin{document}

\title{Bayesian calibration of coupled computational mechanics models under uncertainty based on interface deformation}

\author[1]{Harald~Willmann$^{*}$}
\author[1,2]{Jonas~Nitzler}
\author[1,3]{Sebastian~Brandstäter}
\author[1]{Wolfgang~A.~Wall}

\affil[1]{Institute for Computational Mechanics, 
Technical University of Munich\protect\\
Boltzmannstr. 15, 85748 Garching b. München, Germany\protect\\
e-mail: \email{harald.willmann@tum.de}
$^*$ corresponding author}

\affil[2]{Professorship for Data-driven Materials Modeling, Technical University of Munich,\protect\\ Garching b. München, Germany}
\affil[3]{Institute for Continuum and Material Mechanics, University of Technology Hamburg,\protect\\ Hamburg, Germany}

\maketitle

\begin{abstract}
Calibration or parameter identification is used with computational mechanics models related to observed data of the modeled process to find model parameters such that good similarity between model prediction and observation is achieved. We present a Bayesian calibration approach for surface coupled problems in computational mechanics based on measured deformation of an interface when no displacement data of material points is available. The interpretation of such a calibration problem as a statistical inference problem, in contrast to deterministic model calibration, is computationally more robust and allows the analyst to find a posterior distribution over possible solutions rather than a single point estimate. The proposed framework also enables the consideration of unavoidable uncertainties that are present in every experiment and are expected to play an important role in the model calibration process. To mitigate the computational costs of expensive forward model evaluations, we propose to learn the log-likelihood function from a controllable amount of parallel simulation runs using Gaussian process regression. We introduce and specifically study the effect of three different discrepancy measures for deformed interfaces between reference data and simulation. We show that a statistically based discrepancy measure results in the most expressive posterior distribution. We further apply the approach to numerical examples in higher model parameter dimensions and interpret the resulting posterior under uncertainty. In the examples, we investigate coupled multi-physics models of fluid-structure interaction effects in biofilms and find that the model parameters affect the results in a coupled manner.
\end{abstract}


\section{Introduction}

In this article we present a robust approach for Bayesian calibration \cite{inv_bilionis2013,inv_Kennedy_2001} of coupled computational mechanical models based on the deformation of an interface or boundary.
In the most general case, the search for a set of parameters leading to a desired model result can be understood as an \emph{inverse problem} \cite{inv_tarantola_2005}. Basic elements are a computational mechanics model $\forwardmodelletter$, also called the \emph{forward model}, with model parameters $\paramvec$ that are considered as inputs to the forward problem. The model parameters can be prescribed in the model and they are expected to significantly influence the associated model response. The inverse problem is characterized by the task to find one or multiple model parameters that result in a desired model behavior. In this paper, we are interested in a special case of desired model behavior which is given in form of observed experimental data. Here, we want to find suitable model parameters such that the model response is close to the experimental data in a given metric. This category of inverse problems is also known as a \emph{calibration} or \emph{parameter identification} problem.

We want to distinguish two different viewpoints on the calibration problem. We refer to the first one as the \emph{deterministic} calibration approach, which poses the calibration task as an optimization problem by minimizing a discrepancy function over the forward model parameters, between the forward system response and the experimental data. Following the deterministic optimization approach, the problems are often ill-posed. The second viewpoint is the probabilistic Bayesian calibration approach, that we follow in this article. In contrast to the aforementioned optimization, the Bayesian approach adopts a statistical viewpoint. It seeks the \emph{posterior} probability density for the input parameters. This density quantifies the probability of resulting forward model system outputs to match the experimental data best, in a specified norm. Instead of a single point estimate as in the optimization problem, the posterior distribution returns the unique probability density for all inputs in the input space. This allows to answer a plethora of additional research questions.
A Bayesian, statistical viewpoint does not only provide a powerful mathematical framework for the formulation of the inverse problem but also helps with the design and interpretation of very flexible discrepancy measures between simulation output and experimental observation. Those can be formulated in form of reproducing kernel Hilbert space (RKHS) norms as demonstrated in this article.
The Bayesian setting allows for the incorporation of available prior knowledge which is especially advantageous in the small data regime, induced by expensive simulation runs or limited experimental data. The Bayesian formulation provides a consistent mathematical framework that can naturally deal with different sources of uncertainty such as they might arise from partially unknown experimental conditions, as also studied in this work.

The scenario where a mechanical structure changes its shape under mechanical load, but no displacements of individual material points can be determined is the focus of the presented approach. Such scenarios appear when only the shape and changes in shape of a structure can be observed, e.g., in the form of image data. In that case only information about the shape of a boundary or interface is reliably accessible, without further details on correspondence of material points in simulation and experimental data. For such scenarios that have been studied before \cite{inv_Moireau_2009} for, e.g., cardiac mechanics \cite{Sermesant_2006} or arterial growth \cite{lnm_inv_kehl2016}, we want to investigate and discuss the effect of different definitions for discrepancy measures between simulated and observed interface deformations of objects. Especially regarding bio-materials the main interest is to determine material properties as they act in-situ, i.e., in the natural environment. Traditional material testing often defines standardized testing methods where the specimen must be isolated and installed in a specific testing device. For very sensitive materials the isolation of a specimen can already change the properties of the material of interest significantly. Often such scenarios occur with coupled physics as, e.g., fluid-structure interaction (FSI) problems, as an isolation of the specimen would interfere with the coupling. That is why a comparison between deformations in the computational model and an observed deformation in an in-situ experiment is required to test such sensitive materials.

While the presented approach will be useful for any kind of (coupled) mechanical model, our focus will be on the particularly challenging problem class of \emph{fluid-structure interaction} (FSI). A specific motivation for us is our research on biofilms and respective experiments conducted with such biofilms. Biofilms grow with aggregates of microorganisms that form a structure of extracellular polymeric substances to withstand environmental influences. This is also known as the biofilm matrix \cite{bio_flemming2010}. Amongst others due to its soft consistency, the determination of mechanical properties of biofilms is an open field of research and different intrusive and non-intrusive attempts have been made to quantify the material behavior \cite{bio_boel2013, bio_gloag2020, bio_boudarel2018}. The better understanding and analysis of biofilm material properties is essential to better explain biofilm behavior and to enable the development of reliable and predictive computational modeling. Well parameterized mechanical models enable engineers to make valid predictions of biofilm behavior, e.g., deformation, growth or erosion and use those to develop biofilm-prone systems. This means to either avoid invasive biofilms or improve productive biofilm systems. Biofilms usually develop on surfaces exposed to fluid flows and therefore a mechanical model must include the FSI between biofilm surface and the fluid. As the capability of computational mechanical models of fluid-solid interaction increases, the inclusion of such models in deterministic inverse analysis of biofilms has emerged in recent years \cite{kit_bio_picioreanu2018, me_willmann_2021}. One of the best approaches to acquire image data of biofilms is via optical coherence tomography (OCT) in flow cell experiments, as used in, e.g., \cite{kit_bio_wagner2010, kit_bio_blauert2015, kit_bio_picioreanu2018}. This type of experiments is favorable in means of mechanical testing as it is non-destructive and the biofilm can be kept in the same environment for the whole cultivation and test process. Recent advances in automated biofilm cultivation and design of flow cell experiments \cite{kit_bio_gierl2020} have already shown that a variety in biofilm shapes is inevitable even for reproducible environmental conditions and therefore a flexible method of comparing biofilm shapes is required for conclusive inverse analysis. Recently, Bayesian estimation and uncertainty quantification (UQ) have been used for models of urea hydrolysis by biofilms \cite{inv_jackson2021}.

The rest of the article is structured as follows. In \cref{sec:bayes}, the theoretical concepts of the Bayesian approach for an efficient model calibration under uncertainty will be presented. The technical details and realization of our approach are then outlined in \cref{sec:numerical_realization}. Eventually, we demonstrate and discuss the calibration procedure of numerical models in examples of biofilm motivated fluid-solid interaction models for generated data in \cref{sec:examples} in two to six input dimensions. Results and key aspects of the workflow are then concluded in \cref{sec:concout}.

\section{Continuous formulation of the Bayesian calibration problem under uncertainty}
\label{sec:bayes}

Our Bayesian calibration approach is based on the continuous formulation of Bayes' rule. Therefore first the general formulation of the Bayesian calibration is introduced and subsequently extended to include the effects of uncertainties. As a necessary step, the formulation of a likelihood model is described and specific choices for the measure of discrepancy between interface shapes are introduced.

\subsection{Standard formulation for Bayesian calibration}
\label{sec:standard_bayes}
We assume that we have a computational forward model $\model{\paramvec,\coords}$ for a real-world process, i.e., in our case a single- or multi-field continuum mechanical problem, whose response depends on the choice of inputs $\paramvec$ and the choice of (e.g., spatio-temporal) coordinates $\coords$. In general, observations are compared in more than one location and point in time and therefore the coordinates are written as a matrix $\coords$ with vector entries for each comparison. Consequently, the observations for all coordinates are a matrix $\yobs$ as well with vector entries for each coordinate vector. In our case for the comparison of interface deformations, the observations are locations of the interface that define its shape for one or more points in time. The input parameters $\paramvec$ are the quantities of interest in the calibration process. The vector $\paramvec$ denotes a collection of model parameters that are subject to calibration and might represent, e.g., material parameters, boundary or initial conditions. The observed data $\yobs$ might additionally be subject to unknown measurement noise.

In the Bayesian framework, the calibration problem is interpreted as an update of a prior belief about the parameters, encoded by a \emph{prior density} $\prior$, by a so-called \emph{likelihood model} $\pdenscond{\yobs}{\model{\paramvec,\coords}}$. The likelihood model expresses the probability density to observe the experimental data $\yobs$ at coordinate $\coords$ given a specific choice of model with input $\paramvec$ evaluated at the same coordinates $\coords$. As the expression $\pdenscond{\yobs}{\model{\paramvec,\coords}}$ is only a valid density in $\yobs$ but not in the model inputs $\paramvec$ one mostly refers to it as the so-called \emph{likelihood function} (see \cref{sec:likelihood_model} for more details).
The likelihood function relates the output $\model{\paramvec,\coords}$ of the computational model $\forwardmodelletter$ with the observation $\yobs$, given a specific choice of parameters $\paramvec$. The value of $\pdenscond{\yobs}{\model{\paramvec,\coords}}$ at a specific $\paramvec$ can be interpreted as the probability density of the observations $\yobs$ for the given model choice $\model{\paramvec,\coords}$ or as a score value for the parameter choice $\paramvec$. The product of the likelihood function and the prior can be evaluated point-wise in the parameter space $\domain_{\paramvec}$, yielding a function that we call the \emph{unnormalised posterior} $\pdenscond{\yobs}{\model{\paramvec,\coords}}\prior$. Normalizing this expression by $\int \pdenscond{\yobs}{\model{\paramvec,\coords}}\pdens{\paramvec} \diffd\paramvec$ such that we get a valid density that integrates to one, yields the \emph{posterior} distribution $\pdenscond{\paramvec}{\yobs}$. The posterior distribution can be interpreted as an updated prior distribution, after the knowledge of the experimental data $\yobs$ has been incorporated and related to the forward model $\model{\paramvec,\coords}$. 

An advantage of the Bayesian viewpoint on calibration is the possibility to encode prior knowledge of the unknown parameters or inputs $\paramvec$ in the so-called \emph{prior} distribution $\prior$. Prior knowledge is information about the parameters that is available before seeing the data. Often, at least a vague understanding about which values are possible or realistic is available (e.g., the Young's modulus must be positive $\youngs > 0\sunit{Pa} $). This knowledge and additional valuable expert knowledge can be incorporated as prior and therefore the solution of the calibration complies with it.

We can then pose the calibration problem as a Bayesian inference task by applying Bayes' rule \cite{stat_robert2007} 
\begin{equation}
\underbrace{\pdenscond{\paramvec}{\yobs}}_{\text{posterior}} = \frac{\overbrace{\pdenscond{\yobs}{\model{\paramvec,\coords}}}^{\text{likelihood}}\overbrace{\prior}^{\text{prior}}}{\underbrace{\int \pdenscond{\yobs}{\model{\paramvec,\coords}}\pdens{\paramvec} \diffd\paramvec}_{\text{evidence}}}\propto\underbrace{\pdenscond{\yobs}{\model{\paramvec,\coords}}\prior}_{\text{unnormalized posterior}}.
\label{eq:bayes}
\end{equation}

In \eqref{eq:bayes}, the \emph{posterior} $\pdenscond{\paramvec}{\yobs} $ represents the unique solution of the Bayesian calibration task in the form of a probability density. It is a probability distribution over the input parameters $\paramvec$ assigning a probability density to each input vector of how well the forward model $\model{\paramvec,\coords}$ evaluated with that model parameter combination represents the observation. The interest of the analyst is to find high posterior values and learn for which inputs they occur. 
The posterior density is usually not known in closed form, due to the implicit dependency on $\paramvec$ in the forward model $\model{\paramvec,\coords}$ within the likelihood function. Nevertheless, the posterior density can be evaluated point-wise, which implies a forward simulation run for the particular choice of $\paramvec$. 

For computationally expensive models, a grid-based evaluation of the posterior in the entire input space $\domain_{\paramvec}$ is unfeasible. Due to the curse of dimensionality, this problem becomes especially amplified for $\dim{(\paramvec)} \gg 1$. The numerical approximation of the posterior is hence usually conducted using more advanced algorithms which aim to exploit regions in the input space with high posterior density. In this work, we use the sequential Monte Carlo (SMC) method for this purpose but postpone a more detailed discussion of that algorithm and its numerical realization to \cref{sec:numerical_realization} and specifically \cref{sec:smc}, to first focus on the continuous presentation of the Bayesian calibration problem.

The denominator $\int \pdenscond{\yobs}{\model{\paramvec,\coords}}\pdens{\paramvec} \diffd\paramvec$, respectively \emph{evidence} in \eqref{eq:bayes} acts as a normalizing constant for the posterior, such that it becomes a valid density function in $\paramvec$ which integrates to one on $\domain_{\paramvec}$. Nevertheless, the evidence is mostly not computed explicitly, as it involves potentially high dimensional integration over $\domain_{\paramvec}$. Consequently, most numerical algorithms operate on the \emph{unnormalized posterior} or its logarithm and take care of the normalization in an easier to compute post-processing step (see \cref{sec:smc}).
Given the posterior distribution or a numerical representation in form of samples or an approximating distribution, further statistics or point estimates can be calculated and derived.

\begin{remark}[Marginalization]

Sometimes one might be interested in the density over a subset of variables, averaging over the remaining parameters. This can be achieved by so-called \emph{marginalization}. A \emph{marginal} represent a projection of the high-dimensional density, e.g., $\pdens{\paramvec_i,\paramvec_j}$, for a selected subset of parameters $\paramvec_i$, and reflects the average effect of all other parameters $\paramvec_j$ on the density over parameter subset $\paramvec_i$. The marginal distribution is then expressed by the following integration 
\begin{equation}
\pdens{\paramvec_i} = \int \pdens{\paramvec_i,\paramvec_j} \diffd \paramvec_j = \int \pdenscond{\paramvec_i}{\paramvec_j}\pdens{\paramvec_j}\diffd \paramvec_j = \Expectwrt{\paramvec_j}{\pdenscond{\paramvec_i}{\paramvec_j}}.
\label{eq:marginal}
\end{equation}
One example are projections of a higher-dimensional posterior $\pdens{\paramvec_i,\paramvec_j| \yobs}$ to a marginal posterior $\pdens{\paramvec_i| \yobs}$ such that the analyst can investigate the posterior in $\paramvec_i$ averaged over the effect of $\paramvec_j$.
This enables us to regard and print marginal posterior distributions with $i < 3 $ in result plots.

\end{remark}

\subsection{Bayesian calibration under uncertainty}
\label{sec:calibration_uncertainty}
After the basic concepts of Bayesian calibration have been presented, we want to extend the ideas to the case of non-controllable, uncertain conditions that influence the model $\forwardmodelletter$. Those are summarized in the vector $\btheta$. We assume that $\btheta$ is not part of the model input variables $\paramvec$ that we want to calibrate. Instead, it represents inherently uncertain external conditions, e.g., of the experimental set-up, that we cannot fully control at the time of the analysis. Furthermore, we assume that these conditions are subject to uncertainties expressed by a distribution $\pdens{\btheta}$ and that the computational forward model is also dependent on $\btheta$ in the sense of $\model{\paramvec,\btheta,\coords}$.

The calibration problem under uncertainty can then be formulated in two steps \cite{stat_sternfels2011, inv_Koutsourelakis_2008}. First, we naively compose a Bayesian calibration problem in analogy to \eqref{eq:bayes}, with the only difference that the model is also dependent on $\btheta$. Consequently, the posterior $\pdenscond{\paramvec}{\btheta,\yobs}$ is conditionally dependent on $\btheta$, as demonstrated in \eqref{eq:cond_bayes}. In a second step, we can then average over the effect of the uncertain conditions $\btheta$ by taking the expectation of the previous posterior $\Expectwrt{\btheta}{\pdenscond{\paramvec}{\btheta,\yobs}}$ with respect to the density $p\left(\btheta\right)$ as shown in \eqref{eq:uncertain_bayes}. The resulting modified posterior, which accounts for the average effect of the additional uncertainty introduced by $\btheta$ is then denoted by $q\left(\paramvec|\yobs\right)$ and is different from the former posterior $\pdenscond{\paramvec}{\yobs}$ which did not incorporate these additional uncertainties.

\begin{subequations}
\begin{align}
\label{eq:cond_bayes}
\pdenscond{\paramvec}{\btheta,\yobs} &\propto\pdenscond{\yobs}{\model{\paramvec,\btheta,\coords}}\prior\\
\label{eq:uncertain_bayes}
\begin{split}
q\left(\paramvec|\yobs\right)&=\Expectwrt{\btheta}{\pdenscond{\paramvec}{\btheta,\yobs}}=\int\underbrace{\pdenscond{\paramvec}{\btheta,\yobs} \pdens{\btheta}}_{\substack{\text{extended posterior:}\\ \substack{ }\pdenscond{\paramvec,\btheta}{\yobs}}} \diffd \btheta\\
&\propto\int\underbrace{\pdenscond{\yobs}{\model{\paramvec,\btheta,\coords}} \prior \pdens{\btheta}}_{\text{unnormalized extended posterior}} \diffd \btheta
\end{split}
\end{align}
\end{subequations}

Another interpretation of \eqref{eq:uncertain_bayes} is the marginalization of the \emph{extended posterior} $\pdenscond{\paramvec,\btheta}{\yobs}$ with respect to the uncertain conditions $\btheta$. Later-on, we show that the sequential Monte Carlo (SMC) algorithm allows simple operations on the \emph{unnormalized extended posterior}, analogous to the \emph{unnormalized posterior} from \eqref{eq:bayes} if no additional uncertainties are present. SMC offers the possibility to conduct the necessary marginalization of $\btheta$ as a cheap post-processing step.

\begin{remark}[Point estimates and moments of the posterior]

Given a potentially high dimensional and complex posterior density, there is often the desire to represent its characteristics by simpler, e.g., scalar quantities.
The most intuitive approach, especially coming from the mindset of deterministic optimization, is to look for the maximum a posteriori (MAP) estimate. It represents the combination of parameters that lead to the highest posterior density. Further, the maximum likelihood estimate (ML) is interesting to isolate the model feedback from prior assumptions. It is analog to the MAP for the assumptions of uniform priors.  Also statistics of the posterior as the posterior mean (PM) and variance can be used for its simplified quantification. Other than point estimates, also a region of values with highest posterior density can be determined. This region is then characterized by holding a certain probability mass fraction of the posterior and is often called \emph{percentile}.

\end{remark}

\subsection{Selecting a likelihood model}
\label{sec:likelihood_model}

The evaluation of the likelihood function, respectively conditional density $\pdenscond{\yobs}{\model{\paramvec,\btheta,\coords}}$ represents the computationally expensive part of the Bayesian calibration as a forward simulation run has to be conducted for every evaluation of the likelihood with respect to $\model{\paramvec,\btheta,\coords}$. The likelihood is a probabilistic model for the discrepancy $\dmeas$ between experimental data $\yobs$ and forward model output $\model{\paramvec,\btheta,\coords}$, written as $\dmeas=\dmeas(\yobs, \model{\paramvec, \btheta,\coords})$.
It usually models the statistical behavior of experimental noise and modeling errors. It returns the probability density of the experimental data for a given choice of simulation model. This means that it is centered around the forward model results. Usually, only one experiment is observed.

While many different likelihood models exist \cite{inv_kaipio2004, stat_bishop2011}, we chose the common case of a conditionally independent Gaussian likelihood model with static noise-variance $\sigman^2$. This choice implies that we assume that the scattering of the measured data $\yobs$ can be well-explained by a normal distribution which is centered around the simulation output $\ymod$ with variance $\sigman^2$. Here conditional independence means that the measured noise in $y_{\text{obs},c_1}$ does not influence the noise in $y_{\text{obs},c_2}$, with $c_1$ and $c_2$ being two different coordinates.

The conditionally independent Gaussian static likelihood model is given by

\begin{equation}
\pdenscond{\yobs}{\model{\paramvec, \btheta, \coords}} = \frac{1}{\sqrt{\left(2 \pi \sigman^2\right)^n}} \flatexp{-\frac{\dmeas^2\left(\model{\paramvec,\btheta,\coords},\yobs\right)}{2 \sigman ^2}},
\label{eq:staticlikeli}
\end{equation}
wherein $n$ is the dimension of the measurement $\yobs$. In the simplest case it is the total number of individual measurements.
As most inverse iterators (see \cref{sec:numerical_realization}) operate on the logarithm of the densities for better numerical condition, we also provide the logarithmic version of \eqref{eq:staticlikeli}, which is often abbreviated by $\mathcal{L}(\paramvec,\btheta)$ resulting in
\begin{equation}
\mathcal{L}(\paramvec,\btheta) = - \frac{n}{2}\log(2\pi\sigman^2) - \frac{\dmeas^2}{2\sigman^2}.
\label{eq:loglik}
\end{equation}

\subsection{Discrepancy measures between interfaces}
\label{sec:measure}

In selecting a likelihood model the immediate question arises about how to define a discrepancy measure $\dmeas$ between the forward model and experimental results. This becomes especially intricate when the experimental images do not provide real displacements for individual material points, i.e., in this sense prohibit a point-wise correspondence of the geometrical objects. This constraint almost only allows for comparisons of edges, boundaries and interfaces in the image data that can usually be identified depending on their contrast. This step is called \emph{segmentation} of the image data. More detailed information on segmentation approaches etc. are out of scope for this work. As a result of the segmentation process one obtains a geometric representation of interfaces between different subdomains in the images.

\begin{figure}[htb!]
\label{fig:different_measures}
\centering
\subfloat[\label{subfig:sketchmp}]{
\includegraphics[width=0.3\textwidth]{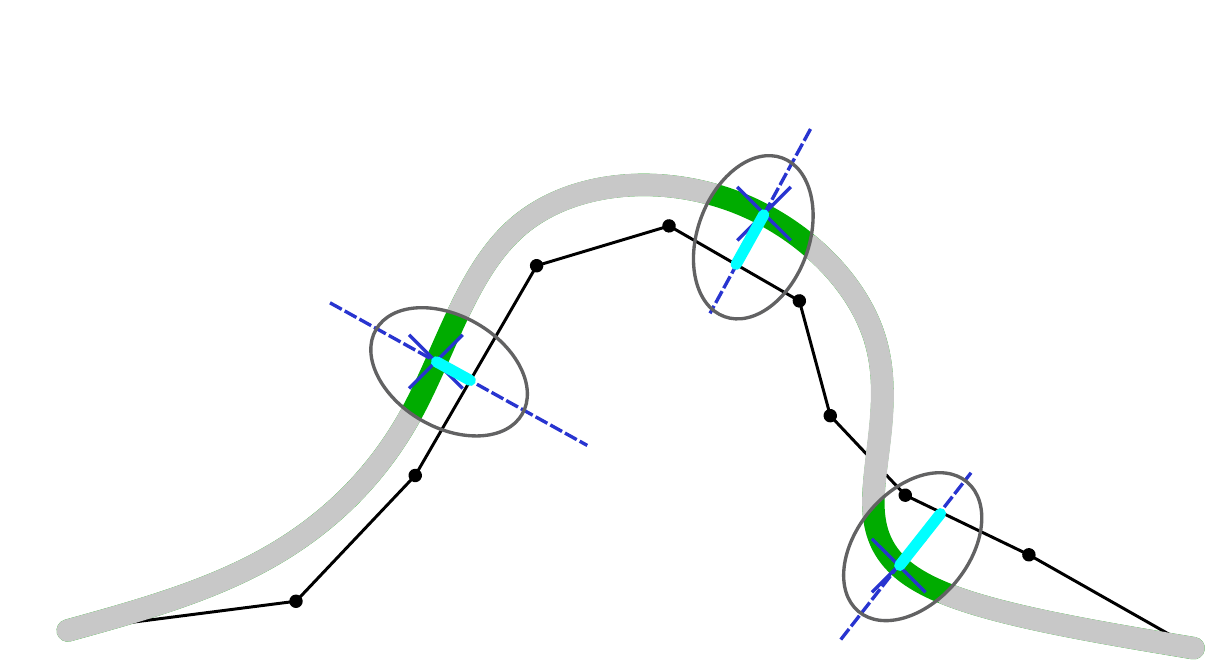}}
\hfill
\subfloat[\label{subfig:sketchcpp}]{
\includegraphics[width=0.3\textwidth]{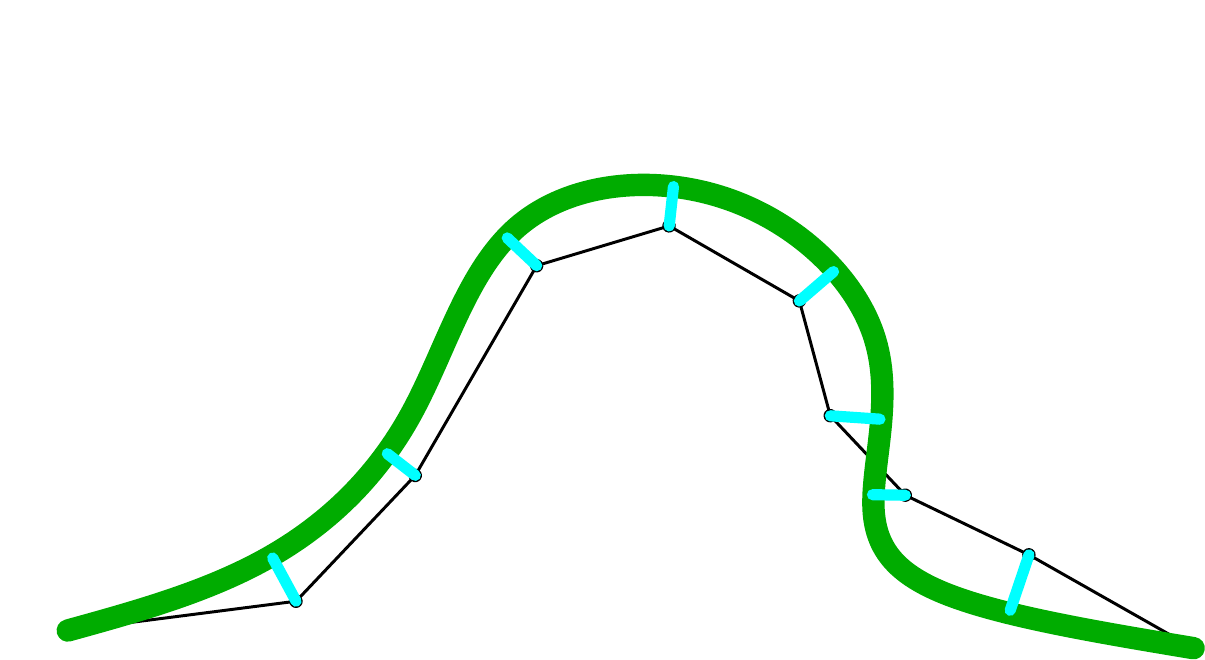}}
\hfill
\subfloat[\label{subfig:sketchsc}]{
\includegraphics[width=0.3\textwidth]{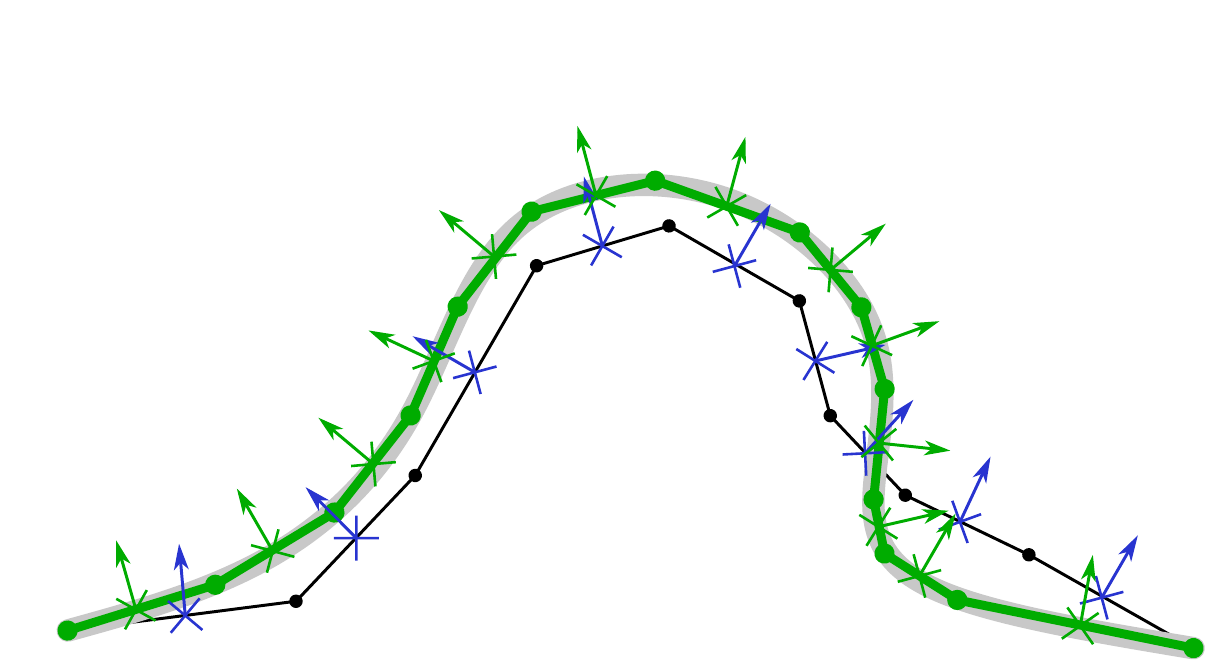}}
\caption{Exemplary 2D sketches for different types of discrepancy measures between experiment (green) and discretized forward model (black). Portions of the observed interface that are accounted for are given in green and unconsidered portions are given in gray. (a) Euclidean distances $\signeddist_\mpindex^i$ at measurement points in preset directions in light blue, (b) closest point projection distances $ \signeddist_\cppindex^i $ for all nodes in light blue, (c) triangulation centers $\tricenter_i$ as crosses and normal vectors $\trinormal_i $ of equal length as arrows for triangulations of the experimentally observed interface in green and the forward model result in blue. Detailed explanation in text below.}
\label{fig:sketchessurfdist}
\end{figure}

In the following, we present three different discrepancy measures that can be used when working with such kind of image data, namely \emph{Euclidean distance at measurement points}, \emph{closest point projection distance} and \emph{reproducing kernel Hilbert space norm}. \Cref{fig:sketchessurfdist} shows sketches of the discussed distance measure definitions in 2D. The sketches in \Cref{fig:sketchessurfdist} are related to the type of problem, of fluid-biofilm interaction, exemplarily analyzed in the numerical examples in \cref{sec:examples}. The methods are not bound to this type of problem and can be applied to different topologies or isolated shape characteristics.

\FloatBarrier
\paragraph{Euclidean distance at measurement points}
\label{par:shape}

The first approach is an Euclidean distance measure as presented in \cite{me_willmann_2021}. At several selected points on the experimentally observed biofilm surface, specific directions are chosen and the distance to the corresponding deformed interface resulting from the forward model simulation is measured. As discussed in \cite{me_willmann_2021} measurement points should be selected in regions of significant, characteristic displacements. Measurement directions should be chosen normal to the observed interface. (See \Cref{subfig:sketchmp} with predefined directions in blue.) These distances are summarized in a vector of length $\lmnmp$. They represent the distance measure between the forward model evaluation and the experimental observation. 
The resulting distance measure $\dmeas_\mpindex$ is then defined via the $\text{L}_2$-norm of the distance vector
\begin{equation}
\label{eq:eucliddist}
\dmeas_\mpindex = \norm{\begin{bmatrix}
\signeddist_\mpindex^1 \\ \vdots \\ \signeddist^\lmnmp_\mpindex
\end{bmatrix}}{2}.
\end{equation}
This procedure is especially well suited if the observed experimental measurements have different reliability throughout the regarded interface, e.g., due to the particular physics or imaging peculiarities as only a point-wise and no full representation of the interface must be measured (i.e., the gray parts in \Cref{subfig:sketchmp} are not part of the analysis). By a meaningful selection of measurement points, where significant deformation takes place and the data is trusted, the analyst has direct control over which data is being processed. 

\paragraph{Closest point projection}
\label{par:pmap}

Another approach for a distance between curves or surfaces is the closest point projection distance for a selected number of points on the interface in the discretized forward model. In the case of finite element models, a suitable choice for the points are the Gauss points or the mesh nodes, as pictured and used here, on the regarded interface. The closest point projection distance to the experimental result is determined based on the segmented image of the interface, as shown in \Cref{subfig:sketchcpp}, and used in the distance vector with the length of the number of interface nodes $\nin$. Afterwards, we define the $\text{L}_2$-norm of the distance vector as the distance measure $\dmeas_\cppindex$ with
\begin{equation}
\label{eq:cppdist}
\dmeas_\cppindex = \norm{\begin{bmatrix}\signeddist_\cppindex^1 \\ \vdots \\ \signeddist^\nin_\cppindex
\end{bmatrix}}{2}.
\end{equation}
We choose the discretization of the finite element model and define the distance vector then by computing the distance of the closest point projection w.r.t. the experimental surface or interface.

\paragraph{Inner product in reproducing kernel Hilbert space (RKHS)}
\label{par:surfcurr}

A third, statistically motivated discrepancy measure is formulated as a \emph{reproducing kernel Hilbert space (RKHS)} norm. It does not only take into account the location of discretization points, but also the orientation of the surface elements in space. Before presenting our specific variant, a more general mathematical foundation of the measure is outlined in the following.

Distance measures $\dmeas$ between two curves (or surfaces) $\boldsymbol{f}_1,\boldsymbol{f}_2$, can be elegantly defined by an inner product of the distance function $\boldsymbol{d_f}(\boldsymbol{s})=\boldsymbol{f}_1-\boldsymbol{f}_2$ in an associated Hilbert space $\mathcal{H}$ as demonstrated in \eqref{eq:inner_prod_hilbert}. Simply speaking, a Hilbert space is a vector space and can be seen as the natural extension of the Euclidean space to arbitrary dimensions. The inner product directly induces a norm according to \eqref{eq:hilbert_norm}.

\begin{subequations}
\begin{align}
\label{eq:inner_prod_hilbert}
    \langle \boldsymbol{d_f}(\boldsymbol{s}),\boldsymbol{d_f}(\boldsymbol{s})\rangle_{\mathcal{H}}&=\int \boldsymbol{d_f}(\boldsymbol{s})\boldsymbol{w}(\boldsymbol{s})\boldsymbol{d_f}(\boldsymbol{s})d\boldsymbol{s}\\
\label{eq:hilbert_norm}
\dmeas_{\mathcal{H}}&=\norm{\boldsymbol{d_f}(\boldsymbol{s})}{\mathcal{H}}=\sqrt{\langle \boldsymbol{d_f}(\boldsymbol{s}),\boldsymbol{d_f}(\boldsymbol{s})\rangle_{\mathcal{H}}}
\end{align}
\end{subequations}
Here, $\boldsymbol{w}(\boldsymbol{s})>0$ is an optional weighting function.
A special case of a Hilbert space $\mathcal{H}$ is the so-called \emph{reproducing kernel Hilbert space (RKHS)}, here denoted by $\mathcal{V}$. We omit a full description of RKHSs and only present the most important concepts in this work but point the interested reader to \cite{stat_aronszajn1950} for a comprehensive derivation and description of the latter. Some further details of the RKHS framework are presented in \cref{asec:rkhsdetails}. An RKHS $\mathcal{V}$ is a Hilbert space that is equipped with a \emph{reproducing kernel} $\boldsymbol{k}(\boldsymbol{x},\boldsymbol{y})$ and the inner product

\begin{subequations}
\begin{align}
\label{eq:inner_prod_rkhs}
    \langle \boldsymbol{d_f}(\boldsymbol{x}),\boldsymbol{d_f}(\boldsymbol{x})\rangle_{\mathcal{V}}&=\int \boldsymbol{d_f}(\boldsymbol{x})\boldsymbol{k}(\boldsymbol{x},\boldsymbol{x}')\boldsymbol{d_f}(\boldsymbol{x}')d\boldsymbol{x}d\boldsymbol{x}'\\
\label{eq:rkhs_norm}
 \dmeas_{\mathcal{V}}&=\norm{\boldsymbol{d_f}(\boldsymbol{x})}{\mathcal{V}}=\sqrt{\langle \boldsymbol{d_f}(\boldsymbol{x}),\boldsymbol{d_f}(\boldsymbol{x})\rangle_{\mathcal{V}}}.
\end{align}
\end{subequations}

The inner product of the difference in the normal vectors of two functions describing two interfaces in \eqref{eq:surface_current} results in a sensitive distance measure that also accounts for the orientation of the interfaces. The distance of the function values is encoded in the kernel function which expresses the statistical correlation between the functions.
For the kernel we choose the radial basis function (RBF) in \eqref{eq:surface_current} which takes the location vectors, respectively the function values $\bs{f}(\bs{s})$ as arguments and returns their correlation. The expression is also known under the term \emph{surface currents} \cite{inv_vaillant2005} and was already successfully applied for inverse analysis in \cite{lnm_inv_kehl2016, inv_imperiale_2013}
\begin{subequations}
\begin{align}
    \bs{d_n}(\bs{s})&=\bs{n}_{\bs{f}_1}(\bs{s})-\bs{n}_{\bs{f}_2}(\bs{s})\\
    \begin{split}
    \dmeas_{\mathcal{V},\scindex}&=\sqrt{\langle \bs{d_n},\bs{d_n} \rangle_{\mathcal{V}}},\\
    \text{with } \kernelscalar{\bs{f}_{1}}{\bs{f}_{2}}&=\flatexp{- \frac{\norm{\bs{d_f}}{2}^2}{2\sigmaw^2}}
    \end{split}
    \label{eq:surface_current}
\end{align}
\end{subequations}
We want to highlight that \eqref{eq:inner_prod_rkhs} allows for many other definitions of $\bs{d}$ and $\bs{k}$ which can be used to emphasize special features of the investigated data. Further examples would be outside the scope of this article. But they might be interesting to emphasize other special features of the investigated data.

The discrete representation of \eqref{eq:surface_current}, e.g., in a finite element simulation, for surfaces or curves are expressed in form of respective elements and meshes. In the simplest case those are interface triangles, resulting from linear tetrahedral elements in 3D, or interface lines, resulting from linear elements in the two-dimensional case. The interface triangles or lines have normal vectors $\bs{n}_{e,\bs{f}_1,i}$,$\bs{n}_{e,\bs{f}_2,j}$ and element-surface centers $\bs{c}_{e,\bs{f}_1,i}$ and $\bs{c}_{e,\bs{f}_2,j}$. Normals and center points can be directly determined from the discretization. The difference vector of the two normal vectors is written as ${\bs{d}_{\bs{n},e}=\bs{n}_{e,\bs{f}_1,i}-\bs{n}_{e,\bs{f}_2,j}}$. The approximation for \eqref{eq:surface_current} follows as a double sum over the elements from $\bs{f}_1$ and elements from $\bs{f}_2$
\begin{align}
\begin{split}
\dmeas_{\mathcal{V}, \scindex}^2\approx&\langle \bs{d}_{\bs{n},e},\bs{d}_{\bs{n},e} \rangle_{\mathcal{V}} = \sum_{i=1} \sum_{j=1} \big[\trinormal_{e,\bs{f}_1,i} \cdot \kernelscalar{\tns{c}_{e,\bs{f}_1,i}}{\tns{c}_{e,\bs{f}_2,j}} \trinormal_{e,\bs{f}_2,j}\\
&- 2 \trinormal_{e,\bs{f}_1,i} \cdot \kernelscalar{\tns{c}_{e,\bs{f}_1,i}}{\tns{c}_{e,\bs{f}_2,j}} \trinormal_{e,\bs{f}_2,j}
+ \trinormal_{e,\bs{f}_1,i} \cdot \kernelscalar{\tns{c}_{e,\bs{f}_1,i}}{\tns{c}_{e,\bs{f}_2,j}} \trinormal_{e,\bs{f}_2,j}\big]
\end{split}
\label{eq:surfcurrontriang}
\end{align}

Using the mechanism of an RKHS for a definition of a surface distance measure has some advantages, that we want to emphasize. First, such a measure allows comparing the whole geometry rather than distances at selected points. It is further a more flexible approach with exchangeable kernel and therefore associated RKHS. It provides a solid mathematical foundation and still gives the analyst the flexibility of choosing an appropriate kernel that will emphasize geometrical features of choice. The kernel parameters that can be referred to as hyperparameters can be made subject to optimization in the calibration approach and therefore the RKHS approach allows seamless integration into a probabilistic framework.

\begin{remark}[RKHS interpretation]
The interpretation of a distance measure as an RKHS with respective norm allows a conclusive mathematical foundation. In fact also the Euclidean distance and the closest point projection distance can be interpreted as different RKHS. These have very specific kernels. Therefore, their re-interpretation is omitted. Generally, a multitude of RKHSs with respective kernels can be designed in an elegant mathematical manner to emphasize different characteristics of the image. One example could be to use a weighting depending on the distance to the closest Dirichlet boundary condition.
\end{remark}

\begin{remark}[Three-dimensional geometry]
Although we will only present two-dimensional applications, the distance measures are equivalently applicable to three-dimensional use cases. As the evaluation of the similarity measures is the only step where geometry is evaluated, this generalization to 3D is obviously also true for the whole approach presented in this work. If accurate three-dimensional images are available from the experiment it would be preferable to use those to increase model accuracy and therefore the quality of the inverse analysis.
\end{remark}
\section{Bayesian calibration -- realization and algorithmic aspects}
\label{sec:numerical_realization}

After the mathematical basis for the inverse problem was presented in \cref{sec:bayes}, we elaborate on the realization and algorithmic aspects as well as computational efficiency of the proposed approach. Several algorithms exist to find an approximation to the posterior distributions in \eqref{eq:bayes} and \eqref{eq:uncertain_bayes}. Common strategies are particle methods that can be achieved through the Markov Chain Monte Carlo method (MCMC) \cite{inv_stat_geyer1992}, specifically by the well-known Metropolis-Hastings algorithm and its variants \cite{inv_stat_Chib_1995}. Further methods are based on Importance Sampling \cite{stat_Glynn_1989, stat_Tokdar_2010} and especially the family of Sequential Monte Carlo (SMC) methods \cite{stat_doucet2001, stat_chopin2020}. Other, more recent strategies are based on Variational Inference \cite{inv_stat_Blei_2017, inv_stat_hoffmann2013}, where a parameterized distribution is optimized to match the true posterior as close as possible in a given norm. Here, we choose an SMC method as an established, state-of-the-art method.

\subsection{Numerical approximation via sequential Monte Carlo (SMC) sampling}
\label{sec:smc}

An efficient approach is needed to approximate the unnormalized posteriors of the Bayesian calibration problem \eqref{eq:bayes}, \eqref{eq:uncertain_bayes} and their normalization. While a grid-based approximation of the posterior is feasible for low dimensional parameters $\bs{x}$, the necessary amount of grid-based evaluations grows exponentially with the dimension of $\bs{x}$ and SMC methods become significantly more efficient as they exploit regions of high density. Sequential Monte Carlo (SMC) methods are popular sampling methods to efficiently explore a probability distribution $\smcppost$ for which no closed expression exists.

The result of the SMC sampler is a particle representation $\dirac_{\paramvec^\smcpindex}$ of $\smcppost$ with associated weights $\smcnormweights^\smcpindex$ in the form of

\begin{equation}
\smcppost\left(\paramvec\right) \approx \sum_{i=1}^\nparticles \smcnormweights^\smcpindex \dirac_{\paramvec^\smcpindex}\left(\paramvec\right).
\label{eq:smcdirac}
\end{equation}
Therein every particle has a location and weight $\left\{\paramvec^\smcpindex, \smcnormweights^\smcpindex\right\}_{i=1}^\nparticles $. It is further known \cite{stat_delmoral2006a, stat_koutsourelakis2009} that with this approximation any integral over an integrable function $h\left(\paramvec^\smcpindex\right)$ can be approximated by a sum that converges to the integral
\begin{equation}
\sum_{i=1}^\nparticles \smcnormweights^\smcpindex h\left(\paramvec^\smcpindex\right) \rightarrow \int h\left(\paramvec\right) \smcppost\left(\paramvec\right) \diffd \paramvec \hspace{0.5cm} \text{almost surely.}
\label{eq:smcintegration}
\end{equation}
Specifically, we use an SMC method \cite{stat_chopin2002} which sequentially blends over from a particle representation of a predefined prior distribution to a particle representation of the actual posterior distribution. A convenient aspect of the particle representation in \eqref{eq:smcdirac} is that they allow for a straightforward approximation of integrals \eqref{eq:smcintegration}, as they appear in marginal distributions \eqref{eq:marginal}. This is especially useful for Bayesian calibration under uncertainty and the associated integration in \eqref{eq:uncertain_bayes}. Here, the numerical integration is conducted by simply ignoring the dependency on $\btheta$ in the particle representation.

The SMC algorithm we use in this work uses an adaptive step length based on \cite{stat_delmoral2006,stat_delmoral2006a,stat_koutsourelakis2009}. Herein, a control parameter $\smczeta$ for the \emph{effective sample size} (ESS) is used to control step length. The ESS is defined as
\begin{equation}
\label{eq:ess}
\ess = \frac{1}{\sum_{i=1}^\nparticles \left(\smcnormweights^\smcpindex\right)^2}
\end{equation}
and is a measure for the current weight distribution. The ESS represents how well the probability density mass is distributed amongst the used particles.

The presentation of the SMC algorithm is compactly demonstrated in the pseudo-algorithm \autoref{alg:smc} and described in \cref{rem:smcwords}. For more details, the interested reader is referred to \cite{stat_doucet2001, stat_chopin2020, stat_koutsourelakis2009}. 

\begin{algorithm}[htbp!]
  \caption{Sequential Monte Carlo algorithm with adaptive step length.}
  \label{alg:smc}
  \begin{algorithmic}
  \State Draw $\nparticles$ particles from prior distribution
  \State Uniformly weight particles $\smcweights_0$
  \State Set step counter $\smcstep = 1$, and control parameter $\smcgamma=0 $
  \While{$\smcgamma < 1.0$}
  \State Find $\smcgamma_\smcstep$ so that $\ess_\smcstep \approx \smczeta \ess_{\smcstep-1}$ by reweighting $\smcweights_\smcstep = \smcweights_{\smcstep-1} \frac{\smcapprox_{(\smcstep-1),\smcgamma_\smcstep}}{\smcapprox_{(\smcstep-1),\smcgamma_{(\smcstep-1)}}}$
  \If{no $\smcgamma_\smcstep < 1.0$ can be found}
  \State Set $\smcgamma = 1.0$ and reweight
  \State Do final step towards posterior $\smcapprox_{\smcnumsteps,1.0} $
  \EndIf
  \If{$\ess_\smcstep< \imin{\ess}$}
  \State Resample
  \EndIf
  \State Rejuvenate with $\smcnumrejsteps$ MCMC iterations on the likelihood model to get $\smcapprox_{\smcstep,\smcgamma}$
  \State Raise step counter $\smcstep = \smcstep +1 $
  \EndWhile
  \State Normalize weights
  \end{algorithmic}
\end{algorithm}

\begin{remark}[Description of SMC algorithm]
\label{rem:smcwords}
An SMC algorithm generally starts with the steps to draw an initial set of particles from the prior distribution and initially equally weigh them. Then it needs to iteratively find a suitable step size and reweigh the particles. Resampling becomes necessary if too few particles carry too much weight of the distribution and likewise many particles lose significance. The particles are sequentially rejuvenated according to the new control parameter $\smcgamma$ and therefore new intermediate distribution \eqref{eq:transition}. The rejuvenation step is carried out with a scaled covariance of the current step as demonstrated in \cite{stat_minson2013} based on the acceptance rate of the Metropolis-Hastings algorithm used for rejuvenation. The Metropolis-Hastings algorithm is a Markov Chain Monte Carlo (MCMC) sampler \cite{stat_robert2004}, of which we omit the presentation for the sake of compactness. It is used to move the particles according to the new intermediate distribution. Reweighting, rejuvenation and possibly resampling are iteratively repeated until the intermediate distribution blends into the posterior, i.e. $\smcgamma =1$, which is the necessary last step if no other suitable $\smcgamma$ can be found. For the sake of a proper probability density, i.e. integration to 1, the resulting weights are finally normalized.
\end{remark}

The so-called \emph{tempering strategy} that defines the transition from the prior to the posterior as a sequence is written as
\begin{equation}
\label{eq:transition}
\smcapprox_{\smcstep,\gamma}\left(\paramvec\right) = \flatexp{\smcgamma\loglikelimodel{\paramvec}} \prior
\end{equation}
with the logarithmic likelihood $\loglikelimodel{\paramvec}$ as introduced in \eqref{eq:loglik} and the prior $\prior$. %
Finally, after reaching $\smcgamma=1$ the particle approximation has blended over into one for the posterior distribution.

\subsection{Approximation of the log-likelihood via Gaussian process regression}
\label{sec:gp_regression}

To efficiently evaluate the likelihood function in \eqref{eq:transition} we use a Gaussian process (GP) regression model \cite{stat_rasmussen2005} for the log-likelihood function \eqref{eq:loglik} that can be trained on a controllable amount of forward model evaluations. The main idea is that for a given computational budget, represented by the number of forward model runs, the approximation error of the surrogate is considerably smaller than the error in the SMC posterior approximation for the same number of forward simulation runs. A similar approach, where the model output is directly used as data to train a Gaussian process regression model as a surrogate for Bayesian calibration was used in \cite{inv_bilionis2013}. Gaussian processes are often used to generate regression models and therefore a brief summary of the core equations is presented in \cref{asec:gp_regression}. The presented overall approach would also work with different regression approaches as linear or spline interpolation or in case of cheap enough forward models would also work directly on the forward model.

We use the log-likelihood $\loglikelimodel{\paramvec}$ of the Bayesian calibration problem as presented in \eqref{eq:loglik} to train our GP upon. This gives the advantage that the regression models do not need to fulfill positiveness constraints. This would be the case if the GP was trained on the unnormalized posterior or likelihood directly. Another advantage of learning the log-likelihood function in contrast to learning the simulation response is that the log-likelihood is a scalar function. 

\emph{Training data} $\Ds$ are input-output pairs of values that are known about the process and the regression model can be based on. In our case, we can evaluate the forward model with any choice of model parameters $\paramvec$, compare it to the observed experiment and determine the logarithm of the log-likelihood $\tns{\loglikeliletter}$ \eqref{eq:loglik}. To simplify notation we specify the training data outputs to a vector of log-likelihoods $\ltrain$ at different coordinates $\xtrain$. Thus, we are free to choose any input batch $\xtrain$ and compute resulting output $\ltrain$ according to our computational budget. The generation of the training data $\Ds$ for the Gaussian process regression causes the highest computational costs in the inverse analysis, as for every training tuple $\{\loglikeliletter_{\text{train},i},\bx_{\text{train},i}\}$ one forward model evaluation is required, due to the dependency of the log-likelihood \eqref{eq:loglik} on the forward model.

Some further advantages of the surrogate approach are an a priori controllable amount of simulation runs, which can also be conducted in parallel, in contrast to the batch-sequential model evaluations that the SMC would impose on the forward model evaluations in a direct application. Once the surrogate is generated, the SMC algorithm can run without the risk of encountering non-converging or failing forward simulations.

To generate the training data $\Ds$ we choose a space-filling design strategy, namely a quasi-random Sobol sequence \cite{stat_soboldistribution,stat_owen1998scrambling}. The Sobol sequence has the advantage to have superior uniformity properties \cite{stat_kucherenko2015} and additionally allows for generating further sequential points that still share the space filling properties.

\begin{remark}[Dimensions in GP approach]
\label{rem:gpdim}
The number of necessary training tuples for desired surrogate accuracy is dependent on the complexity of the function, the type of (space-filling) experimental design and the dimension of the function. The required amount of training data grows exponentially with the dimension of the problem (curse of dimensionality), so that for higher dimensions (a rough guideline might be $\dim(\bx)>15$) direct sampling (using more advanced strategies that can incorporate gradient information of the model w.r.t. the inputs $\bx$) becomes more efficient.
\end{remark}

\subsection{Implementation aspects}
\label{sec:implementation}

The implementation of the algorithm described above was done in the Python software framework \emph{QUEENS} \cite{QUEENS}. Herein, the Gaussian process regression module \emph{GPy} \cite{GPy} was used for the construction of the Gaussian process surrogates. The sequential Monte Carlo algorithm based on \Cref{alg:smc} is also implemented in \emph{QUEENS}. For processing of forward model results the Python package \emph{vtk} for "The Visualization Toolkit" was used \cite{vtk}. Some of the following visualizations were generated with \emph{Seaborn} \cite{seaborn} and \emph{Matplotlib} \cite{matplotlib}. Parallel axis plots are generated with plotly \cite{plotly}. We furthermore use \emph{PyTorch} \cite{python_pytorch} for the generation of the Sobol sequences.
The forward models were solved with the in-house C++ research code BACI \cite{BACI}.

\section{Numerical examples}
\label{sec:examples}

In this section, we present our Bayesian calibration approach for a coupled fluid-structure interaction problem. We demonstrate the approach with generated data to better assess its individual steps. Our focus lies on the presentation of the proposed approach and some general characteristics in the resulting posteriors. In application with real-world data, the procedure can be used without any changes. The computational mechanics models in the examples are schematic models for fluid-biofilm interaction that is the motivation for our research. Therein the fluid-solid interface deforms as a consequence of the interaction. A further description of the experiments is moved to \cref{asec:experimental_setup} as we want to focus on the model here. In the following examples, we calibrate biofilm material properties under partially uncertain experimental conditions. For the numerical demonstrations we use the fluid-solid interaction (FSI) between incompressible Navier-Stokes flow and a hyperelastic nonlinear solid material model which is briefly introduced in \cref{asec:model}. Although the presented calibration approach is equivalently applicable for single field problems with deformable boundaries we take the challenge of a coupled multi-physics model of FSI because we want to advertise the benefit of the approach in such applications. A variety of different models for biofilms are available and also further effects can be included (see, e.g., \cite{lnm_bio_coroneo2014, me_willmann_2021}), which would just lead to different forward models.

\subsection{Problem setup}
\label{subsec:problemsetup}

\begin{figure}[htbp]
    \centering
    \includegraphics[width=0.6\textwidth]{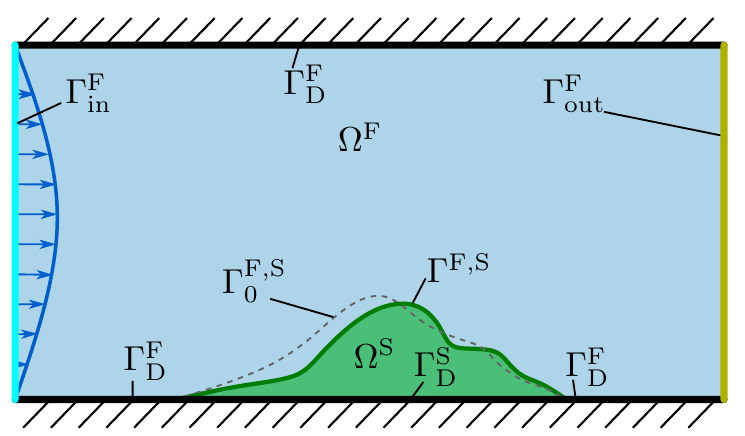}
    \caption{Schematic of problem setup with domain and interface names. A biofilm (green domain) is grown on the flow cell floor and is interacting with a flowing liquid (blue domain) that is providing nutrients to and also deforming the biofilm which in return is changing the flow field. Undeformed (gray dotted line) and deformed biofilm interface. Inflow (teal) and outflow (yellow) boundaries of the flow cell model are also shown.}
    \label{fig:computational_domain}
\end{figure}

 The calibration is performed for hyperelastic material properties of the solid domain for which we will calibrate the two parameters of a Saint-Venant Kirchhoff material model. For the given setup of FSI models for biofilms and the biofilm flow cell data, the location of the fluid-biofilm interface is the primary data available and therefore used for comparison. A schematic sketch of the problem setup is drawn in \Cref{fig:computational_domain}. For easier demonstration we first investigate a two-dimensional calibration problem ($\dim(\bx)=2$) without experimental uncertainties ($\dim(\btheta)=0$) and then move on to more complex examples. 

We chose the same problem setup as presented in \cite{me_willmann_2021}. The biofilm geometry is inspired by analyses done on experimental results in \cite{kit_bio_picioreanu2018, kit_bio_blauert2015}.
The model domain represents a two-dimensional channel with dimensions $1\sunit{mm}\times2\sunit{mm}$, where horizontal fluid flow with parabolic profile is enforced from the left boundary (see $\svggdfin$ in \Cref{fig:computational_domain}) with a maximal volume rate of $ \volinflowrate= 100\sunitfrac{mm^2}{s} $. The solid biofilm (green) is attached to the channel floor. A no-slip condition for the fluid is used on the channel floor and top boundary as well as the biofilm on the channel floor (see $\svgdnsf$ and $\svgdnss$ in \Cref{fig:computational_domain}) and a horizontal outflow is enforced on the right edge (see $\svggdfout$ in \Cref{fig:computational_domain}). These boundary conditions are modeled via Dirichlet boundary conditions on the fluid velocity (and solid displacement accordingly) and respective free outflow in horizontal direction. The horizontal condition represents the ongoing empty channel left and right from the modeled domain.

\begin{figure}[htbp!]
\centering
\subfloat[\label{subfig:FSIveldisp}]{
\includegraphics[width=0.49\textwidth]{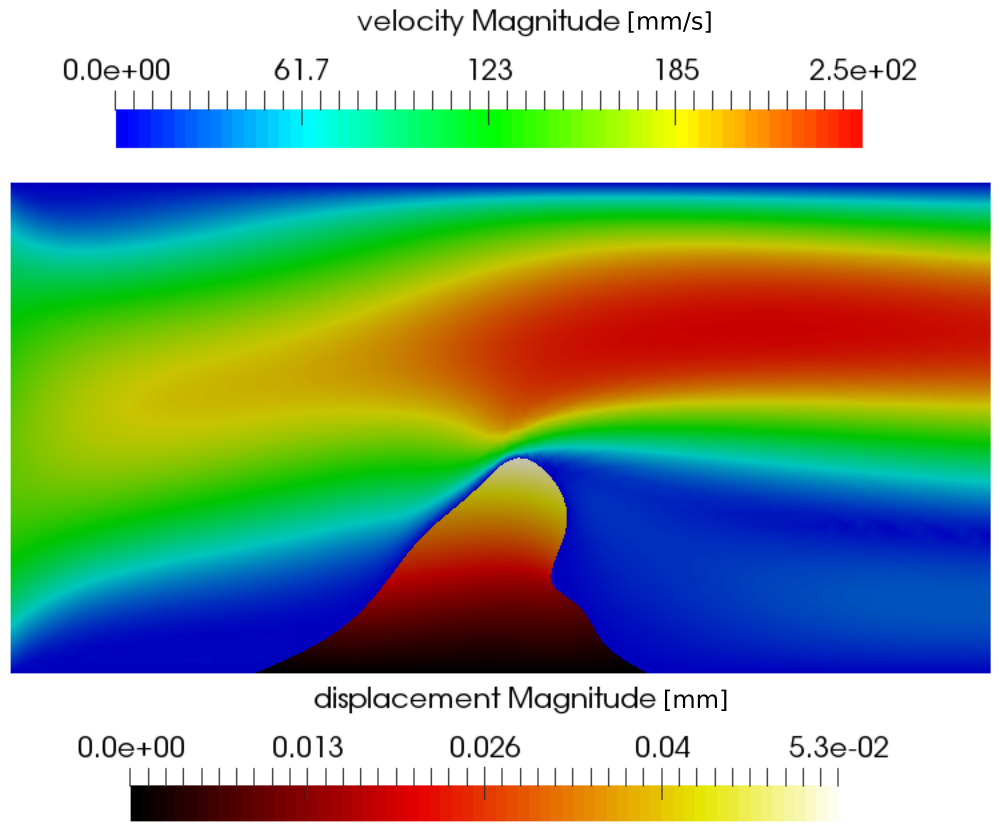}}
\hfill
\subfloat[\label{subfig:FSIpres}]{
\includegraphics[width=0.49\textwidth]{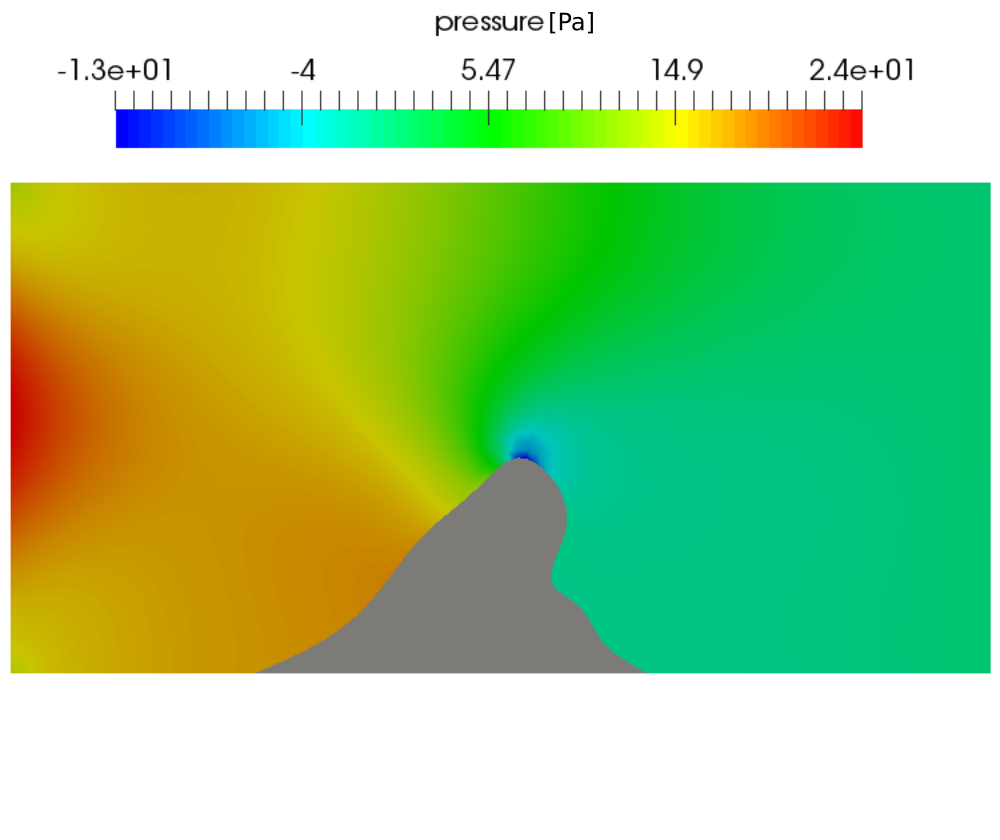}}
\caption{Field solutions of the ground truth simulation with parameters $\gtx $ on the deformed geometry. (a) Fluid velocity magnitude, solid displacement magnitude. (b) Fluid pressure solution.}
\label{fig:FSIresult}
\end{figure}

\begin{figure}[htbp!]
  \centering
  \subfloat[\label{subfig:contours_gt}]{
  \includegraphics[width=0.36\textwidth]{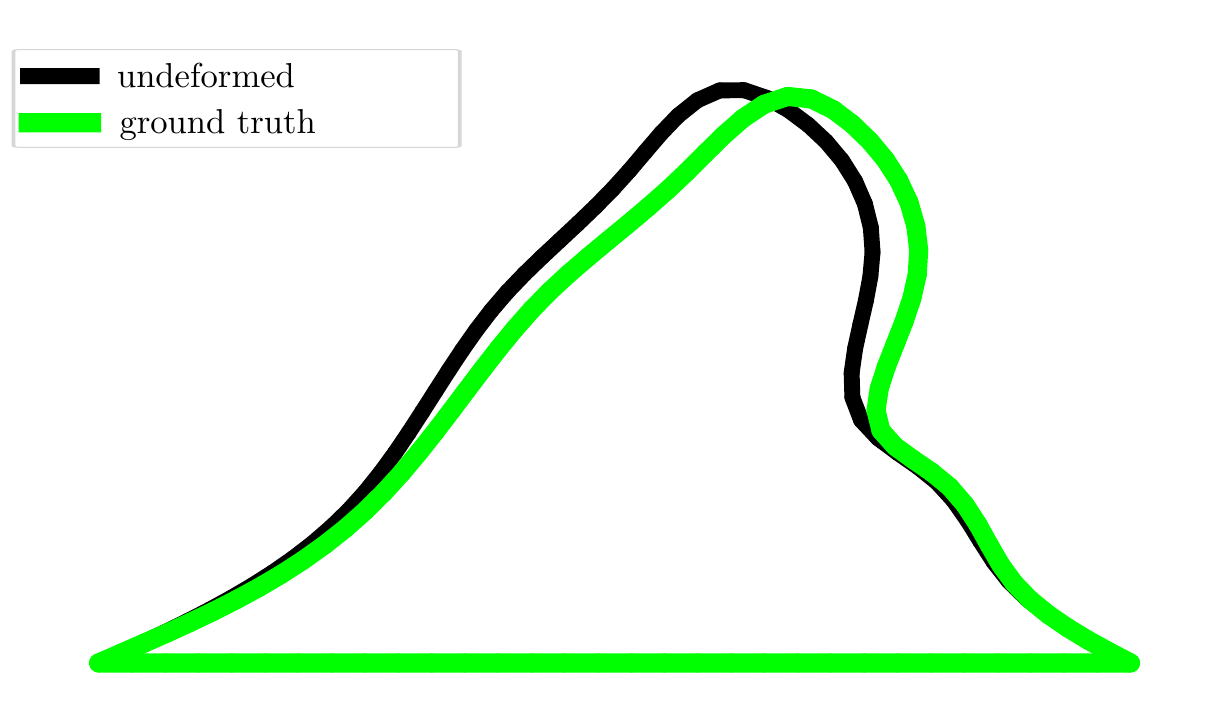}}
  \hfill
  \subfloat[\label{subfig:contours_all}]{
  \includegraphics[width=0.48\textwidth]{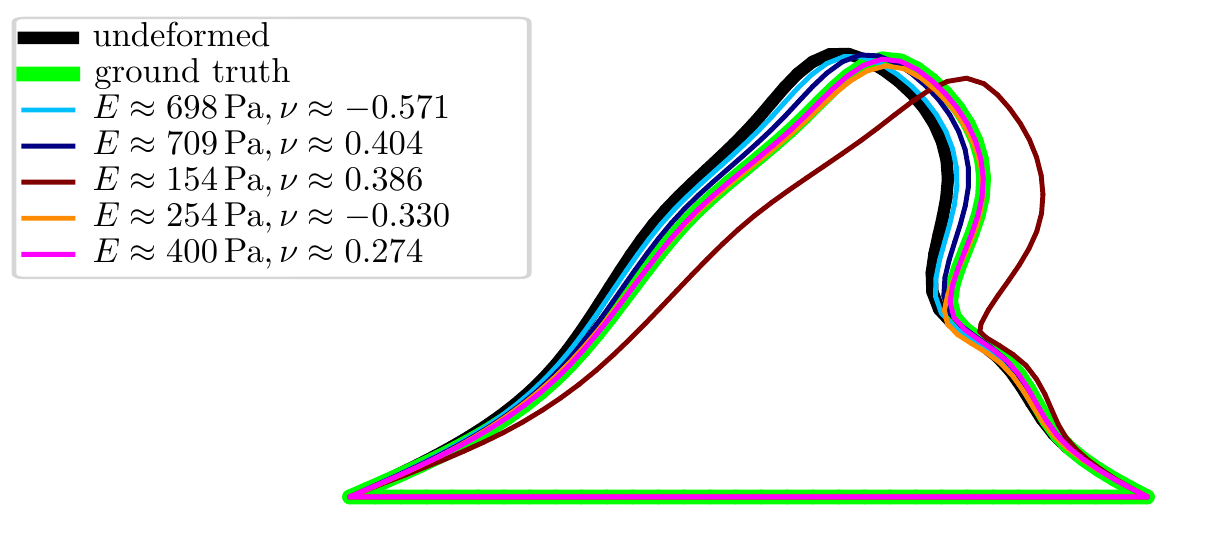}}
  \caption{Exemplary forward model results for deformed interface shapes for different input parameter samples compared to the result with ground truth $\gtx=\transpose{[\poissonratio = 0.3,\;\youngs = 400\sunit{Pa}]}$ as input and the undeformed state.}
  \label{fig:contoursamples}
\end{figure}

To generate the artificial experimental data, we use a forward simulation of the fluid-biofilm interaction with a Saint-Venant-Kirchhoff material model for the solid biofilm domain. The material is characterized by two parameters, namely Young's modulus $\youngs$ and Poisson's ratio $\poissonratio$. We summarize the chosen input parameters, used for the generation of the data, in form of the ground truth vector $\gtx=\transpose{\begin{bmatrix}\poissonratio = 0.3, \youngs = 400\sunit{Pa}\end{bmatrix}}$. The fluid has a dynamic viscosity of $\dynvisf = 10^{-3}\sunit{Pa\, s} $ and a density of $ \densf = 10^3\sunitfrac{kg}{m^3} $ as a model for water. The biofilm has the same density as the fluid. The solution of the velocity and pressure field of the fluid and displacement field of the biofilm is depicted in \Cref{fig:FSIresult} for the regarded quasi-steady deformed state in the ground truth forward model evaluation. As a reaction to the load imposed by the fluid inflow boundary condition, the solid bends towards the right as it is plotted in \Cref{fig:contoursamples}. 
In \Cref{subfig:contours_gt} we see the artificial observation data $\yobs$ as the result of the reference simulation with ground truth values $\gtx$. $\yobs$ represents the deformed location (green) of the interface in this example for one single point in time $\coords$. In \Cref{subfig:contours_all} we plot the results for some exemplary parameter combinations additionally.

\subsection{Likelihood response surface for different discrepancy measures}
\label{subsec:response_surf_measures}

In a first step, we compare the effect of the different discrepancy measures as introduced in \cref{sec:measure}. We directly approximate the log-likelihood by the posterior mean function of a Gaussian process surrogate and use the discussed discrepancy measures. Additionally, we also provide the posterior standard deviation of the GP to quantify the remaining uncertainty in the surrogate. All surrogate models for the log-likelihood function resulting for the different discrepancy measures use the same training input that was sampled with a quasi-random Sobol sequence to yield a space-filling training design (see \cref{sec:gp_regression}). The training inputs are different parameters in the forward model and the resulting forward simulations.

For an estimation of a suitable $ \sigman $ in the likelihood model \eqref{eq:staticlikeli} the known resolution $\approx 8 \sunit{\mu m} $ of OCT \cite{kit_bio_blauert2015, kit_bio_gierl2020} is considered. The noise standard deviation is assumed to be in the same order of magnitude, such that $\sigman = 0.01\sunit{mm}$ is used in the following. OCT resolution and standard deviation in the likelihood model are not expected to be equal. A further discussion is omitted here, as the demonstration here works independent of the choice and an expressive value can only be found related to real data and chosen image segmentation. For the RKHS norm we need two parameters $\sigman$ and $\sigmaw$. An estimation of the length scale $\sigmaw$ for the RBF kernel in the RKHS approach in \eqref{eq:surface_current} is made as $\sigmaw =  0.005\sunit{mm}$, being approximately $10\%$ of the maximal displacement magnitude (see \Cref{subfig:FSIveldisp}).

\begin{figure}[htbp!]
\centering
\subfloat[\label{subfig:likshape}]{
\includegraphics[width=0.3\textwidth]{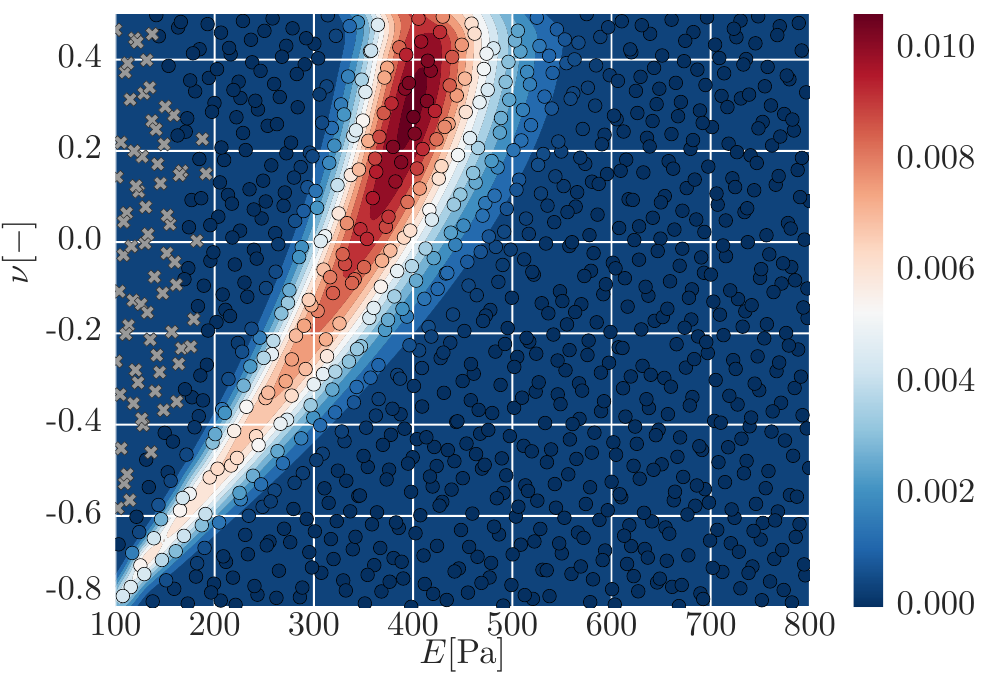}}
\hfill
\subfloat[\label{subfig:likcpp}]{
\includegraphics[width=0.3\textwidth]{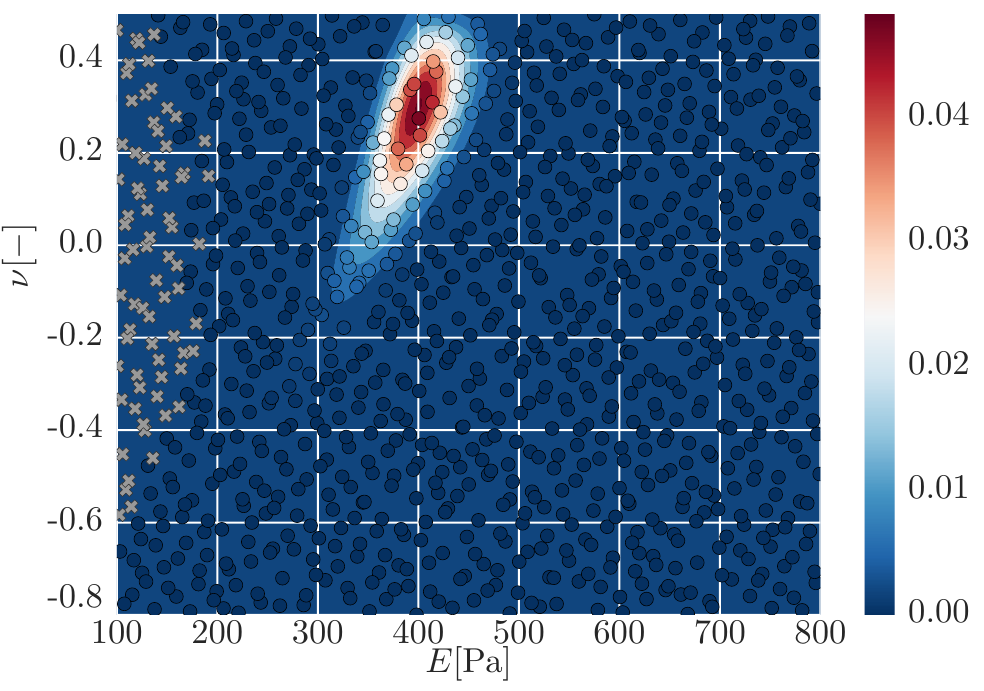}}
\hfill
\subfloat[\label{subfig:liksc}]{
\includegraphics[width=0.3\textwidth]{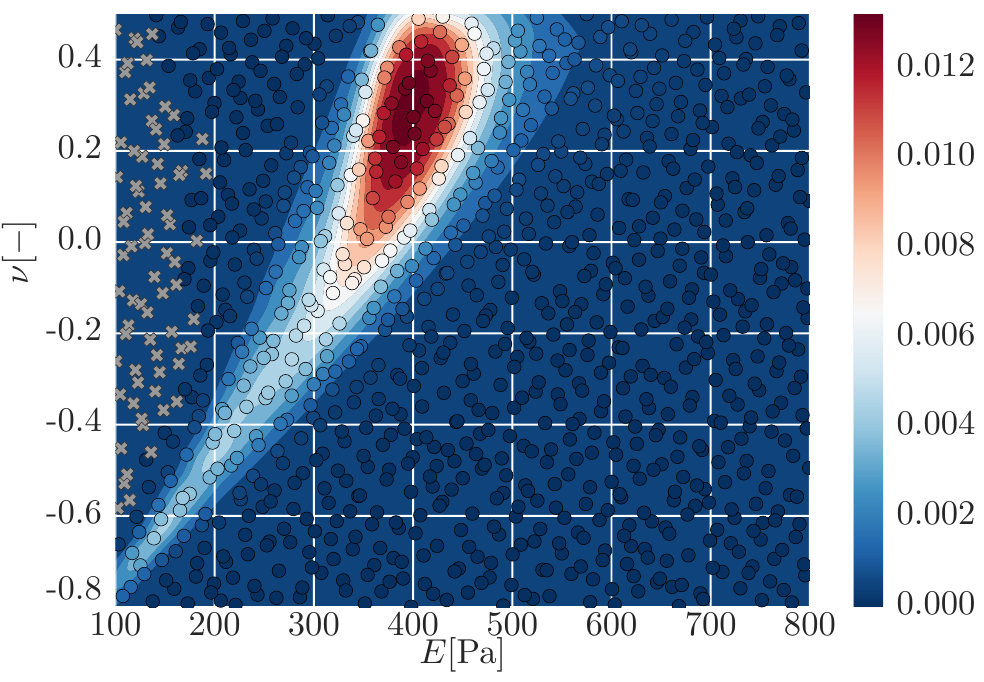}}
\caption{Likelihoods for different surface measure distances with $\ntrain = 1000$. (a) Euclidean distance at measurement points (b) closest point projection distances (c) RKHS norm. Normalized for better comparability. Training samples as circles for successful and gray crosses (on the left side) for failed forward model evaluations.}
\label{fig:likexamples}
\end{figure}

\begin{figure}[htbp!]
\centering
\subfloat[\label{subfig:loglikshape}]{
\includegraphics[width=0.3\textwidth]{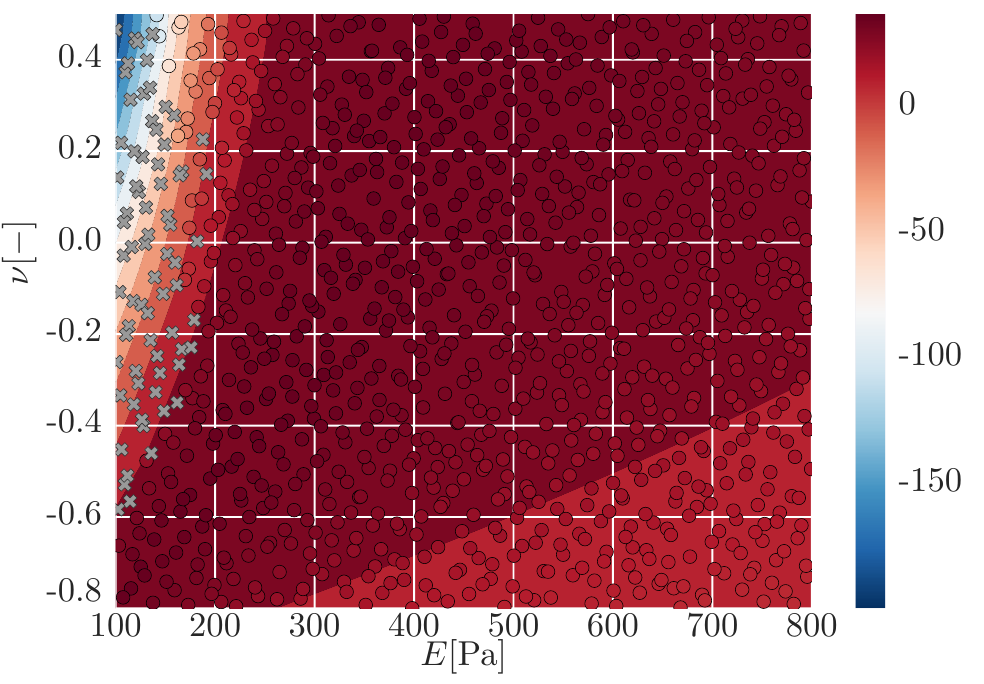}}
\hfill
\subfloat[\label{subfig:loglikcpp}]{
\includegraphics[width=0.3\textwidth]{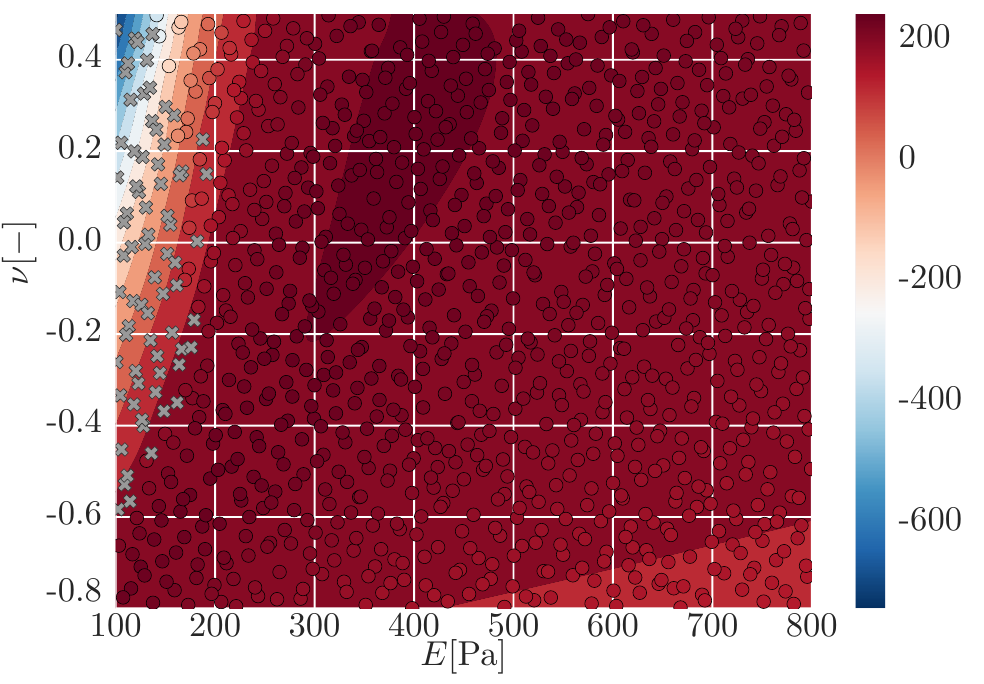}}
\hfill
\subfloat[\label{subfig:logliksc}]{
\includegraphics[width=0.3\textwidth]{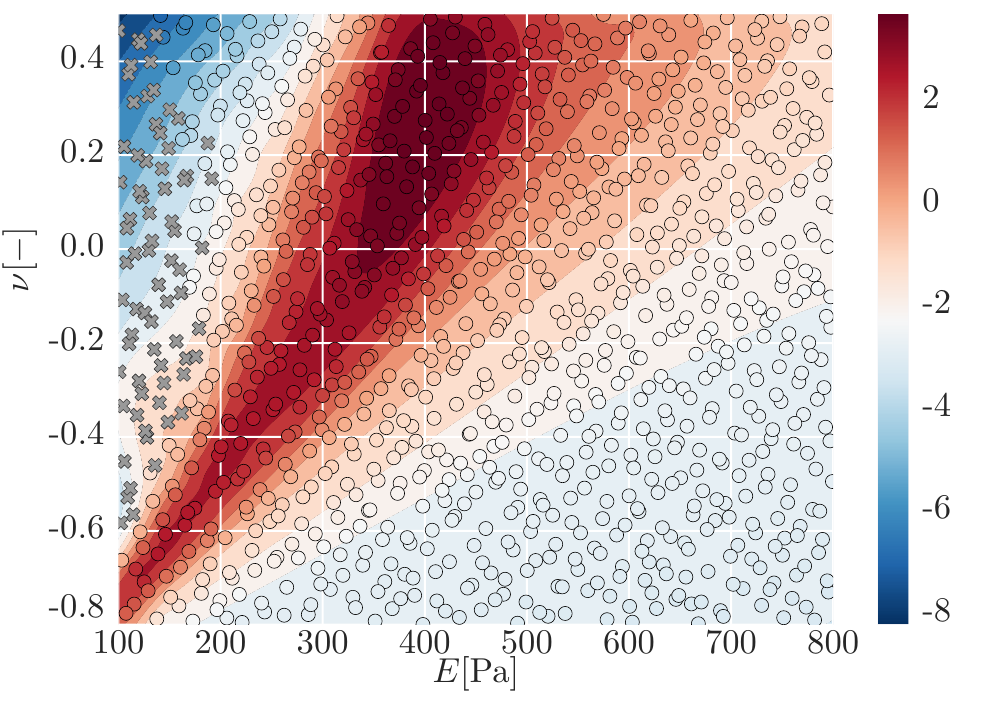}}
\caption{Logarithmic likelihood regression model for different surface measure distances with $\ntrain = 1000$. (a) Euclidean distance at measurement points (b) closest point projection distances (c) RKHS norm. Training samples as circles for successful and gray crosses for failed forward model evaluations.}
\label{fig:loglikexamples}
\end{figure}

\begin{figure}[htbp!]
\centering
\subfloat[\label{subfig:stdloglikshape}]{
\includegraphics[width=0.3\textwidth]{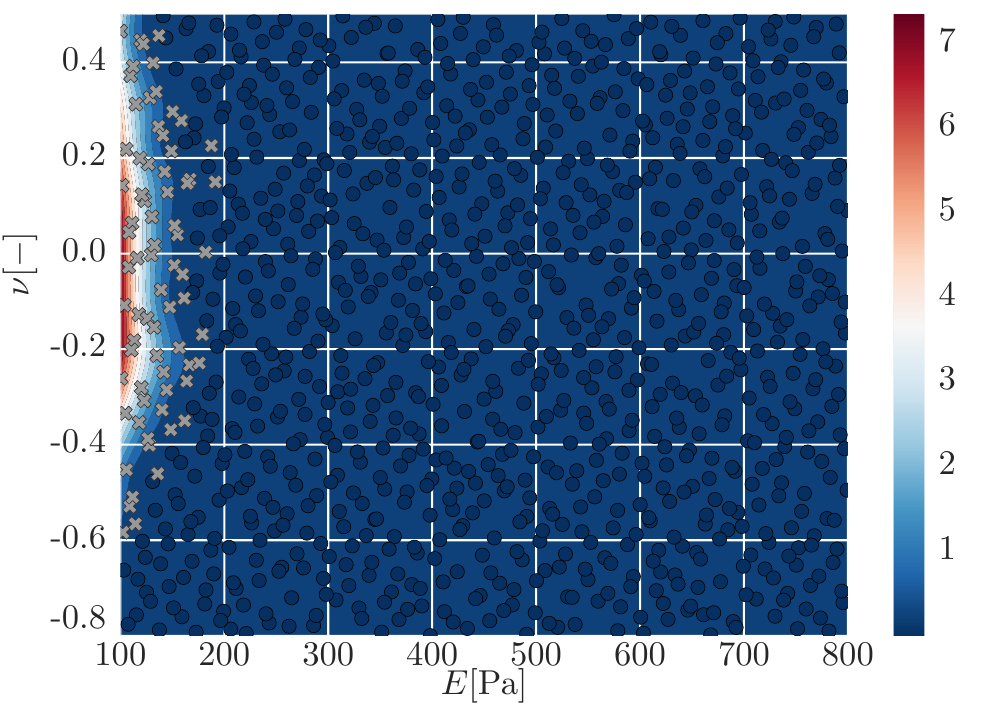}}
\hfill
\subfloat[\label{subfig:stdloglikcpp}]{
\includegraphics[width=0.3\textwidth]{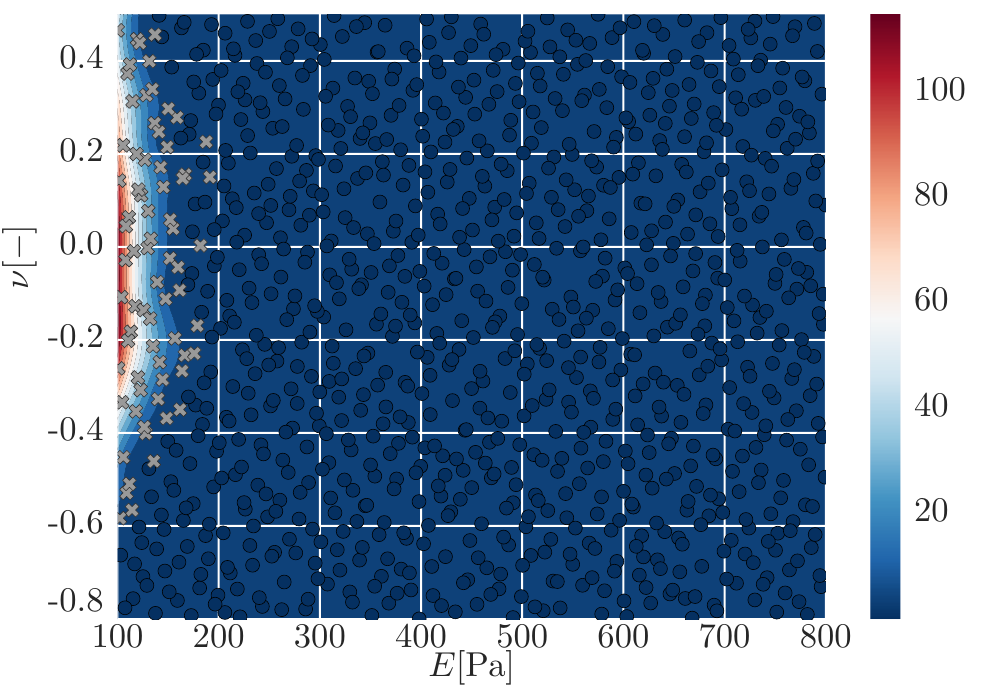}}
\hfill
\subfloat[\label{subfig:stdlogliksc}]{
\includegraphics[width=0.3\textwidth]{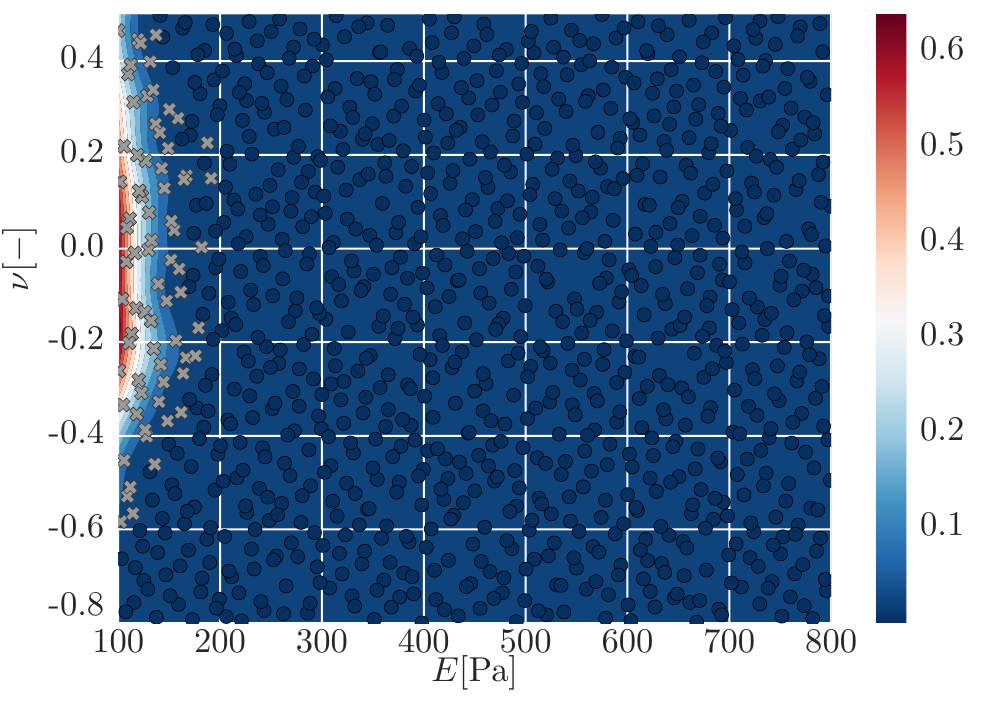}}
\caption{Standard deviation from logarithmic likelihood regression model variance for different surface measure distances with $\ntrain = 1000$. (a) Euclidean distance at measurement points (b) closest point projection distances (c) RKHS norm. Training samples as circles for successful and gray crosses for failed forward model evaluations.}
\label{fig:stdexamples}
\end{figure}

The resulting regression models for the likelihoods are shown in \Cref{fig:likexamples} for $\ntrain = 1000$ training points with a Matérn 3/2 kernel (see \cref{asec:gp_regression}, \cref{eq:matern32} for details). In general, the likelihood over the parameters can be understood as a score value of how well the forward model response with respective parameters leads to similar results compared to the reference data. A discussion and interpretation of the figure will follow below. For this first comparison, we just use this large number of training points and postpone the discussion on the GP convergence over $\ntrain$ to the case only using the RKHS norm measure. For the distribution of the measurement points in this comparison, the reader is referred to \cite{me_willmann_2021}. In \Cref{fig:likexamples} the fields of the likelihoods are determined from the logarithmic likelihoods and those fields are normalized in the plots for better comparability.
Parameter combinations that led to failed forward model evaluations are marked by gray crosses in the following figures. For our setup, the failing simulations occurred for low values of Young's modulus (located on the left side of the plots) representing soft biofilm material, which lead to large mesh distortions in the ALE FSI approach.
For the sake of comparability, the respective logarithmic likelihoods are plotted in \Cref{fig:loglikexamples} and the associated standard deviations in the regression models in \Cref{fig:stdexamples}.

 It can be seen, that the response in the likelihoods show a high likelihood of the parameters in red for a characteristic curved shape with its peak at the expected ground truth parameters $\gtx=\transpose{\begin{bmatrix}\poissonratio = 0.3, &\youngs = 400\sunit{Pa}\end{bmatrix}}$. A high likelihood corresponds to a high probability density of the parameter combination to represent the reference data. This means that the corresponding values for $\paramvec$ in this region of the input space result in simulation outputs that are very close to the observation data under the employed discrepancy measure. The likelihood falls very close to zero when moving away from the high likelihood regions, representing low similarity of the forward model result compared to the reference data. Especially the failed simulations (gray crosses) fall in a region with very low likelihood values which renders them irrelevant for our investigations. For the rest of the input space, it can be seen in \Cref{fig:stdexamples} that the standard deviation in the regression model is low and only rises in regions with little data.

Comparing the Euclidean distance measure in \Cref{subfig:likshape} and the closest point projection in \Cref{subfig:likcpp} it can be seen, that the second one is more peaked around the maximum of the likelihood that is close to the ground truth value $\gtx$. This is also a consequence of the formulation \eqref{eq:staticlikeli} and \eqref{eq:loglik}, as the numbers of distance measurements differ greatly with the number of measurement points $\lmnmp=10 $ and the number of interface nodes $\nin=66$. The higher number of individual single point measurements generally leads to a more peaked likelihood with the same $\sigman $. It must also be stated that no further weighting of the closest point projection distances was done, so all nodal closest point projection distances are considered equally important. This includes the distances for many mesh nodes, that are close to the boundary conditions and therefore there the displacement magnitude is lower. An additional data compression approach, e.g., kernel principal component analysis \cite{stat_bishop2011, stat_Schoelkopf1997} or also a selection of only a subset of the interface nodes could be used to increase comparability, but this is outside of the scope of the current article. The likelihood from the RKHS based distance measure (in \Cref{subfig:liksc}) combines both features to be expressive around the maximum likelihood (ML) point and still have information from more distant points. The RKHS norm measure correlates all discretized interface locations and orientations of the model to all counterparts in the observation (see \eqref{eq:surfcurrontriang}), therefore it is also expected to be the most detailed measure for this comparison.

The likelihood based on the forward model evaluations adequately combined with prior distributions results in the posterior (see \eqref{eq:bayes}), which is a probability density of the parameters leading to the observations. It is favorable to have an expressive posterior and therefore expressive likelihood as a result to be able to compare posterior values for different parameter combinations and with that develop an understanding of the forward model in relation to the observations.
In the case of a very flat likelihood the conclusion is that all parameter combinations are similarly well suited to explain the observations and therefore they are potentially insignificant to the forward model, at least in the range of the regarded parameter intervals. This means that all input parameters $\paramvec$ will lead to forward model results that are very close to the observation under the employed discrepancy measure. The expressive shape of the likelihood based on the RKHS based measure underlines that this measure is well suitable in the given example to be used for Bayesian calibration. Therefore it is used in all following examples.

The presented likelihoods were generated using the different discrepancy measures \eqref{eq:eucliddist}, \eqref{eq:cppdist} and \eqref{eq:surfcurrontriang} and all yield similar characteristics in the shape of the likelihood. Therefore it can be concluded that in this setting and a fluid-biofilm interface-based measurement of a flow cell experiment an underestimation of Young's modulus $\youngs$ is coupled to an underestimation of Poisson's ratio $\poissonratio$. In applications with real experimental data, it would make sense to restrict the training points to the intervals, that are believed to contain the optimum or at least relevant values for the parameters. That could mean to restrict the Poisson's ratio to positive values, as this is what would be expected for most materials. Nevertheless, the resulting shapes representing high likelihood are very smooth for the whole tested range between $\poissonratio = -0.8$ and $\poissonratio = 0.5$ and do not show a distinct border between positive and negative values. This is interesting with regard to biofilm mechanics in flow cell experiments as it shows that the estimation of $\youngs$ and $\poissonratio$ is coupled for all surface comparisons that were tested. For some investigations, this also hints toward the need to incorporate additional measurements or information, e.g., prior knowledge.

\subsection{Convergence over number of training points}

The most costly part of the presented algorithm are the forward model evaluations that are necessary to generate the training data $\Ds$ of the GP log-likelihood regression model. While the convergence of the GP over the number of training data points is dependent on the character of the underlying function as well as on the selected design of experiments, we want to show a short qualitative convergence study for the problem at hand for the distance measure based on the RKHS inner product. In general the convergence study of a GP w.r.t. the number of training points is difficult as usually it is not feasible to increase the data point set size by orders of magnitude. Therefore other strategies, e.g., leave-one-out cross validation can be used as a proxy for the regression error \cite{stat_rasmussen2005}. The requirement towards the number of training points and the regression model is to catch the relevant shape characteristics of the likelihood.

For the application of such an approach, it is crucial to know how many forward model evaluations are required to get results efficiently. To get a first picture the Gaussian process regression model for the logarithmic likelihood is created for a series of training point set sizes $\ntrain$. For this qualitative study, only the RKHS norm based likelihood was used as it uses the most detailed comparison and seems to give the most expressive shape of the posterior. The likelihood parameters were set to $\sigmaw=0.005 \sunit{\mu m}$, $\sigman^2=0.0005\sunit{mm}^2$ in all following examples. The choice $\sigman^2=0.0005\sunit{mm}^2 $ in our case relates to the discretization size, as the rounded squared average interface element length is $\bar{l}^2 \approx 0.0005 \unit{mm}^2 $. It is difficult to interpret $\sigman $ in relation to the measurement error alone, as especially also the image segmentation approach chosen to determine the model fluid-biofilm interface from experimental data influences the choice. We deviated from this parameter choice for $\sigman $ in the comparison of the measures to keep it the same for all measures.

We used a Matérn 2/3 kernel (see \cref{asec:gp_regression}, \cref{eq:matern32}) with one length scale hyperparameter $l$ (in the standardized input space) and one signal variance $\sigma_0^2$ which will be determined via ML estimation using an L-BFGS-B optimizer for the evidence of the GP (see \cref{eq:gploglikeli}). The Matèrn 2/3 kernel appeared to be the best suitable to represent the likelihood field as it resulted for the given example. Especially the great variety in steepness and therefore low smoothness of the likelihood seems to be the challenge for the regression approach. However, at the moment, no general recommendation for the GP kernel can be given. As commonly done \cite{stat_rasmussen2005} we used a fixed nugget noise $ \noisevar=10^{-5}\cdot\left(\maxof{\ltrain}-\minof{\ltrain}\right) $ to stabilize the training of the GP.

\begin{figure}[htbp!]
\centering
\subfloat[\label{subfig:sc_scaling_10}]{
\includegraphics[width=0.3\textwidth]{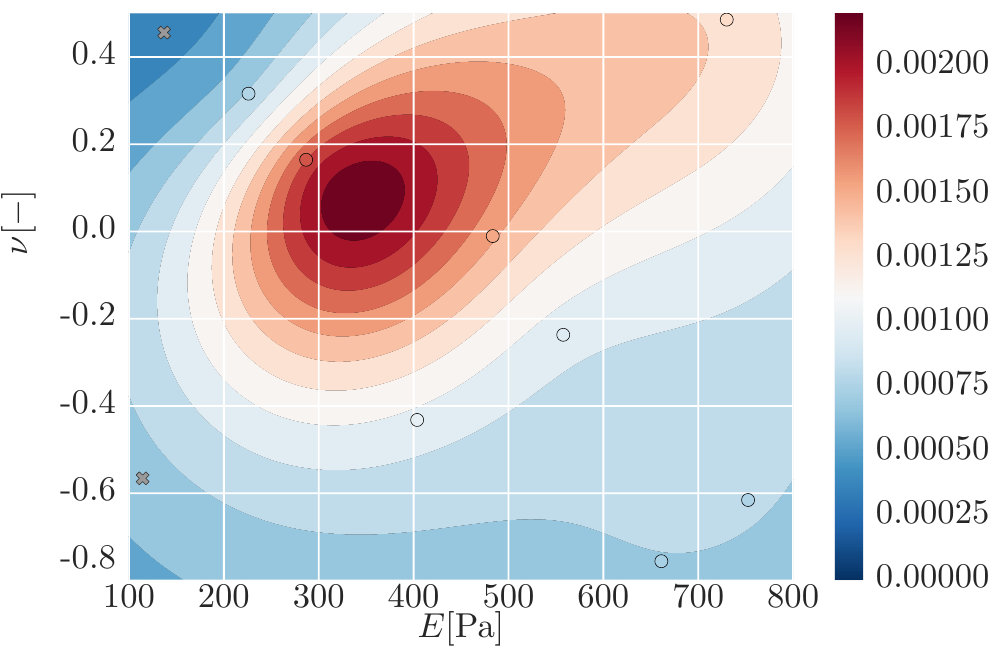}}
\hfill
\subfloat[\label{subfig:sc_scaling_20}]{
\includegraphics[width=0.3\textwidth]{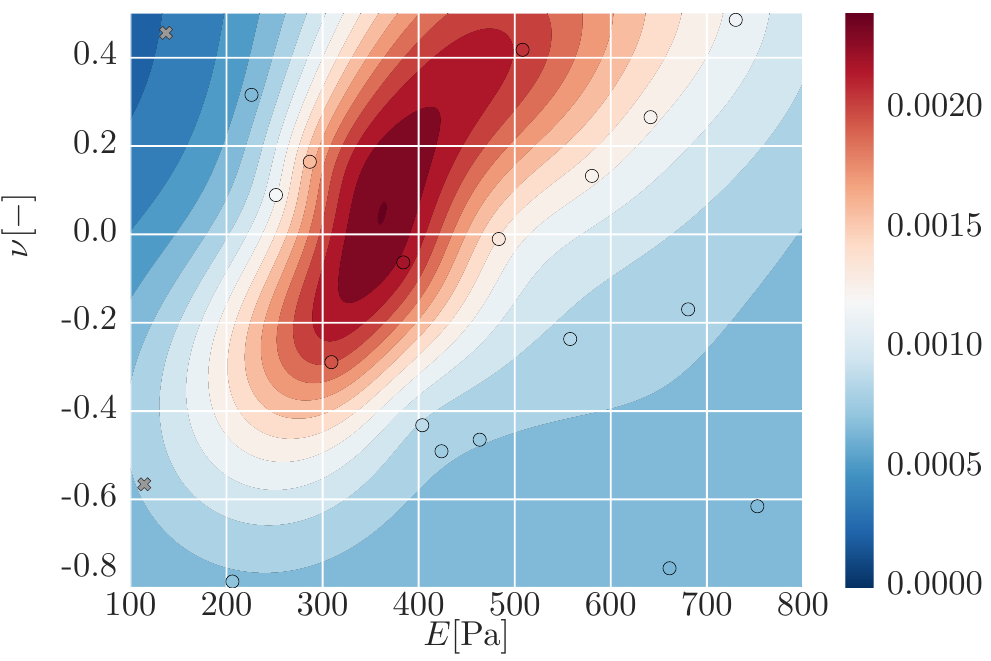}}
\hfill
\subfloat[\label{subfig:sc_scaling_50}]{
\includegraphics[width=0.3\textwidth]{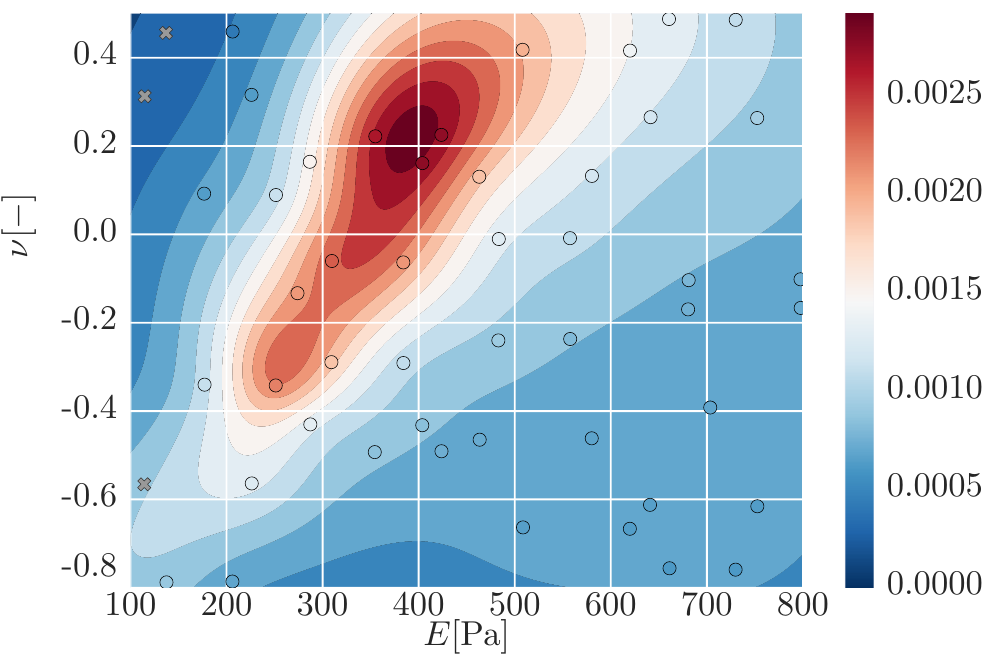}}
\hfill
\subfloat[\label{subfig:sc_scaling_100}]{
\includegraphics[width=0.3\textwidth]{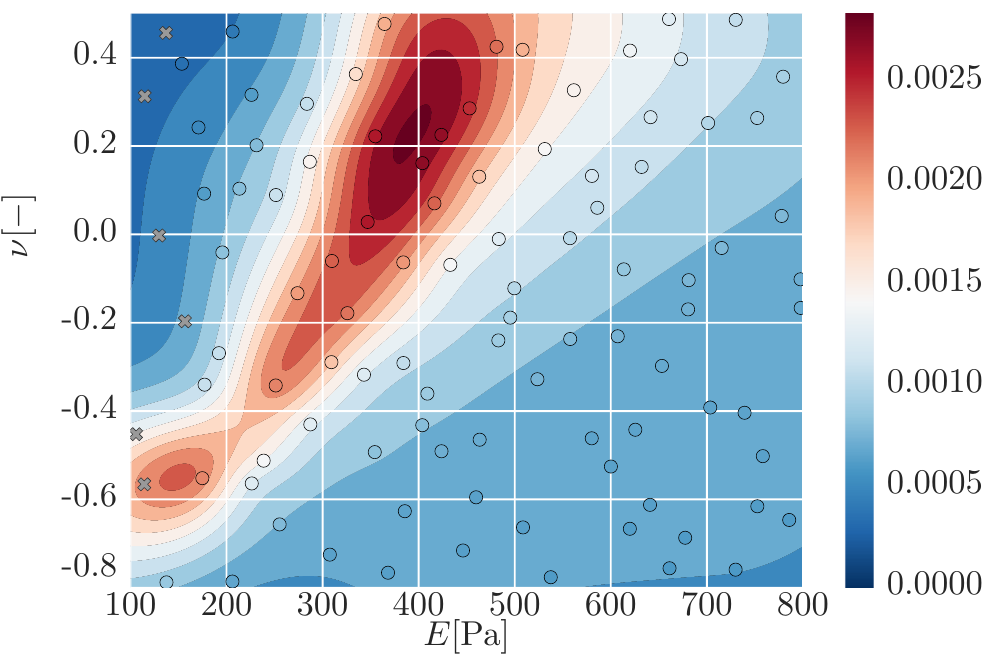}}
\hfill
\subfloat[\label{subfig:sc_scaling_200}]{
\includegraphics[width=0.3\textwidth]{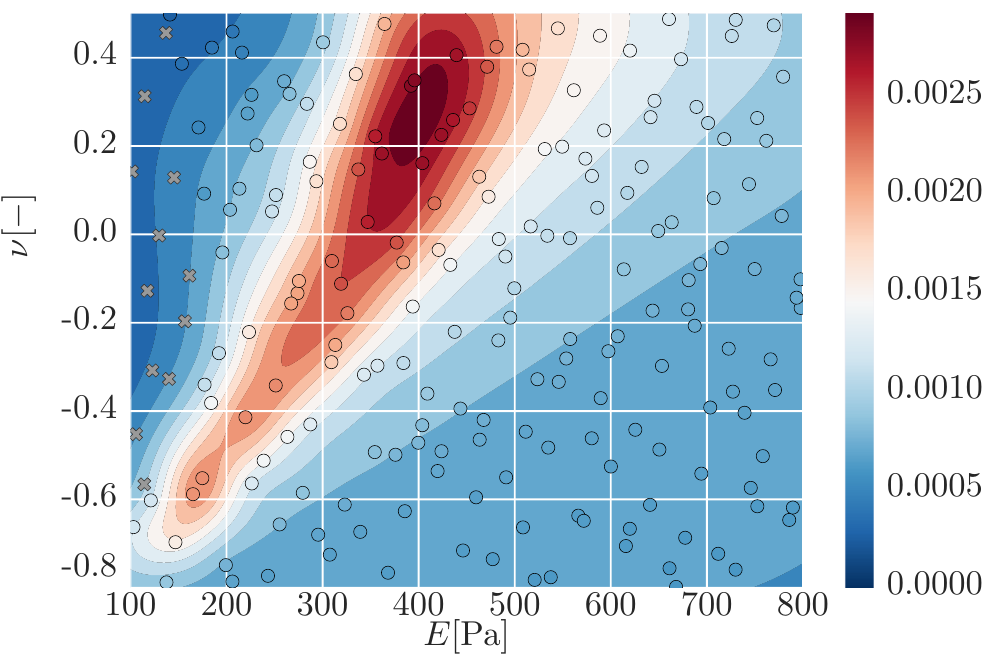}}
\hfill
\subfloat[\label{subfig:sc_scaling_900}]{
\includegraphics[width=0.3\textwidth]{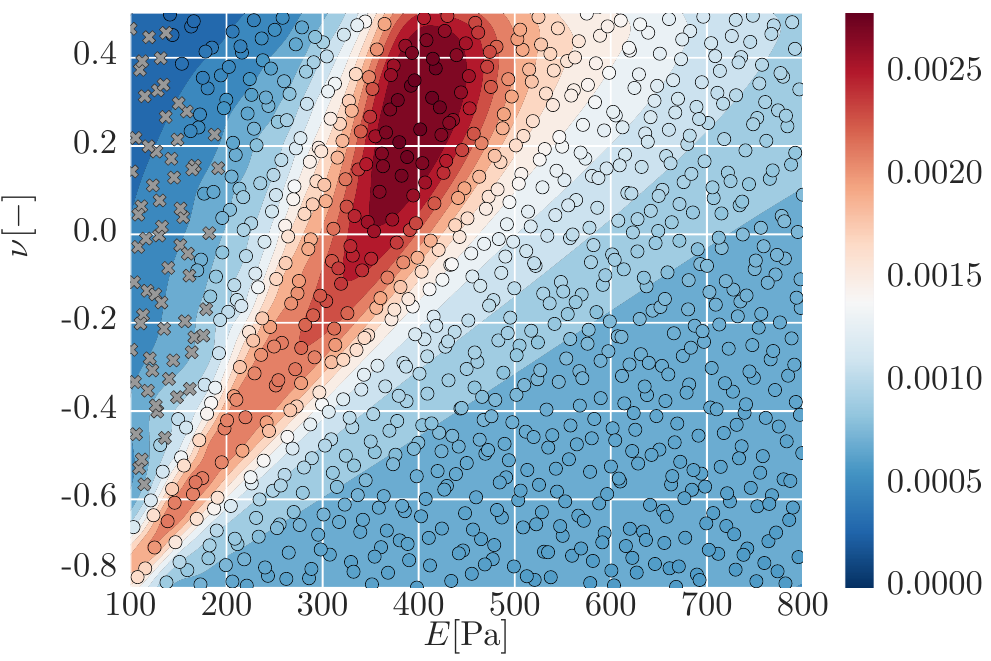}}
\caption{Convergence of the Gaussian process regression model for the likelihood function over number of training points $\ntrain$ (forward model evaluations). (a) $\ntrain=10$, (b) $\ntrain=20$, (c) $\ntrain=50$, (d) $\ntrain=100$, (e) $\ntrain=200$, (f) $\ntrain=900$. In all cases the same Sobol sequence was used to generate a space-filling training data set. Training samples as circles for successful and gray crosses (left side) for failed forward model evaluations.}
\label{fig:sc_scaling}
\end{figure}

In the following, the convergence of the likelihood surrogate model over the number of used training points is qualitatively shown for the problem at hand. For this study, the Sobol sequence sampling property is used as all samples are consecutive samples from the same Sobol sequence. In \Cref{fig:sc_scaling} it is apparent, that only very few training points are necessary to estimate the general shape of the likelihood distribution and have a rough estimate for the optimum. With $\ntrain=200$ samples the ML estimate is already in good agreement with the ground truth $\gtx=\transpose{\begin{bmatrix}\poissonratio = 0.3, &\youngs = 400\sunit{Pa}\end{bmatrix}}$.

It can be concluded that for this two-dimensional example the likelihood surrogate shows relevant features and the global shape well for $\ntrain = 200 $ forward model evaluations and no significant gain in accuracy can be expected for a moderate increase of the sample size. Given that in the following examples a comparison with different priors is made and another problem dimension in form of an uncertain parameter is added, and therefore the complexity is increased, we use the Gaussian process model with one length scale for all (standardized) parameters, a fixed nugget noise and $\ntrain=1000$ training points for the following examples (see \Cref{rem:gpdim} on dimensionality).

\begin{remark}[Distribution of training points]
\label{rem:disttrp}
With the applied approach of a grid-based quasi-random distribution of the samples, there is no compromise between exploration and exploitation, but the emphasis is put on exploration. It is possible to use available prior information for the generation of the samples and therefore have higher density of samples in high prior areas, or use an iterative approach and refine the samples in high posterior regions.
\end{remark}

\subsection{Calibration of constitutive parameters in biofilm models}

In this subsection, the regression model will be generated according to the findings in the previous examples. The GP was constructed with a Matérn 2/3 kernel, a single length scale and signal variance for all parameters and a fixed nugget noise variance $\noisevar$ in \eqref{eq:gploglikeli}.
1000 samples were used for the training of the GP.
As the likelihood model is available in form of a cheap to evaluate surrogate, we use 5000 SMC particles and 20 rejuvenation steps per SMC iteration (which would result in 1 million likelihood calls for 10 SMC iterations). For the adaptive step size of the SMC iterator a control parameter of $\smczeta = 0.995$ was used (see \Cref{alg:smc}).

\begin{figure}[htbp!]
  \centering
  \vfill
  \subfloat[\label{subfig:sc_jointplot}]{
  \includegraphics[width=0.47\textwidth]{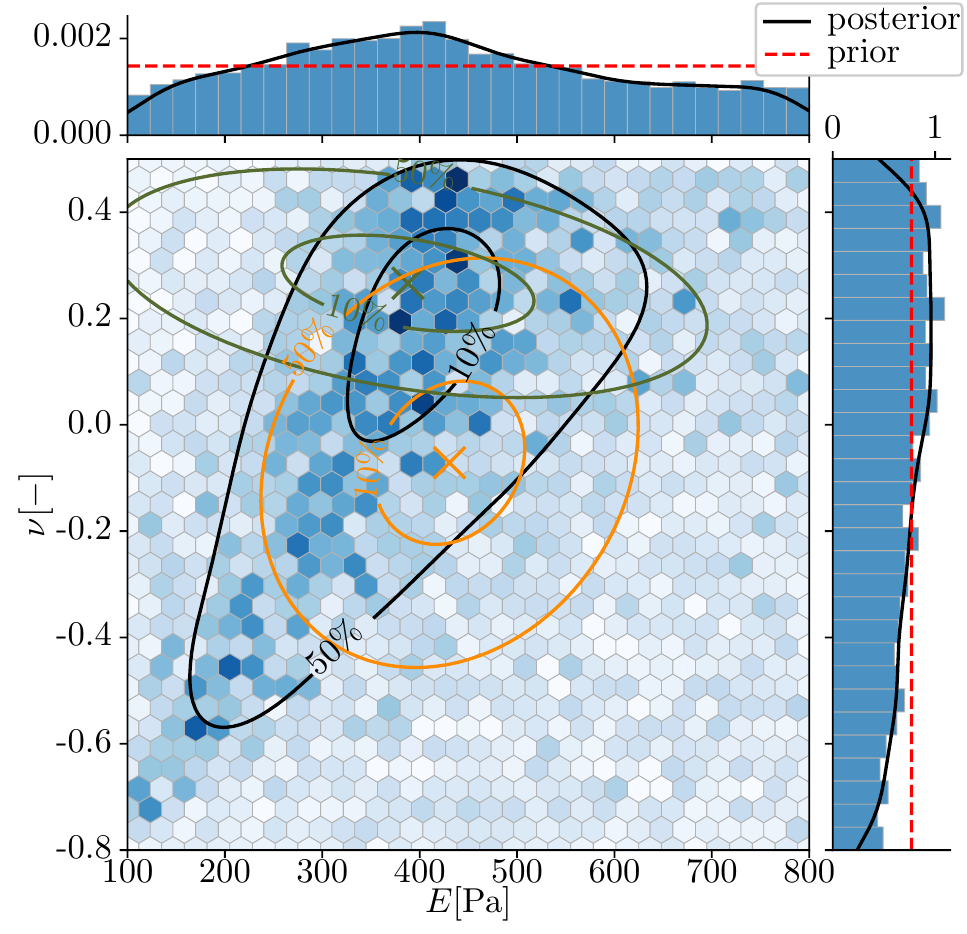}}
  \hfill
  \subfloat[\label{subfig:sc_particles}]{
  \includegraphics[width=0.49\textwidth]{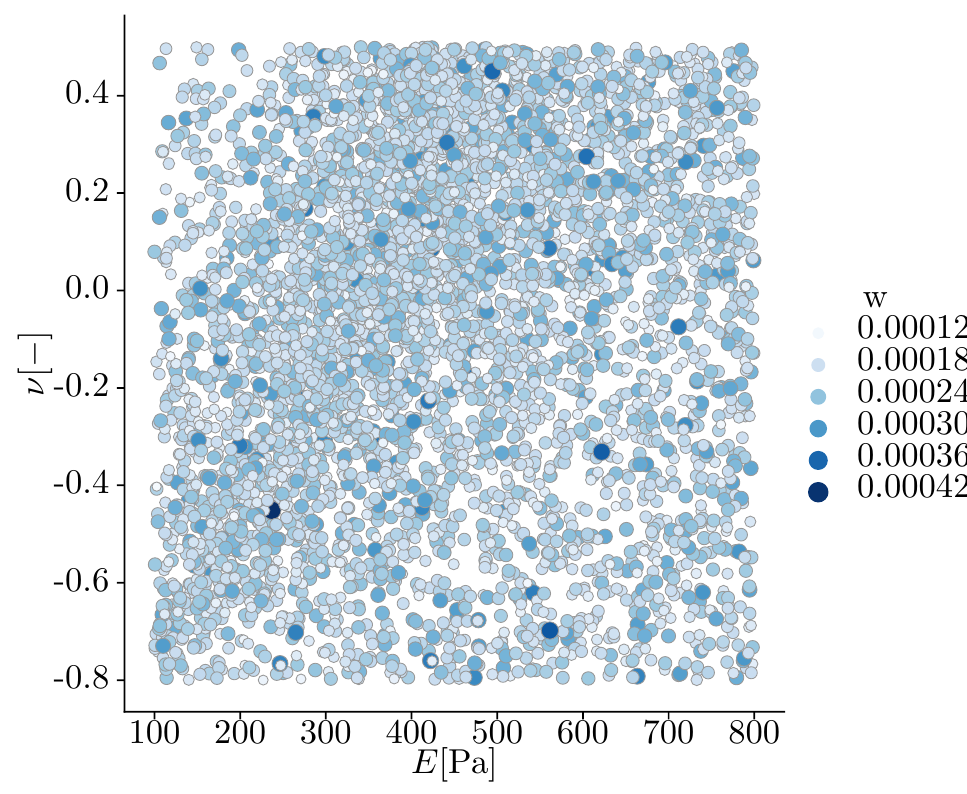}}
  \\
  \centering
  \subfloat[\label{subfig:sc_jointplot_prior}]{
  \includegraphics[width=0.47\textwidth]{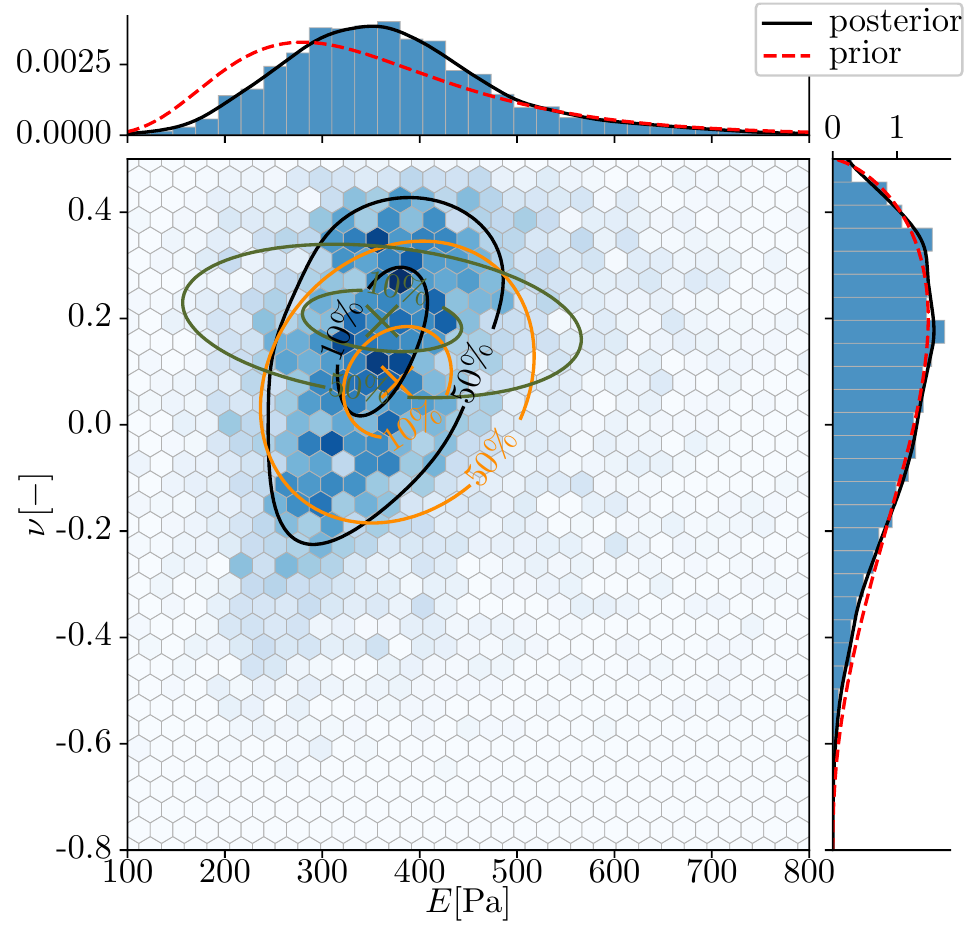}}
  \hfill
  \subfloat[\label{subfig:sc_jointplot_uia}]{
  \includegraphics[width=0.47\textwidth]{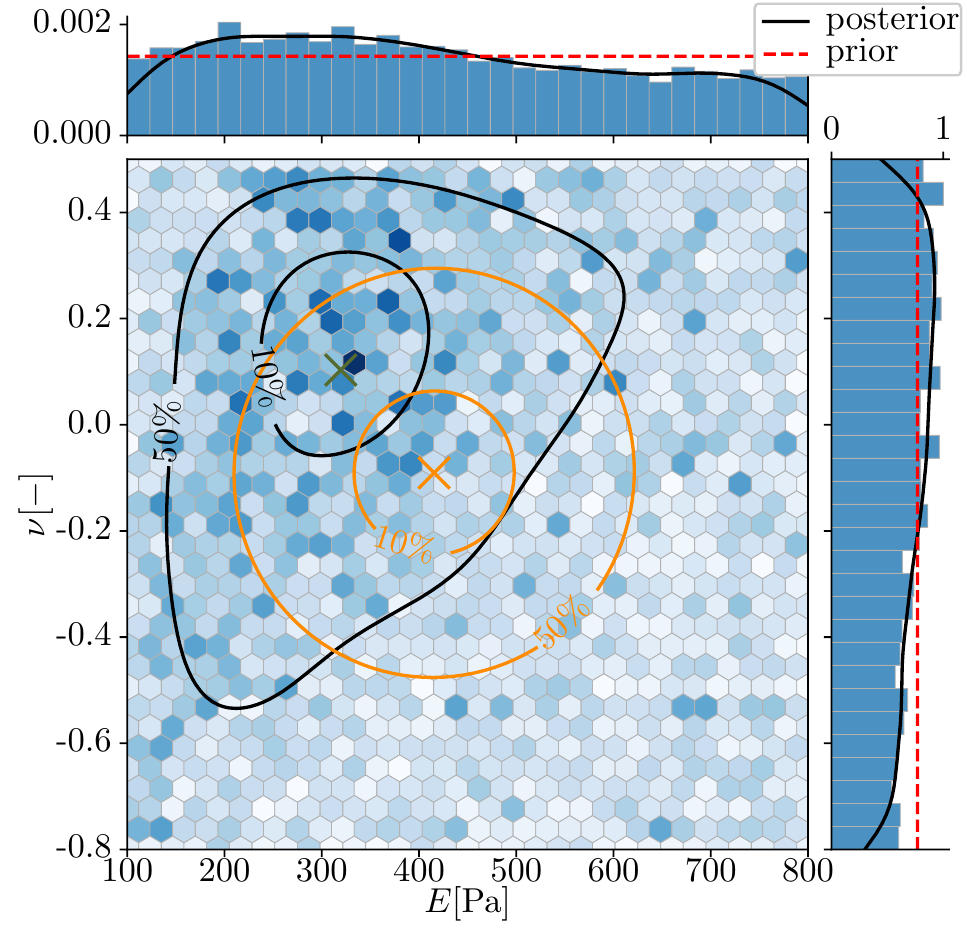}}
  \hfill
  \caption{Representation of the posterior distribution $\pdenscond{\youngs,\poissonratio}{\yobs}$ in histograms. Two-dimensional hexagon histograms for combined posterior and according marginals $\pdenscond{\youngs}{\yobs}$ and $\pdenscond{\poissonratio}{\yobs}$ attached to top and sides. Percentiles of kernel density estimates depicted as black solid lines, global Gaussian approximations as orange solid lines around the posterior mean ($\xpm$ as orange cross) and the Laplace approximations as a green solid lines around the MAP estimate ($\xmap$ as green cross). (a) Posterior with uniform prior and (b) resulting particle and weight distribution from SMC. (c) Posterior with informed priors. (d) Posterior with uniform prior and uncertain inflow.}
  \label{fig:sc_posteriors_twodim}
  \end{figure}

In the following examples, we examine the influence of priors in the parameters and uncertainties on the resulting posteriors.
In \Cref{fig:sc_posteriors_twodim} all results discussed in this section are plotted on the same page for better comparability. The different cases will be described and discussed subsequently in the following paragraphs. \Cref{fig:sc_posteriors_twodim} displays the resulting posteriors of the examples in form of SMC particle approximations. To visualize the character of the data, one particle distribution is shown explicitly in \Cref{subfig:sc_particles}, where the particles are plotted at their respective coordinates. The particle weights are illustrated as circle size and in a color scale. The other three subplots (\Cref{subfig:sc_jointplot}, \Cref{subfig:sc_jointplot_prior} and \Cref{subfig:sc_jointplot_uia}) show two-dimensional hexagonal histograms in a color scale. The particles are sorted into the displayed bins and summed up using their weights. The percentiles of the two-dimensional posteriors are additionally simplified as kernel density estimates (KDE) that are shown as black solid lines. For the KDE a radial basis function (RBF) kernel with bandwidth optimization was used.

On the top and right sides, the marginal distributions depending on both individual input parameters are plotted as histograms. The one-dimensional marginal posterior distributions $\pdenscond{\youngs}{\yobs}$ and $\pdenscond{\poissonratio}{\yobs}$ can easily be approximated by sorting the weighted particles into bins in the respective dimension. The marginals are additionally displayed in form of KDEs as black solid lines. The used priors are indicated as red dashed lines in all following plots.

Characteristic points deduced from the particle approximation are marked as crosses. The maximum a posteriori (MAP) estimate $\xmap$ is marked in green and the posterior mean (PM) $\xpm$ in orange. 

In high dimensions, the posterior cannot be plotted as easily as in the two-dimensional case. The analyst wants to have a comparative overview of where the mean value of the posterior is and how the probability mass is distributed in the posterior. For that and as another common approximation, we also show percentile lines of the global Gaussian approximation to the posterior in orange. The global Gaussian approximation is parameterized by the posterior mean (PM) vector $\xpm$ and the covariance matrix which can both be calculated from the weighted SMC particles very quickly.

We also present the Laplace approximation \cite{stat_bishop2011} of the posterior distribution around the MAP estimate depicted in \Cref{subfig:sc_jointplot} and \Cref{subfig:sc_jointplot_prior} with green solid lines for the two cases without uncertainty. The Laplace approximation can be understood as a local quadratic approximation of the posterior density function around its MAP in the log space in form of a Gaussian distribution. The covariance matrix of the resulting local Gaussian distribution gives an idea of how fast the posterior changes in a certain direction, starting from the MAP estimate. Please note, that the posterior distribution is not known in closed form but only approximated by SMC particles with associated weight. We approximate the MAP estimate by the SMC particle that scored the highest posterior value in the last rejuvenation step of the last iteration. In these examples, the necessary gradients of the posterior distribution were approximated by finite differences w.r.t. the input variables $\paramvec$ on the GP regression model. The Laplace approximation was omitted for the case with uncertain inflow. Laplace approximations are used for approximative Bayesian inverse analysis methods \cite{lnm_inv_kehl2016}, where the MAP is first found through optimization and then the Lagrange approximation is computed for an estimate of the variance.

\subsubsection{Influence of prior assumptions on the posterior distribution}

To show the influence of quantifiable prior knowledge on the posterior, we discuss the same example with uniform prior and a combined beta-distribution and log-normal distribution prior on the model input $\paramvec$.

\paragraph{Uniform prior}
In this first demonstration we choose an uninformative uniform prior distribution for Young's modulus $\pdens{\youngs}=\unidist{\youngs}{100\sunit{Pa}}{800\sunit{Pa}}$ and for the Poisson ratio $\pdens{\poissonratio}=\unidist{\poissonratio}{-0.8}{0.5}$. Those are independent priors on the parameters and can be combined to $\pdens{\youngs, \poissonratio} = \pdens{\youngs} \pdens{\poissonratio}$.

\Cref{subfig:sc_jointplot} displays the approximation of the resulting posterior in form of a hexagonal histogram plot generated with the weighted particles from the SMC run. In \Cref{subfig:sc_particles} the resulting particle distribution of the SMC along with the colored-coded particle weight is shown.

The posterior in \Cref{subfig:sc_jointplot} shows an almost linear band shape of high densities. This shows that it is more crucial to have a good value for Young's modulus $\youngs$ than one for Poisson's ratio $\poissonratio$ to obtain good similarity between model output and reference data for the given parameter ranges. The posterior mean (PM) vector computes to $\xpm=\transpose{[\youngs\approx431\sunit{Pa},\; \poissonratio\approx-0.0071]}$. In \Cref{subfig:sc_jointplot} the Gaussian approximation has the same orientation as the particle approximation to the posterior. Still it can not represent the posterior complexity. The maximum a posteriori (MAP) is approximated to $\xmap = \transpose{[\youngs\approx388\sunit{Pa},\; \poissonratio\approx0.266]}$. This $\xmap$ is close to the ground truth in relation to the number of training points and the resulting resolution of training points in the input space.

Interestingly, the Laplace approximation, represented by its percentile contour lines in \Cref{subfig:sc_jointplot} in green, is almost oriented orthogonal to the actual posterior (solid black line). This means, that the Laplace approximation gives a misleading local approximation in this specific case. That can occur if the posterior has a more complex curvature around the MAP. In our example, this might be partially induced by the GP approximation, as well. It seems that the Laplace approximation cannot live up to the complexity of the given posterior and therefore simplified methods based on the Laplace approximation would work poorly in presented examples without the analyst knowing. This is why we advocate the fully Bayesian treatment for this kind of problems.

As mentioned before, we also plot the marginal posterior distributions of $\pdenscond{\youngs}{\yobs}$ and $\pdenscond{\poissonratio}{\yobs}$, respectively. The marginals show that $\pdenscond{\youngs}{\yobs}$ has a single, stable, global optimum and $\pdenscond{\poissonratio}{\yobs}$ forms a plateau of high densities. Due to the strong coupling of $\youngs$ and $\poissonratio$ and the complex shape of the joint posterior $\pdenscond{\youngs,\poissonratio}{\yobs}$, the marginal posterior distributions alone however are not informative for the coupling effects in the global posterior distribution.

\paragraph{Informed prior}

Physical insight can be incorporated in prior assumptions and can have a great influence on the posterior distribution and should be integrated in the analysis. Besides the uninformative uniform prior that we used in the first example, we now also want to demonstrate the effect of an informed prior.
For the Young's modulus, a log-normal prior is assumed with a mode of $\youngs = 300 \sunit{Pa}$. The log-normal can be parameterized with 
$ \lognormaldist{\youngs}{\mu_{\mathcal{LN}}\approx5.86}{\sigma_{\mathcal{LN}}= 0.4}$ with parameters 
$\mu_{\mathcal{LN}}, \sigma_{\mathcal{LN}}$ that are not standard deviation and mean of the distribution, but 
$\log{(\youngs)}\sim \normaldist{\mu_{\mathcal{LN}}}{\sigma_{\mathcal{LN}}^2}$. This accounts for the fact that the Young's modulus must have positive values and it is very unlikely that it is close to zero but more probable to find values higher than the mode. For the Poisson's ratio a beta-distribution $\betadist{\poissonratio}{a=43/13}{b=22/13} $ between $-0.8$ and $0.5$ is used as prior. This distribution has its mode at $\poissonratio = 0.2$ and accounts for the belief that the Poisson's ratio is more probable to have positive values and is strongly bounded between $-1.0$ and $0.5$ with a decreasing probability towards the boundaries of this interval.

The modes of the presented priors are intentionally chosen to deviate from the ground truth $\gtx$ to show the effect of informative priors. The priors are plotted in \Cref{subfig:sc_jointplot_prior} alongside the respective marginal posterior distributions. It can be easily observed that compared to the uniform priors used in \Cref{subfig:sc_jointplot}, the informed prior assumptions, i.e., distributions that weight specific areas in the input space higher than others, lead to a posterior distribution which is more pronounced around the ground truth $\gtx$ and has less probability mass in regions with low prior density. As the majority of the probability mass of the resulting posterior is now in a more compact area of the design space, also the marginal posterior distributions $\pdenscond{\youngs}{\yobs}$ and $\pdenscond{\poissonratio}{\yobs}$ show a more defined shape with one predominant mode. 

Similar to \Cref{subfig:sc_jointplot}, the joint posterior $\pdenscond{E,\nu}{\yobs}$ has more variance in the $\poissonratio$-dimension, rendering $\poissonratio$ a \emph{sloppy} parameter. This becomes also apparent as the marginal posterior distribution $\pdenscond{\poissonratio}{\yobs}$ almost coincides with the prior assumption $\pdens{\poissonratio}$, indicating a very small influence of this parameter on the likelihood function. This also means that for this type of measurement the exact value of model parameter $\poissonratio$ has less importance for the agreement of mechanical model and observed experiment. In $\pdenscond{\youngs}{\yobs}$ on the other hand the density does not have the same mode as the prior $\pdens{\youngs}$. This underlines that the likelihood contains characteristic information about this parameter.

The MAP estimate has the values $\xmap = \transpose{[\youngs\approx361\sunit{Pa},\; \poissonratio\approx0.195]}$ in this case. This is slightly lower in both parameter values than in the run with uniform prior. With the prior modes deviating from the ground truth, a deviation of $\xmap$ from the ground truth towards lower values is expected.
Furthermore, we see that the global Gaussian approximation of the posterior around $\xpm = \transpose{[\youngs\approx377\sunit{Pa},\; \poissonratio\approx0.08]}$ moves closer to the mode of the posterior distribution, as the low likelihood areas are further weighted with low prior values and therefore lose probability mass compared to the uniform priors. The Laplace approximation is still not a very good local approximation of the posterior.

\subsubsection{Material calibration under uncertain boundary conditions}

\begin{figure}[htpb]
\centering
\includegraphics[width=0.47\textwidth]{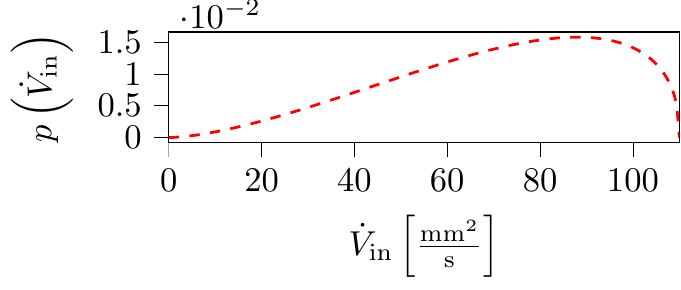}
\caption{Assumed beta-distribution $\pdens{\volinflowrate} $ of assumed uncertain inflow volume rate $\volinflowrate$.}
\label{fig:sc_marg_uq}
\end{figure}

Now we also demonstrate the treatment of additional uncertain influences on the system, denoted as $\theta$ above. We use the example of an uncertain inflow volume rate in the flow cell experiment. As generally the biggest biofilm patches in the channel are analyzed, they take up a significant portion of the cross-section of the channel and force the flow to go around it. Further, only the middle of the channel can be scanned to high quality using OCT. Hence, the distribution of the volume flow rate between outlying parts of the cross-section and the analyzed patch is subject to uncertainties. Overall these considerations are summed into the assumption of an uncertain inflow rate distribution, that has its main mode significantly lower than the average inflow rate and a spread distribution around that, with low density for higher flow rates. This is modeled with an assumption of $\pdens{\volinflowrate} $ as a beta-distribution $\betadist{\volinflowrate}{a=2.6}{b=1.4} $ between $0 \sunitfrac{mm^2}{s} $ and $ 110 \sunitfrac{mm^2}{s} $ with mode at $88 \sunitfrac{mm^2}{s}$ which is plotted in \Cref{fig:sc_marg_uq}. The value of the volume inflow rate used for data generation is kept the same as in all other examples $\volinflowrate = 100\sunitfrac{mm^2}{s} $. The distribution accounts for the belief that only less than average of the fluid volume rate flows over the biofilm as compared to the rest of the channel. For the two material parameters uniform priors $\pdens{\youngs}=\unidist{\youngs}{100\sunit{Pa}}{800\sunit{Pa}}$ and $\pdens{\poissonratio}=\unidist{\poissonratio}{-0.8}{0.5}$ were used.

The resulting joint posterior of the two analyzed parameters under the influence of the uncertain inflow are plotted in \Cref{subfig:sc_jointplot_uia}.
The posterior under uncertainty, denoted by $q(\youngs,\poissonratio|\yobs)=\Expectwrt{\volinflowrate}{\pdenscond{\youngs,\poissonratio}{\volinflowrate,\yobs}}$, was then calculated according to \eqref{eq:uncertain_bayes} by incorporating the average effect of the uncertainty of the inflow $\volinflowrate \sim \pdens{\volinflowrate}$. In \Cref{subfig:sc_jointplot_uia} it can be seen that, as it should be expected, the additional consideration of uncertainty of the inflow made the posterior less expressive, such that an increase in the variance can be especially found along the dimension of the Young's modulus.

The consideration of an uncertain boundary condition has also moved the point estimates. The MAP estimate for both inputs $\paramvec$ moves to lower values of
$\xmap=\transpose{[\youngs\approx322\sunit{Pa},\; \poissonratio\approx0.09]} $ than in the uniform prior example without uncertainties. The posterior mean is $\xpm = \transpose{[\youngs\approx416\sunit{Pa},\; \poissonratio\approx-0.09]}$.
This is expected as the mean of the assumed inflow distribution is lower than the value used for data generation and therefore lower stiffness leads to a better suiting deformation. 

\begin{remark}[MAP estimation after marginalization]
With the consideration of uncertain parameter $\tns \theta (=\volinflowrate) $ the MAP estimate is more difficult to obtain than without uncertainties. MAP determination is more difficult because the extended posterior \eqref{eq:uncertain_bayes} must first be integrated to the posterior under uncertainty. Although this integration can easily be evaluated using the particle representation \eqref{eq:smcintegration}, the maximum of the posterior under uncertainty cannot be found without another assumption. We chose a histogram approach to collect the SMC particles in squared bins with 30 intervals per input variable $\paramvec$ according to the illustration in \Cref{subfig:sc_jointplot_uia}. The trick of picking the particle that scored the highest posterior in the last SMC iteration does not work for the posterior under uncertainty or marginal distributions, as their determination first needs an integration step. 
\end{remark}

The global Gaussian approximation is more isotrop when considering the uncertainty. This is represented as a Gaussian approximation that has more circular percentile lines. In the interpretation of the covariance of the posterior, this means that there is no prevalent direction in this posterior.
Compared to the posterior for the fixed boundary condition, the posterior including the uncertainty $q(\youngs,\poissonratio|\yobs)=\Expectwrt{\volinflowrate}{\pdenscond{\youngs,\poissonratio}{\volinflowrate,\yobs}}$ is less restrictive by means of necessary assumptions and therefore also less stiff in the results. This gives also more flexibility in the results as there is a broader range of parameters to explain the observed experimental results. Nevertheless, better knowledge about uncertainties $\pdens{\volinflowrate}$ can greatly improve the calibration result.

In case non-controllable \emph{aleatory} uncertainty is present, neglecting it will lead to an overconfident, wrong posterior as the analyst introduces a modeling error by neglecting these effects.
Incorporating aleatory uncertainty in the probabilistic model will generally introduce more uncertainty to the posterior (the distribution widens) and might potentially even completely change the characteristics of the posterior distribution.

\subsection{Calibration of heterogeneous biofilm model under uncertain inflow boundary condition}

\begin{figure}[htbp]
\centering
\begin{minipage}[b]{0.3\textwidth}
\includegraphics[width=\textwidth]{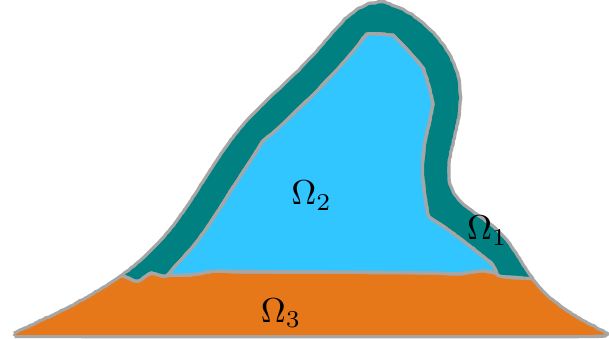}
\vspace{.3cm}
\caption{Subdomains for heterogeneous biofilm model.}
\label{fig:subdomains}
\end{minipage}
\hfill
\begin{minipage}[b]{0.65\textwidth}
\centering
\includegraphics[width=\textwidth]{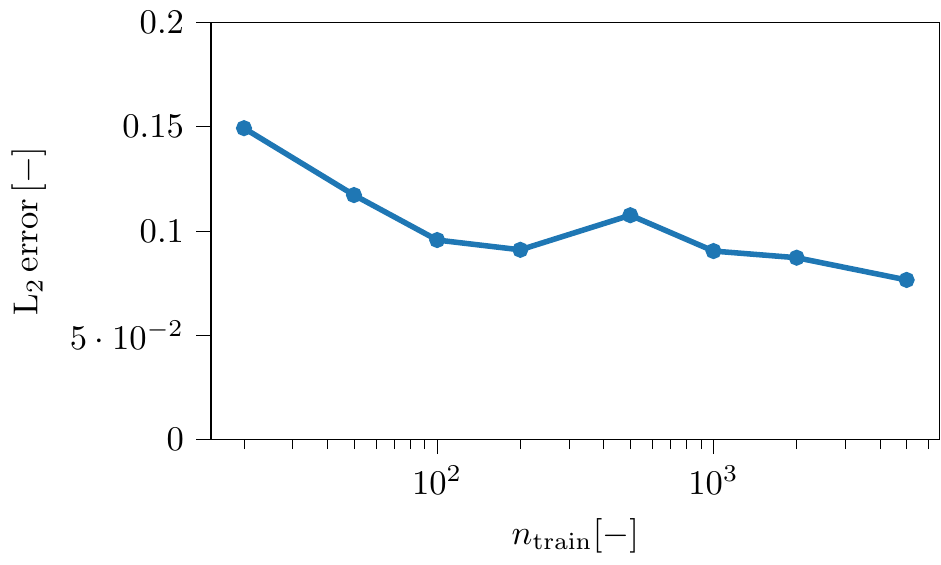}
\caption{Error for GP regression model with $\ntest = 100$ test points over number of training data points $\ntrain$.}
\label{fig:6p_gp_e}
\end{minipage}
\end{figure}

In our last example we want to calibrate material properties of a heterogeneous biofilm FSI model as in \cite{me_willmann_2021}, but here under an uncertain inflow rate boundary condition. Heterogeneity comes into play in such problems due to different age and/or different nutrient availability of different parts of the domain. Hence, it was a natural choice to also shed some light on a more demanding case involving more parameters. The three subdomains are depicted in \Cref{fig:subdomains} and lead to six input parameters, consisting of the Young's modulus and Poisson's ratio of each subdomain.

The parameters $\gtx=\transpose{[\youngs_1 = 500 \sunit{Pa},\; \poissonratio_1 = 0.2,\;\youngs_2 = 200 \sunit{Pa},\; \poissonratio_2 = 0.1 ,\; \youngs_3 = 1000 \sunit{Pa},\; \poissonratio_3 = 0.3]}$ for a hyperelastic Saint-Venant-Kirchhoff material model are used as the ground truth inputs, along with the inflow rate $\volinflowrate=100\sunitfrac{mm^2}{s}$. All other model details are the same as for the previous examples, to then generate the synthetic experimental data $\yobs$ for this heterogeneous case. Please note that in this example we additionally assume noise polluted measurement data according to
\begin{equation}
    \label{eq:noisy_data}
    \begin{split}
    \yobs &= \model{\bx_{\text{gt}},\btheta_{\text{gt}},\coords} + \sigma_{\text{n,obs}}\cdot \boldsymbol{\epsilon}\\
    \text{with } \boldsymbol{\epsilon} &\sim \normaldist{\boldsymbol{0}}{I}
    \end{split}
\end{equation}

For the following example we choose noise with a standard deviation of $\sigma_{\text{n,obs}}=1\sunit{\mu m}$. In \cref{rem:convgp6p} we comment on the determination of the GP surrogate for this example.
\begin{remark}[Convergence of the Gaussian process surrogate]
  \label{rem:convgp6p}
As the GP surrogate model for the likelihood is now dependent on six input variables plus one uncertain variable, more training data is necessary to reach acceptable accuracy of the posterior mean function of the GP. A small convergence study was performed to find an appropriate training size. Therefore we successively increased $\ntrain$ according to the Sobol sequence and then calculated the $\text{L}_2$-error norm between the posterior mean of the GP and the likelihood evaluated with the forward model at $\ntest=100$ testing points that are fixed consecutive samples from later in the Sobol sequence and unused in the training data. This is a representative choice as the good space filling property leads to test points which are well distributed. The result of the convergence study is plotted in \Cref{fig:6p_gp_e}. The tested scenario is a kernel with multiple length scales with a multiplicative coupling \cite{stat_rasmussen2005, stat_duvenaud2014} as used in the following example. The parameters are expected and also observed to have different influence on the likelihood. Therefore different length scales and variances in a multiplicative coupling of Matérn $2/3$ kernels for each parameter dimension is used. As opposed to the L-BFGS-B optimizer used in previous examples, a scaled conjugate gradient optimizer is used for training for stability reasons.

\Cref{fig:6p_gp_e} shows that between $\ntrain=1000$ and $\ntrain=5000$ no significant improvement in the given error measure can be achieved. That is why the evaluation with $\ntrain=2000$ data points is chosen for the following example as a compromise between efficiency and accuracy. In general asymptotic convergence can be expected, meaning that the error measure will go to zero for an infinite amount of training points. For $\ntrain=500$ an increase in the error can be detected. Here a new characteristic in the likelihood function was introduced with the training points between $\ntrain=200$ and $\ntrain=500$, that lead to the increased deviation at the test points.
\end{remark}

For the SMC approach a set of 10000 particles with 30 rejuvenation steps was used with a step size control parameter of $\smczeta = 0.998$. Just as a comparison to the $\ntrain=2000$ training points used, a full evaluation in every rejuvenation step of the SMC algorithm would require three million ($10000\cdot30\cdot10= 3000000$) forward model evaluations for exemplary 10 SMC steps. This is neither desirable nor feasible if directly applied to the expensive forward model. After the proof of concept in previous examples, with two input parameters and potentially one uncertain parameter, it must be emphasized, that volume integrals for respective marginals and especially the normalization of the posterior \eqref{eq:bayes} in six plus one dimensions scale even worse. Just for a moderately sized, grid-based discretization with 100 sample points in every dimension we would get $100^7$ necessary evaluations of the GP-mean. So we make use of the good scalability of the SMC algorithm in higher dimensions in the following examples.

As already indicated above, in this last example we want to show the full capability of the approach and therefore we use an example with uncertain inflow rate $\volinflowrate $ and added noise to the data generated from the ground truth values. The assumed distribution on $\volinflowrate$ is the same as in \Cref{fig:sc_marg_uq} above with a beta-distribution $\betadist{\volinflowrate}{a=2.6}{b=1.4} $ between $0 \sunitfrac{mm^2}{s} $ and $ 110 \sunitfrac{mm^2}{s} $ with its mode at $88 \sunitfrac{mm^2}{s}$. The MAP approximation is found at 
$\xmap=\transpose{[\youngs_1 \approx 345 \sunit{Pa},\; \poissonratio_1 \approx 0.43,\;\youngs_2 \approx 135 \sunit{Pa},\; \poissonratio_2 \approx 0.30 ,\; \youngs_3 \approx 1034 \sunit{Pa},\; \poissonratio_3 \approx -0.47]}$ and the global posterior mean computes to $\xpm = \transpose{[\youngs_1 \approx 443 \sunit{Pa},\; \poissonratio_1 \approx -0.15,\;\youngs_2 \approx 431 \sunit{Pa},\; \poissonratio_2 \approx -0.13 ,\; \youngs_3 \approx 637 \sunit{Pa},\; \poissonratio_3 \approx -0.14]}$.
Analogously to the lower dimensional example, the MAP estimate is obtained with a binning approach with 10 intervals in each input dimension. Therefore it is a rough estimate. Nevertheless, the curse of dimensionality inhibits an excessively fine grid.
The maximum of the posterior under uncertainty qualitatively represents the ground truth with $\youngs_3>\youngs_1>\youngs_2 $ and $\poissonratio_1 > \poissonratio_2$. Only for $\poissonratio_3$, it does not correspond to the ground truth. This can be a consequence of the relevance of the parameter to the overall posterior, as the lateral contraction in the stiffest subdomain, attached to the ground also has little effect on the interface deformation. In challenging examples it can be a good idea to first perform a global sensitivity analysis \cite{lnm_stat_Brandstaeter_2021,lnm_stat_wirthl_2022} and then focus the inverse analysis only on the most sensitive parameters. It also shows again, that the problem setup is challenging as the volume inflow rate is used as an uncertain parameter. But this is also often the case in real-world applications. This has the effect that the MAP parameters are those of a less stiff biofilm material, than the ones used for the ground truth simulation.

\begin{figure}[htpb]
\centering
\subfloat[\label{subfig:heteronoise6p_margE1}]{
\includegraphics[width=0.47\textwidth]{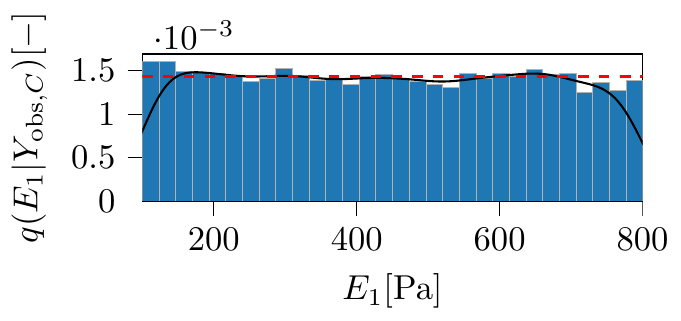}}
\hfill
\subfloat[\label{subfig:heteronoise6p_margnue1}]{
\includegraphics[width=0.47\textwidth]{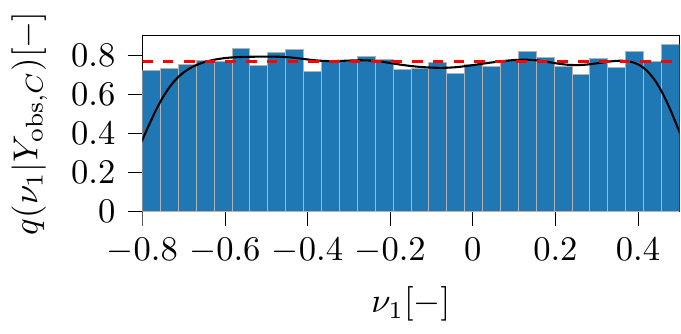}}
\hfill
\subfloat[\label{subfig:heteronoise6p_margE2}]{
\includegraphics[width=0.47\textwidth]{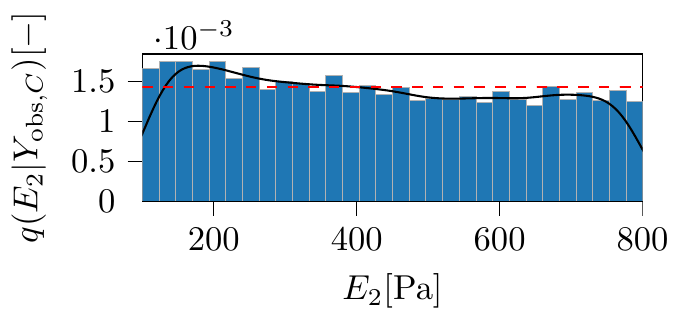}}
\hfill
\subfloat[\label{subfig:heteronoise6p_margnue2}]{
\includegraphics[width=0.47\textwidth]{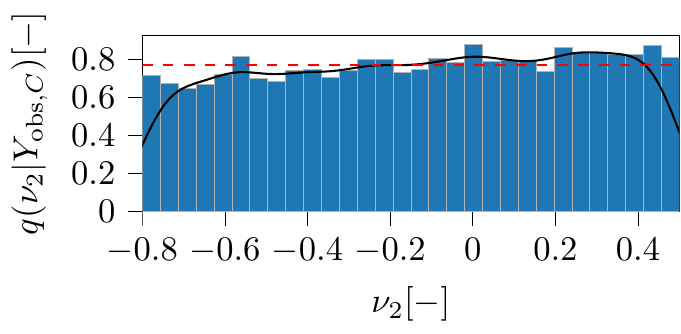}}
\hfill
\subfloat[\label{subfig:heteronoise6p_margE3}]{
\includegraphics[width=0.48\textwidth]{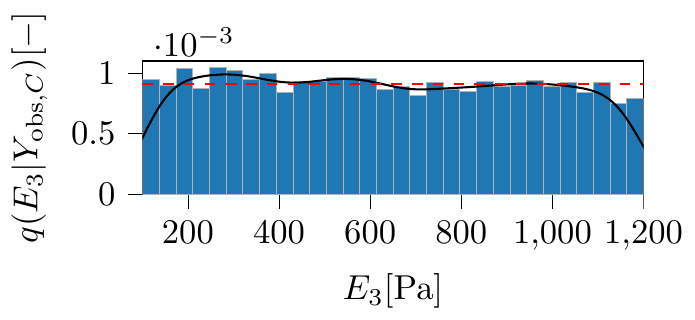}}
\hfill
\subfloat[\label{subfig:heteronoise6p_margnue3}]{
\includegraphics[width=0.47\textwidth]{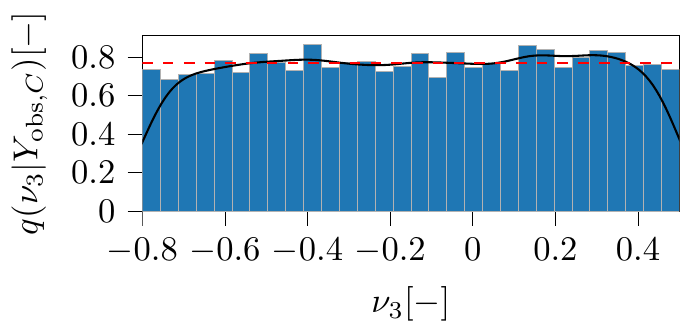}}
\caption{Marginals of the posterior for (a)-(f) individual parameters in heterogeneous example with uncertain inflow condition with added Gaussian noise and uniform prior as red dashed line. Marginals as histograms (blue) with a KDE approximation (black).}
\label{fig:heteronoiseuniuia_margs}
\end{figure}

\begin{figure}[htbp!]
\centering
\subfloat[\label{subfig:heteronoiseuniuia6p_E1E2}]{
\includegraphics[width=0.3\textwidth]{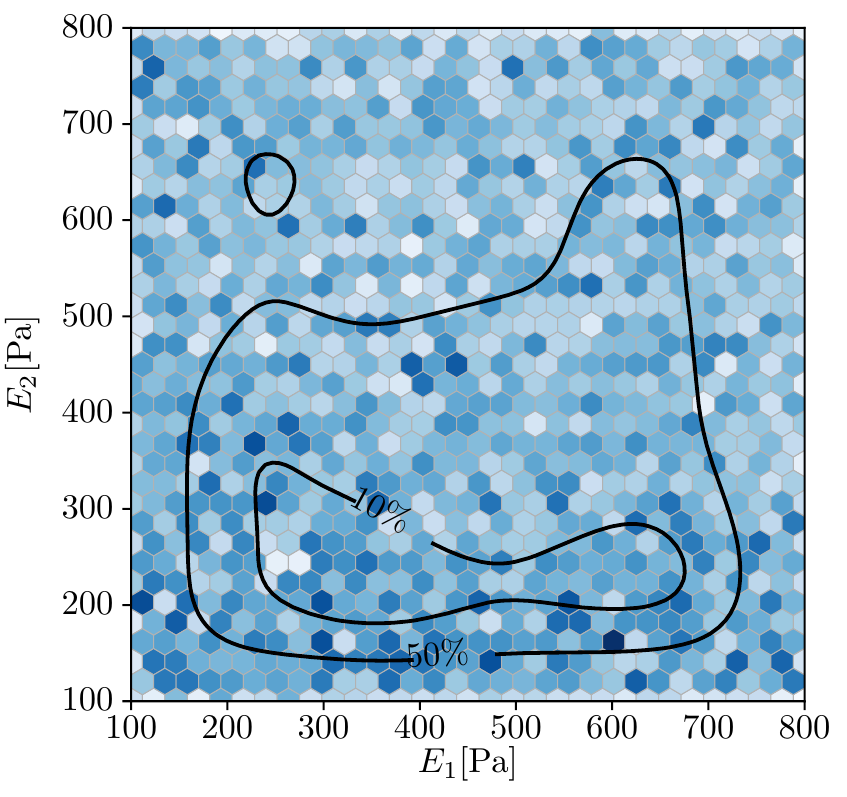}}
\hfill
\subfloat[\label{subfig:heteronoiseuniuia6p_E1E3}]{
\includegraphics[width=0.305\textwidth]{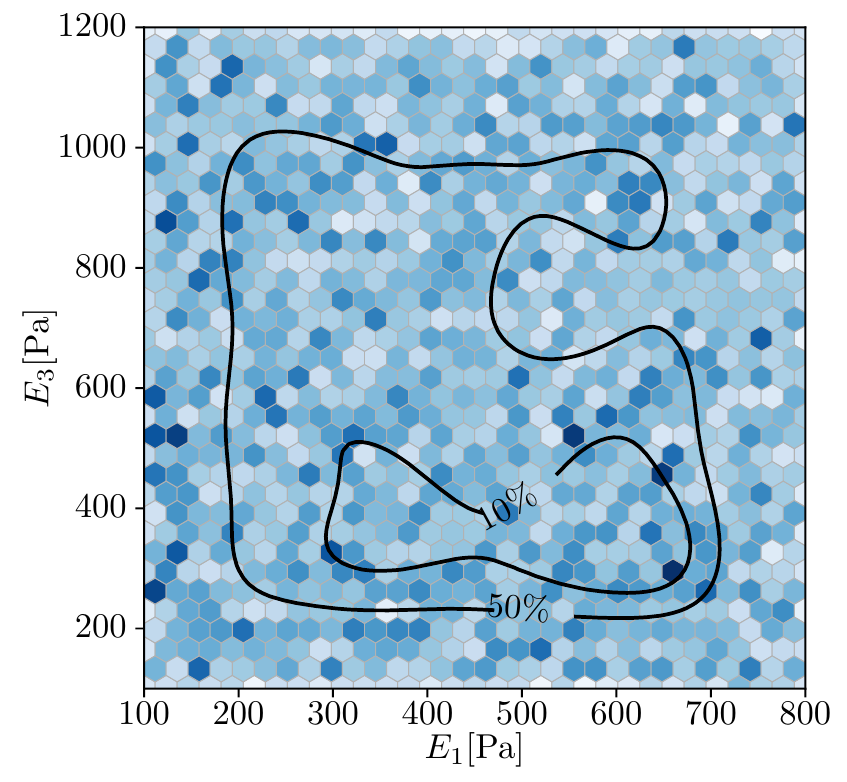}}
\hfill
\subfloat[\label{subfig:heteronoiseuniuia6p_E2E3}]{
\includegraphics[width=0.305\textwidth]{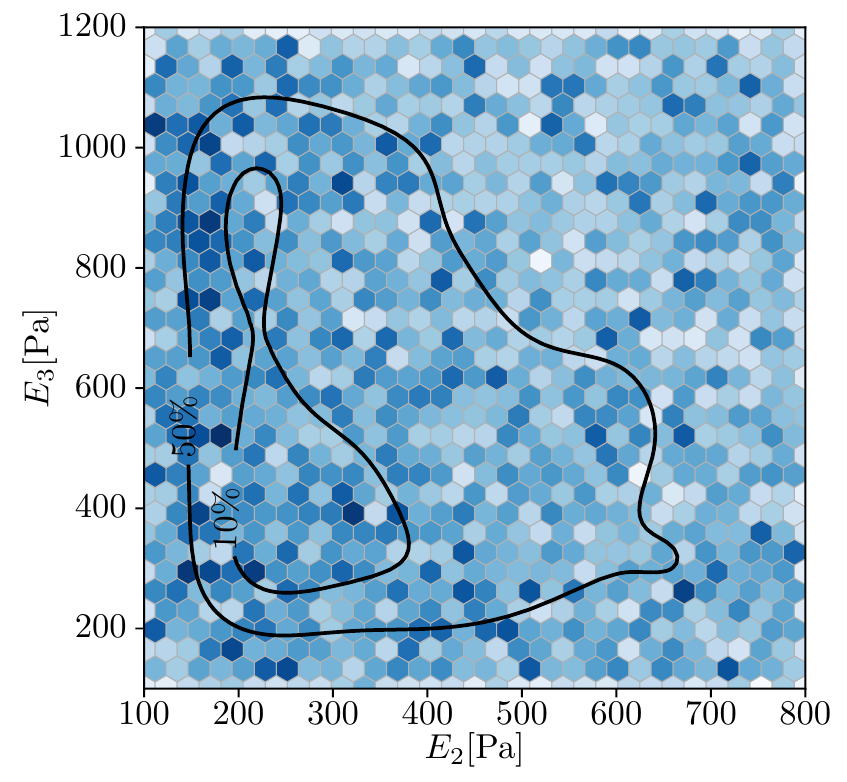}}
\hfill
\subfloat[\label{subfig:heteronoiseuniuia6p_E1nue1}]{
\includegraphics[width=0.3\textwidth]{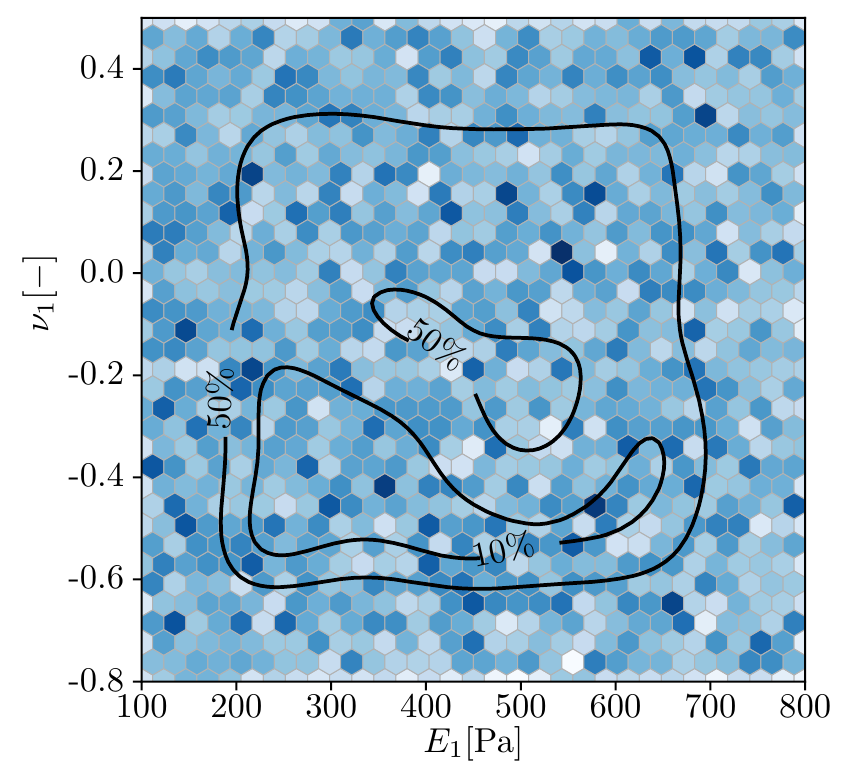}}
\hfill
\subfloat[\label{subfig:heteronoiseuniuia6p_E2nue2}]{
\includegraphics[width=0.302\textwidth]{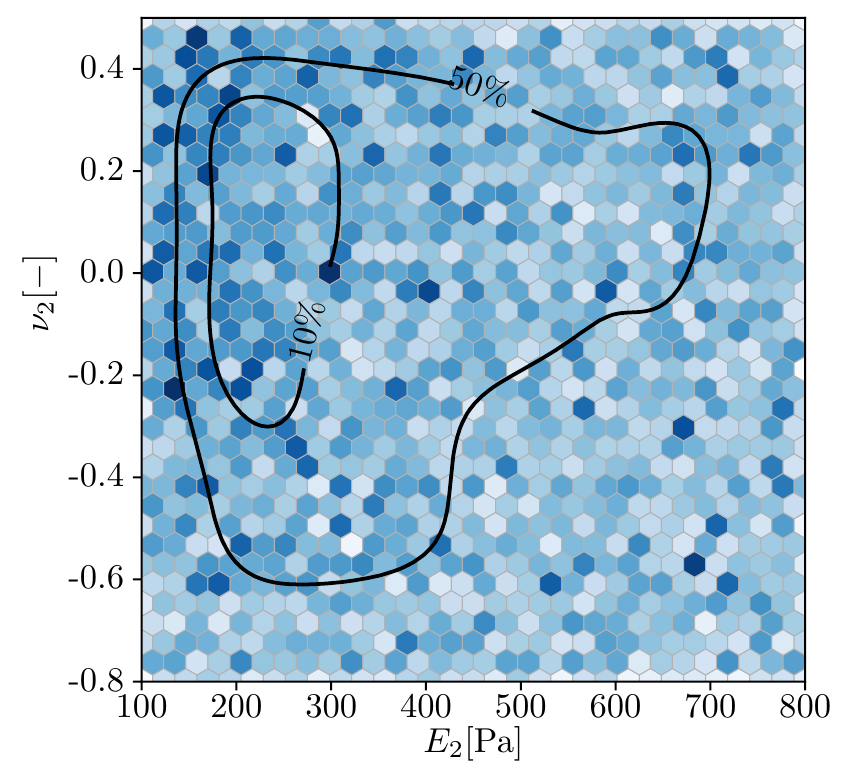}}
\hfill
\subfloat[\label{subfig:heteronoiseuniuia6p_E3nue3}]{
\includegraphics[width=0.305\textwidth]{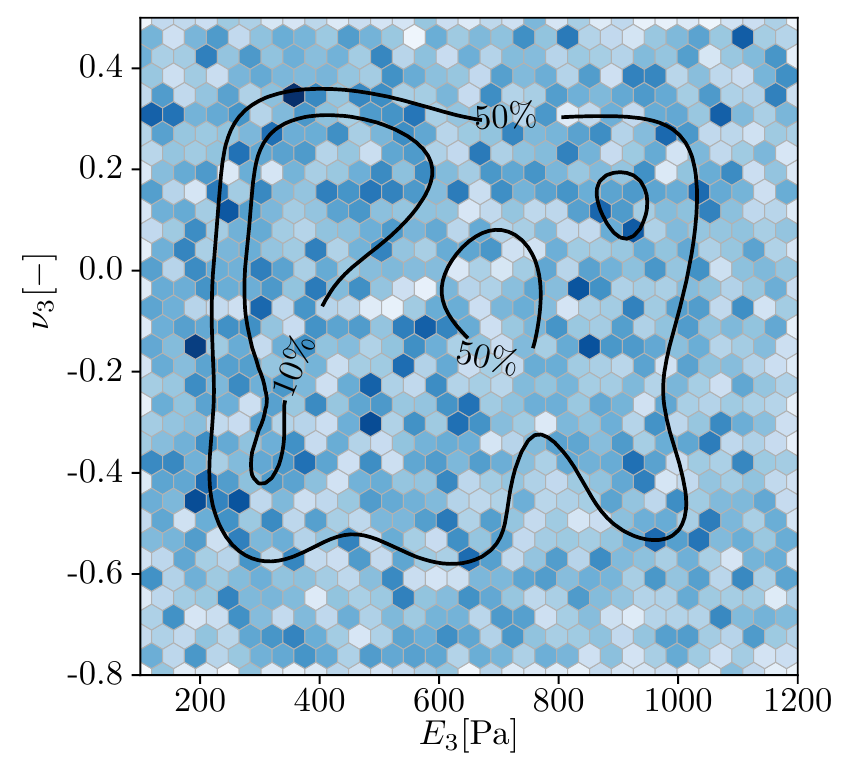}}
\caption{Hexagonal histograms (high weight in dark blue) for marginals for selected pairs of parameters in heterogeneous example with uncertain inflow condition and added Gaussian noise. Percentile lines of KDE approximations in black.}
\label{fig:heteronoiseuniuia2dofmargs}
\end{figure}

The resulting one dimensional marginals over the parameters, for which uniform priors were assumed, are plotted in \Cref{fig:heteronoiseuniuia_margs} as histograms with KDE approximations in black. Here the letter $q(\paramvec|\yobs)$ is used because the distributions represent marginals of the posterior under uncertainty \eqref{eq:uncertain_bayes}.
It can be observed that most of these marginals show a rather uniform distribution. In this example, this happens as they represent the integration of the posterior with respect to all other parameters respectively and therefore the probability mass is accumulated herein. Only for subdomain $\domain_2$ (in \Cref{subfig:heteronoise6p_margE2} and \Cref{subfig:heteronoise6p_margnue2}) the posterior marginals acummulate density around the ground truth $\youngs_2 \approx 200\sunit{Pa}$ and $\poissonratio_2 > 0$. As these single-dimensional marginals are not unveiling the full complexity of the posterior, a further step is made to show two-dimensional marginals for a selected combination of parameters in \Cref{fig:heteronoiseuniuia2dofmargs} as hexagonal histograms with the percentile lines of the KDE approximations. Therein the interaction effects between the respective combination of parameters can be seen. Especially within the regions enclosed by the percentile lines of $10\%$ the most probable parameter combinations for the respective marginals can be found.

The marginal posterior distributions in \Cref{fig:heteronoiseuniuia2dofmargs} show complex densities. As compared to a deterministic calibration approach for the same example \cite{me_willmann_2021}, where even the less complex case with a fixed volume inflow rate $\volinflowrate$ could not be solved, this complex character can be well represented here in the obtained solution. Furthermore, for \Cref{subfig:heteronoiseuniuia6p_E1E2} and \Cref{subfig:heteronoiseuniuia6p_E2E3} it can be seen that $\youngs_2 $ dominates the marginal posteriors that are high around $\youngs_2 \approx 200 \sunit{Pa}$ for a plausible range of the other Young's moduli $\youngs_1,\youngs_3$. Also the shape of the distribution for $\domain_2 $ in \Cref{subfig:heteronoiseuniuia6p_E2nue2} resembles the posterior shape in \Cref{subfig:sc_jointplot_uia}, which is the one for homogeneous example with uniform prior and uncertain inflow. This means that parameters in $\domain_2$ qualitatively have similar influence in this marginal, as the two parameters have in the homogeneous example in the posterior. This is also expected as $\domain_2 $ takes up the highest portion of the biofilm domain (see \Cref{fig:subdomains}) and therefore dominates the deformation.

\begin{figure}[htbp!]
\centering
\includegraphics[width=0.8\textwidth]{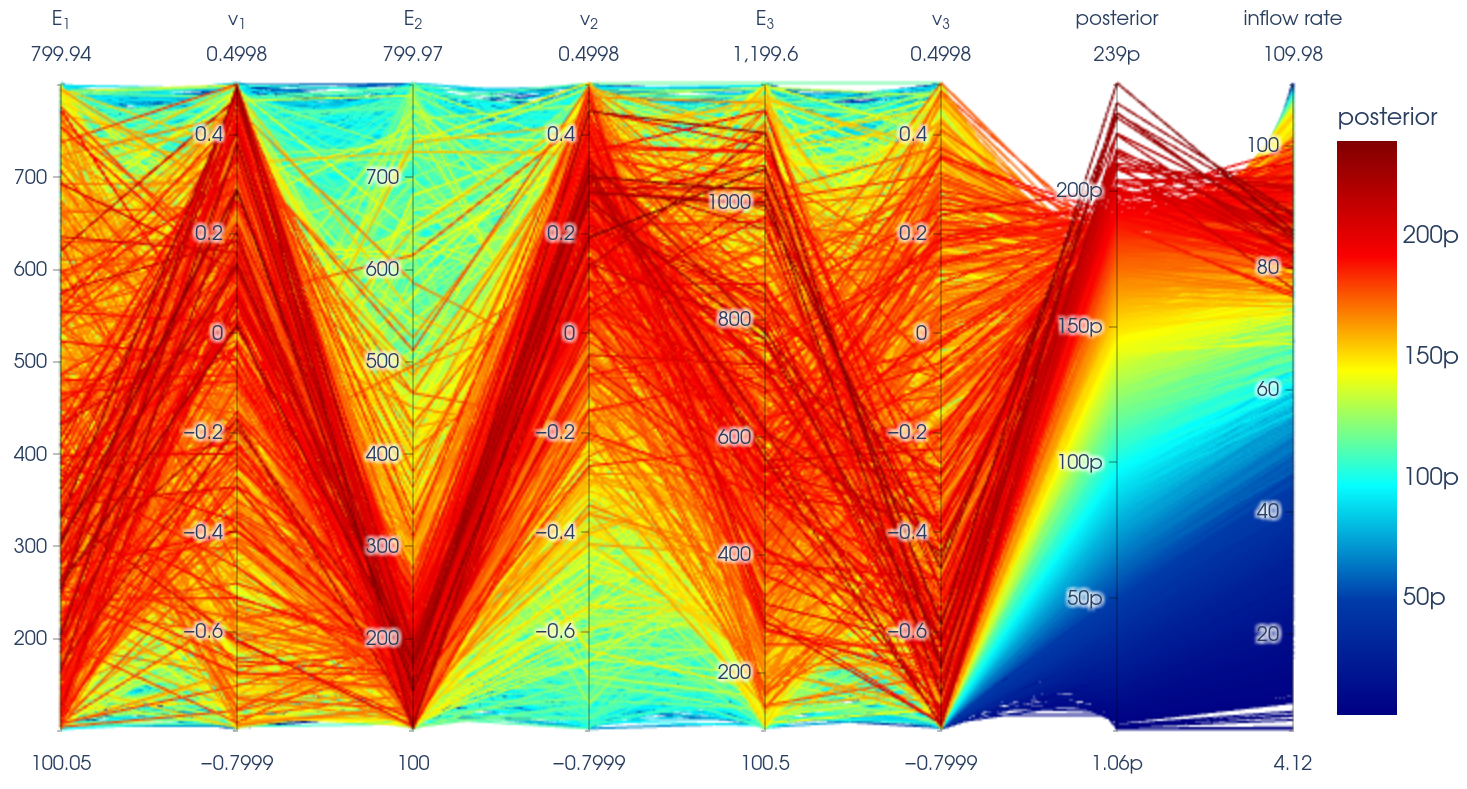}
\caption{Parallel axis plot over all parameters, posterior values $\pdenscond{\paramvec}{\btheta,\yobs}$ and inflow rate for all particles of the SMC particle approximation of the posterior with uniform prior assumptions for $\paramvec$.}
\label{fig:papuni}
\end{figure}

As we have six parameters involved, a full plot of the posterior is impossible. Alternatively, a parallel axis plot as in \Cref{fig:papuni} can be used to get an idea of the underlying posterior shape.
Over the first six axis we see the input parameters for the range of interest. The seventh axis shows the extended posterior values for the 10000 particles from the SMC particle approximation. The last axis represents the uncertain inflow boundary condition $\volinflowrate$. Every line connecting the axis represents one particle of the SMC solution and all lines are colored by their absolute extended posterior density value. The lines are drawn on top of each other, starting with low posterior density values in blue and ending with the highest ones in red. Most of the parameter combinations with high posterior density accumulate around $\youngs_2 \approx 200\sunit{Pa}$ and $\poissonratio_1, \poissonratio_2 > 0$. The other three parameters $\youngs_1, \youngs_3, \poissonratio_3$ seem to score relatively high posterior values for the whole range of interest. That means they are individually less significant to a specific value of the posterior density. Given the subdomain topology, as seen in \Cref{fig:subdomains}, these posterior results can also be expected as $\domain_2 $ takes up the largest portion of the biofilm and therefore can also be expected to have the highest influence on the deformation in this case.

The possibility for such a phenomenological interpretation of the posterior is a very attractive feature of tackling the inverse problem via the proposed approach, as not only good point estimates can be concluded, but more importantly the influence of the individual parameters on the posterior can be observed and interpreted. The information revealed in such a probablistic analysis, as for example displayed in \Cref{fig:heteronoiseuniuia2dofmargs} and \Cref{fig:papuni}, gives indeed a lot of insight into the problem at hand. Among others, it also shows how useful given experimental information is or whether other measurements are needed for the identification of relevant parameters. With the rich information from such an analysis also an estimation of the plausibility and stability of the point estimates can be made. As an example, a deterministic or trial and error approach might end up in identifying negative Poisson ratios for biofilms (easily understandable when looking at \Cref{fig:sc_posteriors_twodim} or \Cref{fig:papuni}) and hence identifying biofilms as auxetic materials, as it has been done in the past. The probabilistic analysis would immediately show the lack of validity for such a conclusion.

\section{Conclusion, Summary and Outlook}
\label{sec:concout}

We presented a robust and efficient Bayesian calibration approach for coupled computational mechanics models with given boundary or interface deformations. We considered the particularly challenging case, where only interface shapes based on images of the domain at different points in time or different experimental conditions are obtainable but no displacements of material points. Lacking such displacements, we also introduced and compared several metrics to calculate the discrepancy between interface shapes. We also considered the additional challenge to incorporate the influence of uncontrollable uncertainties in the experimental setup, such as uncertain boundary conditions

In contrast to deterministic optimization approaches for calibration tasks, Bayesian calibration is mathematically much more robust. This is because it is formulated in a probabilistic manner that describes the problem globally in form of a so-called posterior distribution. An optimization problem only results in one particular point estimate for the parameters instead. The latter is prone to get stuck in local optima which might not lead to satisfactory results.

The Bayesian calibration approach also allows for a more meaningful interpretation of the calibration result. The posterior density can be explored using a sequential Monte Carlo approach. This also allows for a convenient computation of the involved high-dimensional integrals or point estimates such as the maximum a posteriori (MAP) estimate. The posterior distribution gives rise to several further expressive point estimates, uncertainty and robustness measures, which help to get a deeper understanding of the parameters' effect on the solution. 

We have observed, that already for very few parameters, approximations, e.g., based on point estimates as the MAP and the Laplace approximation, cannot live up to the complexity of the resulting posteriors. Therefore, a full approximation of the posterior appears to be a necessary approach to studied inverse problems. That allows to gain further insight into the relevance of the parameters and the interaction of parameters in the model. Thereby a better understanding of the computational forward model in relation to the observed results of an experiment can be obtained.

To control the computational cost of the approach we proposed to build a Gaussian process (GP) surrogate model on the log-likelihood. The computationally most expensive step in the presented approach is the generation of the training data for the construction of the GP surrogate. The required forward model evaluations can however be computed in parallel and their exact amount is open to the analyst's choice. By constructing the log-likelihood surrogate, the approach is robust against single failing forward model evaluations, as it can be built on the remaining successful runs.

We presented and tested three different types of discrepancy measures between interfaces when no comparison of displacements of material points is possible. At first, an Euclidean distance measure at selected measurement points at characteristic locations on the interface was used. It is expected to be the easiest measure to apply to real data, as no full representation of the deformed geometry is required. Second, we used closest point projection distances for all interface nodes from the finite element discretization as a comparative measure. Eventually, the RKHS norm measure led to the most expressive posterior and is in general appealing because of its solid mathematical foundation and flexibility. Nevertheless, all measures yielded suitable likelihood distributions. A proper choice should be made based on the characteristics and quality of the experimental data. The choice of the distance measure is also dependent on the type of calibration task and on the question of which aspect of the deformation the analyst wants to emphasize.

For the demonstration of our calibration approach, we chose the material parameter identification of a spatially two-dimensional fluid-structure interaction (FSI) problem including a homogeneous hyperelastic solid biofilm model. We used artificial reference data such that we know the ground truth in order to be able to focus the attention on the characteristics of the proposed approach. We showed that the point estimates in the examples with two calibrated parameters were close to the ground truth $\gtx$ and the approximation of the full posterior allowed to observe the strong coupling between the input parameters in their influence on the posterior density. The posterior solution allowed us to observe the interaction of the two parameters in the model which in the present case was an almost linear relationship between $\youngs$ and $\poissonratio$ for which high posterior densities occurred in this setting. We furthermore studied the influence of uncertain experimental conditions on the posterior density. Such uncertainty led to a slightly flatter and therefore less expressive posterior such that the variance of the posterior increased and the posterior density filled the design space more equally. Besides the expected increase in variance also the expectation and MAP estimate deviated. Compared to the previous case, where the additional uncertainty was neglected, this shows that it is important to consider existing uncertainties. Otherwise the posterior might be overconfident and might have entirely different characteristics.

As a further example, we investigated a more complex calibration problem of a heterogeneous biofilm model with six unknown material parameters under uncertainty that also included artificial measurement noise. In this challenging example, the MAP estimate deviates more from the ground truth because the problem has more degrees of freedom and additional uncertainty and noise. This example, that could not be solved at all for an easier setting without uncertainty with a deterministic approach in \cite{me_willmann_2021}, could be solved and interpreted with the approach presented herein, hinting to higher robustness of the presented approach.

While the examples have been chosen from a particular field of application, the approach is general and can be applied to all sorts of single-field or coupled multi-field calibration problems with given boundary or interface deformations, where boundary deformations are available, but no point-to-point correspondence between simulation results and experimental observations can be found. This means that the presented approach can be used without limitations also for spatially three-dimensional models and more complex and expensive forward models.

In high parameter dimensions, other methods like variational Bayes approaches or multi-fidelity approaches such as \cite{lnm_stat_nitzler2020} are promising alternatives.

\appendix

\section{Flow cell experiments with biofilms and optical coherence tomography}
\label{asec:experimental_setup}

\begin{figure}[htb]
    \centering
    \includegraphics[width=0.8\textwidth]{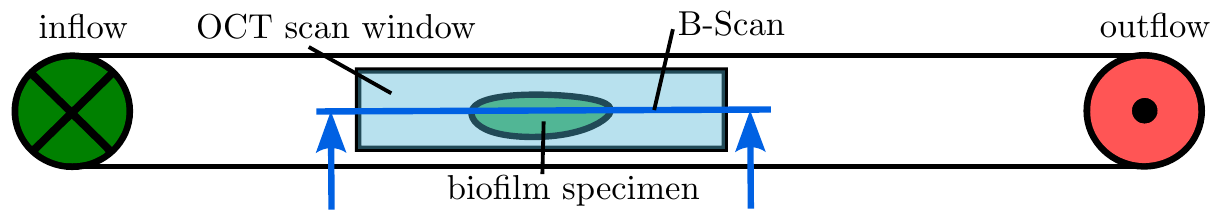}
    \caption{Schematic setup of a flow cell experiment with biofilms for measurement with optical coherence tomography (OCT).}
    \label{fig:experiment_scheme}
\end{figure}

This is a short introduction to flow cell experiments with biofilms. In a flow cell experiment, the biofilm is grown on the bottom of a channel with a flowing liquid that is providing nutrients to the microorganisms. The channel has dimensions like $50\sunit{mm}\times5\sunit{mm}\times0.45\sunit{mm} $ in recent experiments \cite{kit_bio_gierl2020} or $124\sunit{mm}\times2\sunit{mm}\times1\sunit{mm} $ in previous experiments \cite{kit_bio_blauert2015}. An exemplary flow cell channel is depicted in \Cref{fig:experiment_scheme} in top-down view. To simplify the illustration only one single bigger patch of biofilm is drawn. After a pronounced patch of biofilm is identified in the channel, a deformation experiment is conducted with the specimen \cite{kit_bio_blauert2015}. Such experiments are typical in biofilm research and hence we use them for parameter identification. Parameter identification is known to be very tricky for biofilms due to the soft consistency of the material and the necessity to keep the biofilm material in natural conditions. In an deformation experiment, mechanical load is applied onto the biofilm via the flowing liquid and then the deformation of the biofilm domain is measured.

To obtain accurate measurements of the biofilm boundary in undeformed reference configuration, the latter is first scanned using optical coherence tomography (OCT) without a significant fluid flow through the channel. Subsequently, a fluid flow is introduced into the channel at the inlet via a fluid pump, for which the volume flow rate can be controlled. We are not simply transferring this to defined load on the biofilm surface, e.g., constant shear stress, but take the full interaction between liquid and biofilm into account. This means that the resulting fluid forces acting at the immersed biofilm boundary deform the biofilm, which in return changes the surrounding fluid flow. This phenomenon is known as \emph{fluid-structure interaction (FSI)} or in this specific case as \emph{fluid-biofilm interaction (FBI)}.

The deformed biofilm is then again scanned via OCT. OCT can only be done in a \emph{scan window} as sketched in \Cref{fig:experiment_scheme}. This is important information for our approach as it also contributes to the uncertainty of conditions. This is because upstream and downstream from the analyzed patch as well as in the attached tubes and inflow and outflow areas there will be "invisible" and unregistered biofilm patches. The resulting snapshots of the deformed and undeformed biofilm represent the data of the flow cell experiments. If required, the experiment and snapshots are repeated for different flow settings or multiple points in time. In the literature (e.g., \cite{kit_bio_blauert2015, kit_bio_gierl2020, kit_bio_wagner2017}) one planar scan with OCT is labelled \emph{B-scan} as it is a combination of one dimensional \emph{A-scans} along the flow direction. By a combination of several \emph{B-scans} to \emph{C-scans} a three-dimensional volumetric representation of the biofilm is measured. This type of image data can then be used in the presented material parameter calibration approach.

Also due to the reduction of the whole channel to a planar view of a portion of the center of the channel, the fluid inflow boundary conditions in the planar consideration becomes prone to uncertainty. First, and as already mentioned, it is unknown if there are different unregistered biofilm patches upstream or downstream of the regarded domain. Practically the inverse analysis and biofilm modeling will focus on the biggest biofilm patch in the channel. Second, if the biofilm patch does not occupy the full channel width uniformly, fluid flow will partly pass the biofilm and therefore the average fluid flow rate cannot be assumed to flow over the patch. These considerations lead to the assumption that the flow rate in the two-dimensional model is uncertain and this type of uncertainty must be quantifiable in our approach.

\section{Fluid-solid interaction approach for modeling biofilm mechanics}
\label{asec:model}
In this work we model the previously described fluid-biofilm interaction of the flow cell experiments with biofilms as a coupled mechanical two-field problem. The experiments can result in significant deformations, displacements and rotations of the initial biofilm configuration, such that a nonlinear kinematic description of the biofilm displacement field is required.
The strong form of the associated differential equations for the fluid-solid interaction are briefly summarized for the fluid domain, the solid domain and the coupling of the two. For the sake of compactness, the presentation is limited to the field equations. The respective boundary conditions are implicitly assumed to be well defined. The interested reader is referred to the literature (e.g. \cite{lnm_fsi_gee_2010, lnm_bio_yoshihara2014}) for a methodological discussion of fluid-solid interaction models and their numerical discretization with a monolithic arbitrary Lagrangian-Eulerian (ALE) approach.

\subsection{Continuum description of the fluid field}
\label{asec:fluid_field}
The fluid field in the channel is well described by the incompressible Navier-Stokes equation for Newtonian fluids. We apply an arbitrary Lagrangian-Eulerian (ALE) description, which uses a moving fluid mesh and accounts for the resulting mesh displacements and velocities in the fluid domain, due to compatibility with the moving solid domain. The velocity of the fluid relative to the fluid mesh is then expressed by the ALE convective velocity $\alevel$.
The fluid equations in ALE formulation read as
\begin{subequations}
\label{eq:fluid_eq}
\begin{align}
\dens^\fluidletter\partialdt{\velf}+\densf(\alevel \cdot \grad) \velf -2 \dynvisf \div \tns{\epsilon} (\velf)+\grad \presf &= \densf \bodyforce^\fluidletter&& \text{in } \domain^\fluidletter\times(0,T)\\
\div \velf &= 0 &&\text{in } \domain^\fluidletter\times(0,T).
\end{align}
\end{subequations}
Wherein the variables of interest are the fluid velocity $ \velf $, the fluid pressure $ \presf $, the fluid density $ \densf $, the fluid dynamic viscosity $ \dynvisf $ and the body force $ \bodyforce^\fluidletter $ in the fluid domain $ \domain^\fluidletter\times(0,T) $.
The strain rate tensor $\eastrain(\velf)$ in \eqref{eq:fluid_eq} furthermore expands to the expression
\begin{equation}
\eastrain(\velf)= \frac{1}{2}\left( \grad \velf + \transpose{ \left( \grad \velf \right)}\right).
\end{equation}

\subsection{Continuum description of the solid field}
\label{asec:structure_field}
In this work, the biofilm is modeled as a nonlinear solid. In the solid domain, the continuum field is described by the nonlinear balance of momentum in reference configuration 
\begin{equation}
\initial{\dens}^\structletter \ddtq{\disps} =\divref (\defgrad \cdot \secpk) +\initial{\dens}^\structletter \initial{\bodyforce}^\structletter \hspace{4mm}\text{in } \initial{\domain}^\structletter\times(0,T).
\end{equation}
Here, the solid displacements $ \disps$ are used as primary variable. The other quantities are the deformation gradient $\defgrad $, the second Piola-Kirchhoff stress tensor $\secpk$ and the body force in reference configuration ${\bodyforce}^\structletter $. $\initial{\dens}^\structletter$ is the solid density in reference configuration. The balance equation is formulated for the initial structure domain in reference configuration $\initial{\domain}^\structletter\times(0,T) $.

\subsection{Fluid-solid interface and coupling}
\label{asec:coupling}

The coupling of the fluid domain and the solid domain is achieved by the interface conditions on the fluid-solid interface $\interface^{\fluidletter, \structletter} \times (0,T) $. Here, the balance of tractions $\tns{h}^\structletter_\interface $ between fluid and solid phase and a no-slip interface condition on the respective primary variables need to hold
\begin{subequations}
\begin{align}
\tns{h}^\structletter_\interface &=-\tns{h}^\fluidletter_\interface &&\text{on }\interface^{\fluidletter, \structletter} \times (0,T),\\
\partialdt{\disps}&=\velf_\interface &&\text{on }\interface^{\fluidletter, \structletter} \times (0,T).
\end{align}
\end{subequations}
Consequently, in \Cref{fig:computational_domain} the initial biofilm interface is labeled $\interface^{\fluidletter, \structletter}_0 $ which is the same as the interface in reference configuration.

\subsection{Numerical discretization of the fluid-solid interaction problem}

For all of the discussed problems in this article, the numerical discretization of the governing equations is conducted with the finite element method (FEM). We use a monolithic approach to solve the coupled FSI equations in ALE formulation \cite{lnm_fsi_gee_2010, lnm_bio_yoshihara2014}, as the approach was shown to be well suited for biomedical problems of similar type \cite{lnm_fsi_kuettler2010} and has also been applied in the biofilm setting before \cite{lnm_bio_taherzadeh2010}.

\section{Reproducing Kernel Hilbert Space Details}
\label{asec:rkhsdetails}

The \emph{reproducing kernel} property implies that the inner product of the kernel and a function reproduces the original function, according to
\begin{equation}
\label{eq:repod_property}
\bs{f}(\bs{y})=\langle \bs{f}(\bs{x}),\bs{k}(\bs{x},\bs{y})\rangle_{\mathcal{H}}.
\end{equation}

The kernel $\bs{k}(\bs{x},\bs{y})$ must be a positive definite function and is unique for the associated RKHS. For our investigations we furthermore require the kernel to be symmetric such that $\bs{k}(\bs{x},\bs{y})=\bs{k}(\bs{y},\bs{x})$, with $\bs{k}(\cdot, \bs{y})\in \mathcal{H}$. 
Symmetry allows us to interpret the kernel function as a correlation function in a statistical sense, which is commonly done in probabilistic machine learning \cite{stat_rasmussen2005,stat_duvenaud2014}.
A kernel can be generated by an associated \emph{feature map} $\bs{\phi}:\mathcal{X}\rightarrow\mathcal{H}$ in the sense of $\bs{k}(\bs{x},\bs{y}):=\langle \bs{\phi}(\bs{x}),\bs{\phi}(\bs{y})\rangle_{\mathcal{H}}$. The feature map $\bs{\phi}$ can be an infinite dimensional vector, or an infinite series, respectively. Often, only a resulting valid reproducing kernel $\bs{k}$ is known and the associated feature map $\bs{\phi}$ remains unknown (kernel trick). Loosely speaking, a desirable property of an RKHS is that a small value of the inner product of the distance function $\bs{d}$ implies also point-wise closeness of associated functions $\bs{f}_1$ and $\bs{f}_2$ \cite{stat_aronszajn1950,stat_berlinet2004}.
The choice of kernel function encodes the smoothness and complexity assumptions about the underlying functions in the inner product. The latter are usually given in a discrete point representation and the continuous function is represented via the kernel function according to the well known \emph{representer theorem}, which directly demonstrates the interpretation of the kernel as basis functions for the underlying function or curve
\begin{equation}
    \label{eq:representer_theorem}
    \bs{f}=\sum_{i=1}^{n}\alpha_i \bs{k}(\cdot,\bs{x}_i).
\end{equation}
Therefore it can be represented by linear combinations with factors $\alpha_i $. An extensive overview of common kernels and kernel algebra can be found in \cite{stat_duvenaud2014}. We only briefly demonstrate the \emph{radial basis function (RBF)} kernel, which results in smooth, infinitely differentiable functions and is also used in our investigations
\begin{equation}
\kernelscalar{\bs{x}}{\bs{y}} = \flatexp{- \frac{\norm{\bs{x}-\bs{y}}{2}^2}{2\sigmaw^2}}.
\label{eq:surfcurrkernel}
\end{equation}

The RBF kernel with variance $\sigmaw^2 $ describes a mapping $\mathbb{R}^{n\times n} \rightarrow \mathbb{R}$, such that it is treated as a scalar $k(\bs{x},\bs{y}) $ in the following. After the presentation of the general concepts of inner products in RKHS, we will now discuss possible definitions of distance measures. For the sake of simplicity, we investigate two-dimensional parameterized curves $\boldsymbol{f}(\boldsymbol{s})$, but the concepts generalize easily to three spatial dimensions and parameterized representation of surfaces. An arbitrary curve can be fully described by the parameterized vector representation in \eqref{eq:param_fun}. We can calculate the unit-normal vectors of the curve by differentiation w.r.t. the parameter $\boldsymbol{s}$ as demonstrated in \eqref{eq:param_normal}.
\begin{subequations}
\begin{align}
\label{eq:param_fun}
    \boldsymbol{f}(\boldsymbol{s})&=
    \begin{bmatrix}
    g(\boldsymbol{s}), &  h(\boldsymbol{s})\\
    \end{bmatrix}^T\\
    \label{eq:param_normal}
  \boldsymbol{n}_{\boldsymbol{f}}(\boldsymbol{s})&=\frac{1}{\sqrt{\left(\frac{d h(\boldsymbol{s})}{d\boldsymbol{s}}\right)^2+\left(\frac{d g(\boldsymbol{s})}{d\boldsymbol{s}}\right)^2}}
  \begin{bmatrix}
    -\frac{d h(\boldsymbol{s})}{d\boldsymbol{s}},& \frac{d g(\boldsymbol{s})}{d\boldsymbol{s}}\\ 
  \end{bmatrix}^T
\end{align}
\end{subequations}

Given the parameterization in \eqref{eq:param_fun} a natural choice for a distance is given by the inner product
\begin{subequations}
\begin{align}
\label{eq:f_rkhs}
\dmeas_{\mathcal{V},\bs{f}}&=\langle \bs{d_f},\bs{d_f}\rangle,\\
\text{with } \kernelscalar{\bs{f}_{1}}{\bs{f}_{2}}&=\flatexp{- \frac{\norm{\bs{d_f}}{2}^2}{2\sigmaw^2}}.
\end{align}
\end{subequations}
In contrast to the closest point projection or $\text{L}_2$ norms, inner products in RKHS correlate all discretized points of curve $\bs{f}_1$ with all discretized points of curve $\bs{f}_2$, such that it is not important to identify specific point pairs for which the measure is calculated. While, \eqref{eq:f_rkhs} is a valid distance measure, it might not account enough for different complexities of $\bs{f}_1$ and $\bs{f}_2$ if the overall distance in the Hilbert space is small. This is why we do not follow this approach in this paper.

To put more emphasis on the difference in functional complexity, one can incorporate derivatives of the curves in the measure. Hence, another special choice of distance measure, that will also be used in this paper, can be constructed with the difference in the vector function of the normal vectors $\bs{n}_{\bs{f}_1}$ and $\bs{n}_{\bs{f}_2}$ from \eqref{eq:param_normal}, instead of the difference of the functions themselves. 

\section{Brief presentation of used Gaussian process regression model}
\label{asec:gp_regression}

A Gaussian process (GP) is fully defined by its mean function $ \meanf{x}$ and a correlation function $\kernelscalar{\bullet}{\bullet}$ (or kernel, see \cref{sec:measure} or \cite{stat_duvenaud2014, stat_rasmussen2005}) and describes a distribution over functions $f(\bx)$, such that for a fixed $\hat{\bx}$, the function value $f(\hat{\bx})$ is normally distributed according to $f(\hat{\bx}) \sim  \mathcal{N}\left(f|\meanf{\hat{\bx}},\kernelx{\hat{\bx}}{\hat{\bx}}\right)$. A realization of a Gaussian process (GP) can be written as
\begin{equation}
\label{eq:prior_gp}
f(x) \sim \GP{\meanf{\bx}}{\kernelscalar{\bx}{\bx'}}.
\end{equation}
As the name suggests, the mean function $ \meanf{\bx}$ represents the statistical mean for an infinite amount of function realizations $f_i(\bx)$. The kernel or covariance function $\kernelscalar{\bx}{\bx'}$ encodes the correlation of two function values $f(\bx)$ and $f(\bx')$ at different inputs $\bx$ and $\bx'$. In analogy to the RKHS in \cref{sec:measure}, the choice of kernel hence encodes smoothness, complexity and characteristics assumptions of the underlying function. 

In case we did not account for any data $\mathcal{D}=\{\Xtrain,\ytrain\}$ in \eqref{eq:prior_gp}, this expression can be interpreted as the so-called \emph{prior} GP. The specific selection of $ \meanf{x}$ and $\kernelscalar{\bx}{\bx'}$ can be used to integrate prior knowledge when GPs are used for regression tasks. In this work we will restrict the considerations to a constant prior mean function $\meanf{\bx}=const$.

\begin{remark}[Prior mean for logarithmic likelihood]
For our application, the prior mean function $\meanf{\bx}$ cannot be neglected, i.e., set to zero, as the log-likelihood for Bayesian calibration \eqref{eq:loglik}, that we want to build the regression model for, cannot just randomly be set zero, as then the associated likelihood tends towards a finite value ($\sim \exp{(0)}$) far away from the training points $\xtrain_i$. Strictly the log-likelihood should tend towards negative infinity for irrelevant regions, which is unfeasible, so that the likelihood tends towards zero. Therefore we define an auxiliary mean that is lower than any occurring log-likelihood in the training data $\meanf{\bx} = \minof{\ltrain} - 1.0\cdot\left( \maxof{\ltrain}-\minof{\ltrain}\right) $ with the log-likelihoods $\ltrain$ as training outputs.
\end{remark}

In this work we choose a Matérn $3/2$ kernel function \cite{stat_rasmussen2005,stat_stein1999, stat_matern1960} for the regression model of the log-likelihood function. In contrast to the infinite differentiable radial basis function kernel used in \eqref{eq:surfcurrkernel}, the Matérn $3/2$ kernel is a one time differentiable covariance function \cite{stat_rasmussen2005} that does hence result in samples $f(\bx)$ with lower smoothness requirement. As the log-likelihood might potentially be peaked in areas with high posterior probability density or might also evince abrupt functional changes, we want to relax the smoothness requirements on the regressor. The Matérn $3/2$ kernel function is defined as
\begin{equation}
\kfun{\mathrm{Matern},\eta = 3/2}{\tns{x}}{\tns{x}'} = \kernelvar \left(1+\frac{\sqrt{3} \norm{\tns{x}-\tns{x}'}{2}}{\kernellen}\right)\exp{\left(-\frac{\sqrt{3} \norm{\tns{x}-\tns{x}'}{2}}{\kernellen}\right)}.
\label{eq:matern32}
\end{equation}
The hyper-parameters $\kernelvar$ and $\kernellen$ control the magnitude of the covariance and its length scale, respectively.

Typically there is \emph{training data} $\mathcal{D}=\{\Xtrain,\ytrain\}$ available and the prior GP can be \emph{conditioned} on $\mathcal{D}$ to yield the \emph{posterior} GP, whose posterior mean function $\meanftest(\xtest)$ is then used as a regression model. Training data are input-output pairs of values that are known or can be determined systematically. The posterior variance function $\variance{\ftest}(\xtest)$ serves as a measure for the uncertainty in the regression model. Here it is assumed that $\ytrain=\transpose{\begin{bmatrix}y_1,\dots,y_{\ntrain}\end{bmatrix}}$ consists of $\ntrain$ scalars (log-likelihood $\ltrain$ in the presented approach) at potentially vector-valued training inputs $\Xtrain=\{\xtrain_{,i}\}\big\vert_{i=1}^{\ntrain}$ (forward model parameters $\xtrain$ in presented approach). The test point $\xtest$ denotes a new input for the prediction of the regression model. The posterior mean and variance function can be calculated from the prior GP and the training data $\mathcal{D}$ using the following expressions \cite{stat_rasmussen2005}
\begin{subequations}
\label{eq:posterior_gp}
\begin{align}
\meanftest &= \transpose{\ktest}\inverse{\left(\gpcovmat + \noisevar \identity\right)} \left(\ytrain-\meanfvec\right), \label{subeq:gpmean}  \\
\variance{\ftest} &= \kernelx{\xtest}{\xtest}-\transpose{\ktest}\inverse{\gpcovmat}\ktest.
\label{subeq:gpvar} 
\end{align}
\end{subequations}
In \eqref{eq:posterior_gp} we use the abbreviations $\gpcovmat = \kernelscalar{\Xtrain}{\Xtrain}$ for the covariance matrix at training inputs and the vector $\ktest = \kernelscalar{\Xtrain}{\xtest}$ for kernel evaluations at test point and training data. $\meanfvec$ is a vector of suitable length $\ntrain $ populated with the constant prior mean $\meanf{\bx}$ in all entries. The so-called nugget noise variance $\noisevar$ is used for numerical stability of the GP \cite{stat_rasmussen2005}.

The optimization of the so-called \emph{marginal likelihood} or \emph{evidence} of the GP w.r.t. the hyper-parameters $\kernelvar, \kernellen$ in \eqref{eq:matern32} is known as the \emph{training} of the Gaussian process. The log-marginal likelihood expresses the likelihood of the training data under the chosen model parameterization and it is given by
\begin{equation}
\begin{gathered}
- \log{\pdenscond{\mathcal{D}}{\kernelvar, \kernellen}} =\\= \frac{1}{2} \transpose{\left(\ytrain-\meanfvec\right)}\inverse{\gpcovmat} \left(\ytrain-\meanfvec\right) + \frac{1}{2} \log{\left|\gpcovmat\right|} + \frac{\ntrain}{2}\log{2 \pi}
\label{eq:gploglikeli}
\end{gathered}
\end{equation}
For numerical reasons one usually minimizes the negative log-marginal likelihood instead of maximizing the marginal likelihood directly.

\section*{Acknowledgements}
Funding of the project with project number WA 1521/22 for this work by the German Research Foundation (DFG) is gratefully acknowledged. Furthermore, the research was partly funded by the German Research Foundation (DFG) under the priority program SPP 1886 \emph{Polymorphic uncertainty modeling for the numerical design of structures}. SB gratefully acknowledges funding from the German Research Foundation (Deutsche Forschungsgemeinschaft, DFG) - Projektnummer 386349077. A base version of the software QUEENS was provided by AdCo Engineering\textsuperscript{GW} GmbH, which is gratefully acknowledged. The first implementation of Gaussian processes and further infrastructure in QUEENS done by B. Wirthl is gratefully acknowledged.

\printbibliography

\end{document}

%% file: symbols.tex
\newcommand{\tns}[1]{\boldsymbol{#1}} 

\newcommand{\norm}[2]{\ensuremath{\left\|#1\ensuremath\right\|_{#2}}}

\newcommand{\grad}{\boldsymbol{\nabla}}
\renewcommand{\div}{ \grad \! \cdot \!}
\newcommand{\divref}{ \initial{\grad} \! \cdot \!}

\newcommand{\identity}{\tns{I}}

\newcommand{\flatexp}[1]{\mathrm{exp}\left(#1\right)}
\newcommand{\diffd}{\mathrm{d}}

\newcommand{\dirac}{\delta}
\newcommand{\maxof}[1]{\mathrm{max}\left(#1\right)}
\newcommand{\minof}[1]{\mathrm{min}\left(#1\right)}

\global\long\def\transpose#1{#1^{\mathsf{T}}}
\global\long\def\inverse#1{#1^{\mathsf{-1}}}

\newcommand{\initial}[1]{{#1}_0}
\newcommand{\imin}[1]{{#1}_{\mathrm{min}}}

\newcommand{\paramvec}{{\tns{x}}}

\newcommand{\lmnmp}{{n_\mathrm{\mpindex}}}

\newcommand{\nin}{{n_\mathrm{in}}}
\newcommand{\mpindex}{\mathrm{mp}}
\newcommand{\cppindex}{\mathrm{cpp}}
\newcommand{\scindex}{{\mathrm{sc}}}

\newcommand{\obsindex}{\mathrm{obs}}

\newcommand{\defgrad}{\tns{F}}

\newcommand{\secpk}{\tns{{S}}}

\newcommand{\eastrain}{\tns{\epsilon}}

\newcommand{\bodyforce}{\hat{\tns b}}

\newcommand{\youngs}{E}
\newcommand{\poissonratio}{\nu}

\newcommand{\domain}{\Omega}

\newcommand{\interface}{\Gamma}

\newcommand{\disp}{d}
\newcommand{\vel}{v}
\newcommand{\pres}{p}

\newcommand{\dens}{\rho}
\newcommand{\dynvis}{\mu}

\newcommand{\unit}[1]{\mathrm{#1}}

\newcommand{\sunit}[1]{\,\mathrm{#1}}
\newcommand{\sunitfrac}[2]{\,{\unit{#1}/\unit{#2}}}

\newcommand{\ddtq}[1]{\dfrac{\mathrm{d}^2 {#1}}{\mathrm{d} t^2}}
\newcommand{\partialdt}[1]{\dfrac{ \partial #1}{\partial t}}

\newcommand{\structletter}{{\mathrm{S}}}

\newcommand{\disps}[1][]{\tns{\displacements}^{\structletter #1}}
\newcommand{\displacements}{\disp}

\newcommand{\fluidletter}{{\mathrm{F}}}

\newcommand{\velf}{\tns{\vel}^\fluidletter}
\newcommand{\presf}{\pres^\fluidletter}

\newcommand{\densf}{\dens^{\fluidletter}}
\newcommand{\dynvisf}{\dynvis^{\fluidletter}}

\newcommand{\volinflowrate}{\dot{V}_{\mathrm{in}}}

\newcommand{\alevel}{\tns{c}^\fluidletter}

\newcommand{\dmeas}{\mathfrak{D}}

\newcommand{\coords}{C}
\newcommand{\yobs}{Y_{\obsindex,\coords}}
\newcommand{\ymod}{Y_{\coords}}

\newcommand{\forwardmodelletter}{\mathfrak{M}}
\newcommand{\model}[1]{\forwardmodelletter\left(#1\right)}
\newcommand{\signeddist}{d}
\newcommand{\sigman}{\sigma_{\mathrm{N}}}
\newcommand{\btheta}{\boldsymbol{\theta}}
\newcommand{\groundtruthindex}{\mathrm{gt}}
\newcommand{\postmeanindex}{\mathrm{PM}}
\newcommand{\mapindex}{\mathrm{MAP}}
\newcommand{\gtx}{\paramvec_{\groundtruthindex}}
\newcommand{\xpm}{\paramvec_{\postmeanindex}}
\newcommand{\xmap}{\paramvec_{\mapindex}}

\newcommand{\tricenter}{\tns{c}}
\newcommand{\trinormal}{\tns{n}}

\newcommand{\kernelscalar}[2]{k\left(#1,#2\right)}

\newcommand{\sigmaw}{\sigma_\mathrm{W}}

\newcommand{\GPletter}{\mathcal{GP}}
\newcommand{\GP}[2]{\GPletter \left(#1,{#2}\right)}
\newcommand{\kfun}[3]{\mathrm{k}_{#1}\left(#2,{#3}\right)}
\newcommand{\kernelx}[2]{\kfun{x}{#1}{#2}}
\newcommand{\trainindex}{\mathrm{train}}
\newcommand{\xtrain}{{\tns{x}_\trainindex}}
\newcommand{\ytrain}{\tns{y}_\trainindex}
\newcommand{\ltrain}{\tns{\loglikeliletter}_\trainindex}
\newcommand{\xtest}{\tns{x}_{*}}
\newcommand{\meanftest}{\bar{f}_{*}}
\newcommand{\ftest}{f_{*}}
\newcommand{\ktest}{\tns{k}_{*}}
\newcommand{\noisestd}{\sigma_\mathrm{n}}
\newcommand{\noisevar}{\noisestd^2}
\newcommand{\gpcovmat}{\tns{K}}
\newcommand{\kernelvar}{{\sigma_\mathrm{k}^2}}
\newcommand{\kernellen}{{l_\mathrm{k}}}
\newcommand{\ntrain}{n_\trainindex}
\newcommand{\ntest}{n_{\mathrm{test}}}
\newcommand{\Xtrain}{X_\trainindex}
\newcommand{\Ds}{\mathcal{D}}

\newcommand{\pdens}[1]{p\left(#1\right)}
\newcommand{\pdenscond}[2]{\pdens{#1|#2}}
\newcommand{\Expectsymbol}{\mathbb{E}}

\newcommand{\Expectwrt}[2]{{\Expectsymbol}_{#1}\left[#2\right]}
\newcommand{\meanf}[1]{\mathrm{m}\left(#1\right)}
\newcommand{\meanfvec}{\tns{\mathrm{m}}}
\newcommand{\Variancesymbol}{\mathbb{V}}
\newcommand{\variance}[1]{\Variancesymbol\left[{#1}\right]}
\newcommand{\normaldist}[2]{\mathcal{N}\left(#1,{#2}\right)}
\newcommand{\betadist}[3]{\mathcal{B}\left(#1|#2,#3\right)}
\newcommand{\lognormaldist}[3]{\mathcal{LN}\left(#1|#2,#3\right)}
\newcommand{\unidist}[3]{\mathcal{U}\left(#1|#2,#3\right)}

\newcommand{\prior}{\pdens{\paramvec}}
\newcommand{\loglikeliletter}{\mathfrak{L}}
\newcommand{\loglikelimodel}[1]{\loglikeliletter \left(#1\right)}

\newcommand{\smczeta}{\zeta}
\newcommand{\nparticles}{N}
\newcommand{\ess}{\mathrm{ESS}}
\newcommand{\smcstep}{s}
\newcommand{\smcapprox}{\pi}
\newcommand{\smcweights}{w}
\newcommand{\smcgamma}{\gamma}
\newcommand{\smcnumrejsteps}{n_\mathrm{rej}}
\newcommand{\smcnumsteps}{n_\mathrm{SMC}}
\newcommand{\smcppost}{\smcapprox}
\newcommand{\smcnormweights}{w}
\newcommand{\smcpindex}{{\left(i\right)}}

\newcommand{\svggdfin}{\interface^\fluidletter_\mathrm{in}}
\newcommand{\svggdfout}{\interface^\fluidletter_\mathrm{out}}
\newcommand{\svgdnsf}{\interface^\fluidletter_\mathrm{D}}
\newcommand{\svgdnss}{\interface^\structletter_\mathrm{D}}

\newcommand{\bs}[1]{\boldsymbol{#1}}
\newcommand{\bx}{\bs{x}}